\documentclass[AMA,STIX1COL]{WileyNJD-v2}

\usepackage{lineno,hyperref}
\usepackage{amsmath}
\usepackage{amsmath}
\usepackage{amssymb}
\usepackage{amsfonts}
\usepackage{graphicx}
\usepackage{subfigure}
\usepackage{algorithm}

\articletype{Article Type}%

\received{26 April 2016}
\revised{6 June 2016}
\accepted{6 June 2016}

\begin{document}

\title{Sampling-based 3-D line-of-sight PWA model predictive control for autonomous rendezvous and docking with a tumbling target}

\author[1]{Dongting Li}

\author[2]{Rui-Qi Dong}

\author[1]{Yanning Guo}

\author[1]{Guangtao Ran}

\author[3]{Dongyu Li}

\authormark{D. Li \textsc{et al}}

\address[1]{\orgdiv{Department of Control Science and Engineering}, \orgname{Harbin Institute of Technology}, \orgaddress{\state{Harbin 150001}, \country{P.R.China}}}

\address[2]{\orgdiv{Department of Precision Instrument}, \orgname{Tsinghua University}, \orgaddress{\state{Beijing 100084}, \country{P.R.China}}}

\address[3]{\orgdiv{School of Cyber Science and Technology}, \orgname{Beihang University}, \orgaddress{\state{Beijing 100083}, \country{P.R.China}}}

\corres{Yanning Guo, Department of Control Science and Engineering, Harbin Institute of Technology, Harbin 150001, P.R.China. \email{guoyn@hit.edu.cn}}


\abstract[Abstract]
{
This article introduces a line-of-sight (LOS)-Euler rendezvous and docking (RVD) framework to dock with a tumbling target under several RVD constraints. 
By a double-loop control scheme, the chaser's position is controlled to track the target's docking port which is coupled with its rotation. The chaser's attitude is driven to track the target's rotation, while satisfying the field of view constraint which is coupled with the LOS range.  These complex couplings are linearly described in the proposed framework.
To this end, the 6 DoF information interactions among the sensor measurements, states, and RVD constraints need no transformation and linearization.
Consider the online piecewise affine (PWA) model predictive controller (MPC) may be unsolvable under complex constraints, which is caused by the accumulated prediction error,
a sampling-based method is proposed. The linear predictions are driven to the closer neighborhoods of the actual nonlinear states by constructing directional sampling intervals.
Besides, a singularity free strategy is provided to realize continuous tracking with crossing the singularities of angle states.
Detailed numerical simulations illustrate the validity of the proposed methods.
}



\maketitle


\section{Introduction}
\label{INTRO}
Autonomous rendezvous and docking (RVD) is indispensable for on-orbital service, which contributes to extending the service period of target spacecraft, improving the functions of service spacecraft, and removing debris or obstacles.  Several existing papers have studied to drive a chaser spacecraft to dock with a cooperative target. \cite{Davis2004, Kawano2001, Whelan2000}  However, there is no prior information during docking with a noncooperative tumbling target, which requires the visual sensor equipped on the chaser to point to the target in real time. Besides, the desired position of the chaser, i.e., the target's docking port, is time-varying and coupled with its rotation. To this end, the complex position-attitude couplings leads to RVD with a noncooperative tumbling target is more challenging. 
  
Several control methods are proposed to drive a chaser to dock with a target spacecraft. Guo et al. adopted a finite-time controller to realize RVD with a noncooperative target with considering collision avoidance.\cite{Guo2021}   Dong et al. developed an adaptive controller based on a time-varying sliding manifold to realize a safety control by proposing a potential function to describe safety requirements.\cite{Dong2017}
Gao et al. provided a time-synchronized stability controller and a fixed-time-synchronized stability controller for a 6-DoF RVD.\cite{Gao2021} 
Liu et al. studied the fixed-time cooperative tracking problem for delayed disturbed systems which could be applied in the multi-spacecraft system.\cite{Liu2022, Liu2022a}
By employing dual-quaternion dynamics, Dong et al. proposed a 6-DoF control method to guarantee cooperative control during the chaser arrives at the target's docking port.\cite{dong2018dual} 
All these papers achieved promising control performance. However, the study on RVD with a tumbling target with position-attitude coupled RVD constraints is still insufficient. 

Due to the advantage of achieving optimal performance while responding to various constraints, model predictive control (MPC) \cite{Mayne2014Model, mayne2000survey} is widely employed in different engineering problems. Several studies have also employed the MPC scheme in RVD missions.\cite{li2017model, li2018model, gavilan2012chance-constrained, weiss2015model, di2012model, Dong2022, Guo2021a, liu2022robust}
Consider the information interaction between the relative navigation information and the controlled states,
the vision or laser information (e.g., relative distance, elevation angle, azimuth angle, et al.) needs nonlinear transformation to the controlled states.\cite{li2017model, li2018model, weiss2015model}
Such complex transformations usually increase the computational cost and time delay. 
To simplify such transformation, Li et al. adopted the 2-D line-of-sight (LOS) dynamics to control an active spacecraft to rendezvous with a stable attitude spacecraft.\cite{Li2017} 
However, the research on combining 3-D LOS dynamics with MPC to realize RVD with a tumbling target is still insufficient.  For attitude control problem,  the quaternion-based attitude dynamics is combined with the MPC framework.\cite{GOLZARI2020105677, 8267197} 
However, the transformation between the field of view and unit quaternion is too complex to apply in the MPC framework.

Consider the information interaction between the RVD constraints and the controlled states, which eventually relate to the quadratic programming (QP) problem in MPC.
Since a convex QP problem can always be solved to guarantee real-time performance, which requires the index to be quadratic and the constraints are described by linear inequality or equality, most papers convert their optimization index into a QP problem.
Thus the nonlinear RVD constraints were linearized in the papers mentioned above. 
The quadratic collision avoidance constraint was linearized since the relative distance was represented by the 2-norm of the coordinates.\cite{li2017model}
The azimuth angle was linearized for nonlinear trigonometric operations.\cite{weiss2015model} 
The entry cone was linearized by an inscribed pyramid, and the entry cone constraint was achieved by keeping the inner product of the position vector and the normal external vectors of the pyramid's side is negative.\cite{gavilan2012chance-constrained, weiss2015model} Note that the constraints are related to the physical states measured by the sensors directly, which means the measurements are transformed to the states, and then inverse-transformed to the RVD constraints.

Additionally, the QP index function subject to a nonlinear dynamics is not always convex.
Constrained nonlinear programming problems are solved in nonlinear MPC through several optimization algorithms (e.g., sequential quadratic programming\cite{kelman2011bilinear},  genetic algorithm\cite{du2016development} et al.).
However, using these algorithms in spacecraft control can not guarantee real-time performance.
Thus most papers convert the nonlinear system to a piecewise affine (PWA) system\cite{lazar2006stabilizing, petsagkourakis2020stability} or a linear time-varying (LTV) system.\cite{falcone2008linear, de2004mpc, wen2009analytical}
Similarly, Korda M. et al. adopted the Koopman operator-based data-driven method to convert the nonlinear model to a lifted linear system.\cite{2020Optimal, 2016Linear, zhang2022robust} 
However, the model deviations between prediction models and control models caused by the accumulated predictive error may result in unsolvable optimization under complex constraints.

Moreover, there exist singularities of angle states,\cite{chaturvedi2011rigid} which means different values represent the same physical position.
For a continuous tracking problem, the mathematical expression of the desired state switches once it reaches a singularity. However, it challenges an input constrained system to overcome such unnecessary switching for the tracking error since the error does not jump physically.
Although there exist non-singular attitude descriptions (e.g., unit-quaternion, Modified Rodriguez Parameters, et al.), there still exists a double cover characteristic in unit-quaternion called unwinding,\cite{Dong2021, dong2021anti} and complex nonlinear transformations exist between the field of view constraint and the controlled states.
Besides, even the singularities of attitude angles could be avoided, the singularities of the elevation angle and the azimuth angle still can not be ignored.

\textbf{\emph{Contributions:}} As discussed above, the information interaction from the sensor measurements to states, and from states to RVD constraints in the literature is a process of "transformation to inverse-transformation." Such process increases the computational cost and time delay. Even the control accuracy of online PWA MPC is reduced once the transformations are highly nonlinear. Inspired by these limitations, the contributions of this article are summarized as follows:

\noindent (i) A novel LOS-Euler RVD control framework is proposed. In contrast to existing papers, complex position-attitude couplings are linearly described, and the 6 DoF information interactions among the sensor measurements, controlled states, and RVD constraints need no transformation and linearization, which contributes to reducing the computational cost and matching better with PWA MPC.
 
\noindent (ii) The problem that PWA MPC may be unsolvable under complex constraints caused by the accumulated prediction error is considered and analyzed. 
Then a sampling-based idea is proposed to construct directional sampling intervals to drive linear predictions to the closer neighborhoods of the actual nonlinear states.
Theoretical analysis of the proposed method applied in the SISO and MIMO system are also provided. Besides, the control inputs are imposed on the actual nonlinear dynamics rather than the PWA models.

\noindent (iii) A singularity free strategy is proposed to realize continuous tracking by crossing the singularities of the angle states.

\textbf{\emph{Organization:}} The remainder of this study is arranged as follows. The proposed LOS-Euler RVD framework is formulated in Section 2. Section 3 presents the proposed sampling-based approach and analyzes the effectiveness theoretically, then the PWA MPC is designed based on the proposed approach. Section 4 illustrates the singularity free strategy. In Section 5, three different cases are provided to show the effectiveness of the proposed approaches. Section 6 sums up the conclusions.  

\textbf{\emph{Notation:}} Variable definition rules in this article are introduced. The subscript "p" defined in this article represents "position",  the subscript "a" represents "attitude". For $i = a,\ p$, denote $\pmb{x}_i(k) \in \mathbb{R}^{6}$ as the controlled state at instant $k$, $\pmb{x}_i^{+} (k) \in \mathbb{R}^{6}$ as the ahead prediction at instant $k+1$, $\pmb{x}^{\ast}_i(k) \in \mathbb{R}^{6N_p}$ as the augment vector of the predictions during the instant interval $[ k+1,\ \cdots, k+N_p]$,  $\pmb{x}^{\ast}_{d,i} (k)$ as the desired states of $\pmb{x}^{\ast}_i(k)$, $\pmb{u}^{\ast}_{i}(k) \in \mathbb{R}^{3N_c}$ as the augment vector of the control sequence during the instant interval $[ k,\ \cdots, k+N_c-1]$, $\Delta \tilde{\pmb{u}}_i(k) \in \mathbb{R}^{{3N_c}} $ as the augment vector of the increment of control sequence, where $N_p$ is the prediction horizon and $N_c$ is the control horizon.

\section{Formulation of LOS-Euler RVD framework}
In this section, the LOS-Euler RVD framework is developed.  Section \ref{Coordinate reference frames} introduces the coordinate reference frames. Section \ref{6-DOF dynamical model} presents the dynamics, including the 3-D LOS dynamics and the Euler dynamics with a specific rotation sequence. Section \ref{Construction_of_constraints} re-constructs the RVD constraints formulated in the LOS-Euler frame. Section \ref{Advantages of LOS-Euler RVD framework} analyzes the characteristics of the proposed framework and presents the control objective.
\subsection{Coordinate reference frames}
\label{Coordinate reference frames}
\begin{figure}[tbh]
\centering
\includegraphics[width=0.35\textwidth]{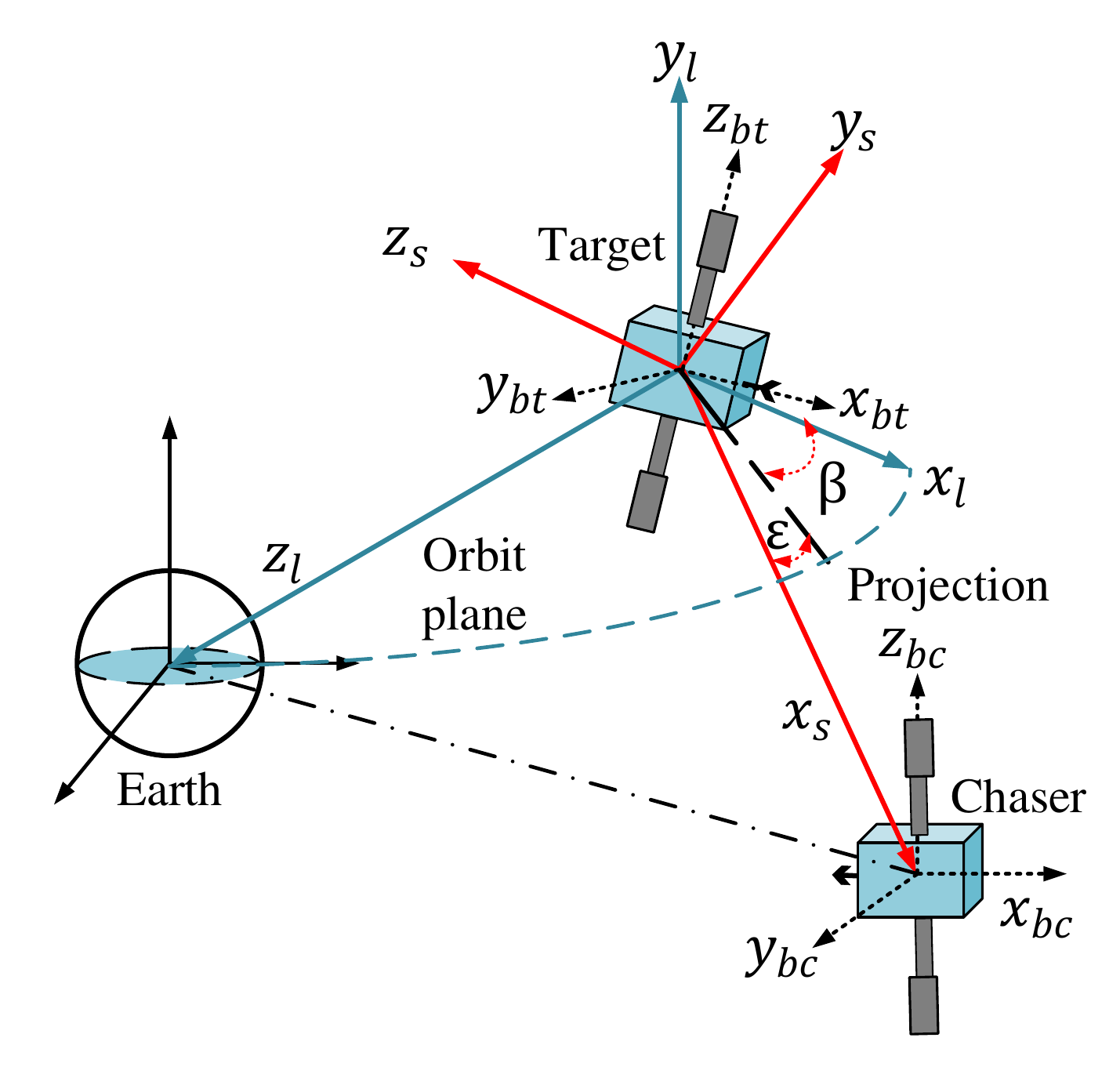}
\caption{Coordinate reference frames}
\label{frame}
\end{figure}

As illustrated in Fig. \ref{frame}, related coordinate frames are firstly defined as follows.
\begin{enumerate}[\textbullet]
\item $\mathcal{F}_{l} = \{ O_{t},\ \hat{x}_{l},\ \hat{y}_{l},\ \hat{z}_{l} \}$: the local-vertical-local-horizontal (LVLH) frame, its origin is fixed at the center of mass of the target $O_t$, $\hat{z}_l$-axis points towards the center of the Earth, $\hat{y}_l$-axis is along the direction of the orbital angular rate, $\hat{x}_l$-axis completes the triad.
\item $\mathcal{F}_{bt} = \{ O_{t},\ \hat{x}_{bt},\ \hat{y}_{bt},\ \hat{z}_{bt} \}$ and $\mathcal{F}_{bc} = \{ O_{c},\ \hat{x}_{bc},\ \hat{y}_{bc},\ \hat{z}_{bc} \}$: the body-fixed coordinate frames of the target and the chaser.
\item $\mathcal{F}_s = \{ O_{t},\ \hat{x}_s,\ \hat{y}_s,\ \hat{z}_s \}$: the LOS frame, its origin is fixed at $O_t$, $\hat{x}_s$-axis points towards the chaser, $\hat{y}_s$-axis is along the direction of angular momentum of $\hat{x}_s$, $\hat{z}_s$ completes the triad. Denote $\varepsilon \in ( - \pi/2,\ \pi/2)$ as the elevation angle between $x_s$ and its projection on $ O_{t} \hat{x}_{l} \hat{z}_{l}$, denote $\beta \in ( - \pi,\ \pi)$ as the azimuth angle between the projection of $x_s$ and $x_l$. According to Fig. \ref{frame}, it can be concluded that $\mathcal{F}_s$ is obtained by the Euler rotation of $\mathcal{F}_{l}$ with $2\ \beta$-$3\ \varepsilon$-$1$ sequence.
\end{enumerate}

\subsection{Dynamical models of relative motion} 
To the best knowledge of the authors, despite the LOS dynamics and Euler dynamics already exist, existing papers have not integrated them with the same rotation sequence to deal with the complex position-attitude couplings and simplify the 6 DoF information interactions analyzed in Section \ref{INTRO}.
\label{6-DOF dynamical model}

\subsubsection{3-D LOS dynamics} 
The relative translation formulated in the LOS frame $\mathcal{F}_{s}$ is described as follows \cite{Yoon2009,wang2019integrated},
\begin{equation}
\left\{
\begin{array}{l}
 \ddot{\rho} - \rho \dot{\varepsilon}^2 - \rho \left( \dot{\beta} - \omega \right)^2
   \cos^2 \varepsilon = \frac{-\mu}{R_t^3} \left(\rho - 3 \rho \cos^2
   \varepsilon \sin^2 \beta \right) + u_{p1} + d_{1},\\
    \rho \ddot{\varepsilon} + 2 \dot{\rho} \dot{\varepsilon}  +  \rho
   \left( \dot{\beta} - \omega \right)^2 \sin \varepsilon \cos \varepsilon = \frac{-3
   \mu}{R_t^3} \rho \sin \varepsilon \cos \varepsilon \sin^2 \beta   +
   u_{p2}+ d_{2},\\
 \rho \left( \ddot{\beta}  - \dot{w} \right) \cos \varepsilon +  2 \dot{\rho} (\dot{\beta}
   -  \omega) \cos \varepsilon - 2 \rho \dot{\varepsilon} \left(\dot{\beta}  -
   \omega\right) \sin \varepsilon  
 = \frac{3\mu}{R_t^3}  \rho \cos \varepsilon \sin
   \beta \cos \beta - u_{p3}- d_{3},
\end{array}
\right.  
\label{position dynamics}
\end{equation}
where $R_t = a \left( 1-e^2 \right) / \left( 1+e \cos \nu \right)$ denotes the distance between the target and the Earth, $a$ denotes the semi-major axis, $ \nu$ donotes the true anomaly, $e$ represents the eccentricity, $\omega= \sqrt{\mu a \left(1 - e^2 \right)} / R^2_t$ denotes the time derivative of $\nu$, $\dot{\omega}= -2 \mu e \sin \nu / R_t^3$ is the time derivation of $\omega$, $\mu$ is the gravity constant. $\rho$ is the LOS range, $\varepsilon $ and $\beta$ represent the elevation angle and the azimuth angle, respectively. For $i= 1,\ 2,\ 3$, $d_i$ denotes the disturbance. Note that $\rho$, $\varepsilon $, and $\beta$ can be measured by the navigation system directly without any coordinate transformation.

By denoting $\pmb{x}_p$ = $\left[ x_{p1},\ x_{p2},\ x_{p3},\ x_{p4},\ x_{p5},\ x_{p6} \right]^{\mathrm{T}}$ =$\left[ \rho,\ \varepsilon,\ \beta,\ \dot{\rho},\ \rho \dot{\varepsilon},\ \rho\dot{\beta} \right]^{\mathrm{T}}$,  then the nominal system of (\ref{position dynamics}) is converted to the following first-order continuous differential equation,
\begin{eqnarray}
  \pmb{\dot{x}}_p(t) = A_p^{\mathrm{c}} \left( \pmb{x}_p \right) \pmb{x}_p(t) + B_p^{\mathrm{c}} \left( \pmb{x}_p\right) \pmb{u}_p(t)  , 
\label{position state}
\end{eqnarray}
where $\pmb{u}_p$ denotes the control input acting on the chaser, described by
\begin{equation}
\pmb{u}_p = \left[ u_{p1},\ u_{p2},\ u_{p3} \right]^{\mathrm{T}}.
\label{up}
\end{equation}
$A_p^{\mathrm{c}} (\pmb{x}_p)$ and $B_p^{\mathrm{c}} (\pmb{x}_p)$ are time-varying state-dependent matrices, i.e.,
\begin{equation*}
A_p^{\mathrm{c}} = \left[\begin{array}{cccccc}
     0 & 0 & 0 & 1 & 0 & 0\\
     0 & 0 & 0 & 0 & \frac{1} { x_{p 1}} & 0\\
     0 & 0 & 0 & 0 & 0 & \frac{ 1}  {x_{p 1}}\\
     a_{41} & 0 & 0 & 0 & \frac{x_{p 5}}  {x_{p 1}} & -a_{64}\cos ^2 x_{p2}\\
     \frac{\left(- \omega^2 R_t^3 - 3 \mu \sin^2 x_{p 3}\right)\sin 2 x_{p2} } {2 R_t^3} & 0 & 0 &  \frac{-x_{p 5}}  {x_{p 1}} & 0 & \frac{a_{64}\sin 2 x_{p2} }{2}\\
      \frac{2 \dot{\omega} R_t^3 + 3 \mu \sin 2x_{p 3}} { 2R_t^3}  & 0 & 0 &  a_{64}  & \frac{2 \tan x_{p 2} \left( -\omega x_{p 1} + x_{p 6} \right)}{x_{p 1}}  & 0
   \end{array}\right],\
   B_p^{\mathrm{c}} = \left[\begin{array}{ccc}
     0 & 0 & 0\\
     0 & 0 & 0\\
     0 & 0 & 0\\
     1 & 0 & 0\\
     0 & 1 & 0\\
     0 & 0 & -\frac{1} { \cos x_{p2}}
   \end{array}\right],
\end{equation*}
where $a_{41} = \omega^2 \cos^2 x_{p 2} - \frac{\mu}{R_t^3}  \left(1 - 3 \cos^2
  x_{p 2} \sin^2 x_{p 3} \right)$, and $a_{64} = 2 \omega - \frac{x_{p 6}}{x_{p1}}$.

\subsubsection{Euler attitude kinematics and dynamics}

Common description of the chaser's attitude is by the rotation from chaser's orbital frame to $\mathcal{F}_{bc}$. However, in close RVD ($\rho \ll R_t$), it can be approximated that the chaser's orbital frame is equivalent to the target's orbital frame $\mathcal{F}_{l}$. Then define the chaser's kinematics as the Euler rotation from $\mathcal{F}_{l}$ to $\mathcal{F}_{bc}$ using the same $2$-$3$-$1$ sequence with the LOS frame, i.e.,
\begin{equation}
 \left[\begin{array}{c}
     \dot{\phi}_c (t) \\
     \dot{\theta}_c (t)\\
     \dot{\psi}_c (t)
   \end{array}\right] =  \left[\begin{array}{ccc}
     1 & - \cos \left(\phi_c\right) \tan \left(\psi_c\right) & \sin \left(\phi_c\right) \tan \left(\psi_c\right)\\
     0 & \cos \left(\phi_c\right) \sec \left(\psi_c\right) & - \sin \left(\phi_c\right) \sec \left(\psi_c\right)\\
     0 & \sin \left(\phi_c\right) & \cos \left(\phi_c\right)
   \end{array}\right] \left[\begin{array}{c}
     \omega_{c1} \\
     \omega_{c2} \\
     \omega_{c3} 
   \end{array}\right]
\label{attitude kinematics},
\end{equation}
where $ \phi_c (t)$, $\theta_c (t)$, and $\psi_c (t) \ (\mathrm{rad})$ denote the chaser's roll, pitch and yaw angles, respectively, $ \omega_{ci}$ ($i=1,\ 2,\ 3$) are the chaser's angular velocities. 
To avoid the gimbal lock phenomenon, the pitch is set within  $( - \pi/2,\ $ $ \pi/2)$, roll and yaw are set within $[ - \pi,\ \pi]$.
\begin{remark}
If approximating the chaser's orbital frame as $\mathcal{F}_{l}$ can not be neglected, since the rotation matrix from chaser's orbital frame to $\mathcal{F}_{l}$ is known, the attitude definition in this paper is still valid.
\end{remark}

The attitude dynamics is formulated as follows,
\begin{equation}
\left\{
\begin{array}{l}
  J_1 \dot{w}_{c1} = (J_2 - J_3) \omega_{c2} \omega_{c3} + M_1,\\
  J_2 \dot{w}_{c2} = (J_3 - J_1) \omega_{c1} \omega_{c3} + M_2,\\
  J_3 \dot{w}_{c3} = (J_1 - J_2) \omega_{c1} \omega_{c2} + M_3,
\end{array}
\right.
\end{equation}
where $ J_1$, $J_2$, and $J_3 $ denote the chaser's principal moments of inertia, and $ M_1$, $M_2$, and $M_3 $ denote the input moments. 

Consider the reaction wheels equipped along each principal body axis as the actuators, the relation between the wheels' dynamics and the moments of the chaser is defined as
\begin{equation}
\left\{
\begin{array}{l}
  M_1 = - \tilde{J}_1 (\dot{\omega}_{c1} + \ddot{\alpha}_1 + \dot{\alpha}_3
  \omega_{c2} - \dot{\alpha}_2 \omega_{c3}) \backsimeq - \tilde{J}_1 (\dot{\omega}_{c1}
  + \ddot{\alpha}_1),\\
 M_2 = - \tilde{J}_2 (\dot{\omega}_{c2} + \ddot{\alpha}_2 + \dot{\alpha}_1
  \omega_{c3} - \dot{\alpha}_3 \omega_{c1}) \backsimeq - \tilde{J}_2 (\dot{\omega}_{c2}
  + \ddot{\alpha}_2),\\
 M_3 = - \tilde{J}_3 (\dot{\omega}_{c3} + \ddot{\alpha}_3 + \dot{\alpha}_2
  \omega_{c1} - \dot{\alpha}_1 \omega_{c2}) \backsimeq - \tilde{J}_3 (\dot{\omega}_{c3}
  + \ddot{\alpha}_3),
\end{array}
\right.
\end{equation}
where $ \tilde{J}_1$, $\tilde{J}_2$, and $\tilde{J}_3$ denote the wheels' moments of inertia, $ \dot{\alpha}_1$, $\dot{\alpha}_2$, and $ \dot{\alpha}_3 $ denote the speed of wheels.
Consider the following linearized relationship between the chaser's angular velocities and the wheels' acceleration,
\begin{equation}
 \dot{\omega}_{ci} = -\frac{\tilde{J}_i}{J_i} {\ddot{\alpha}}_i,\ i = 1,\ 2,\ 3 .
\label{wheels}
\end{equation}

By denoting $\pmb{x}_a$ = $\left[x_{a1},\ x_{a2},\ x_{a3},\ x_{a4},\ x_{a5},\ x_{a6}\right]^{\mathrm{T}}$ = $\left[ \phi_c  ,\ \theta_c  ,\ \psi_c  ,\ \omega_{c1} 
,\ \omega_{c2}  ,\ \omega_{c3}  \right]^{\mathrm{T}}$, then (\ref{attitude kinematics})-(\ref{wheels}) are converted to the following first-order differential equation,
\begin{equation}
\dot{\pmb{x}}_a(t) = A_a^{\mathrm{c}} \left( \pmb{x}_a \right) \pmb{x}_a(t) + B_a^{\mathrm{c}} \pmb{u}_a(t),
\label{attitude state}
\end{equation}
where $\pmb{u}_a$ denotes the control input acting on the chaser, described by
\begin{equation}
\pmb{u}_a = \left[ u_{a1} ,\ u_{a2} ,\ u_{a3} \right]^{\mathrm{T}}.
\label{ua}
\end{equation}
$A_a^{\mathrm{c}} (\pmb{x}_a)$ is a time-varying state-dependent matrix with
\begin{equation*}
A_a ^{\mathrm{c}} = \left[\begin{array}{cccccc}
     0 & 0 & 0 & 1 & - \cos x_{a1} \tan x_{a3} & \sin x_{a1} \tan x_{a3}\\
     0 & 0 & 0 & 0 & \cos x_{a1} \sec x_{a3} & - \sin x_{a1} \sec x_{a3}\\
     0 & 0 & 0 & 0 & \sin x_{a1} & \cos x_{a1} \\
     0 & 0 & 0 & 0 & \frac{x_{a6} (J_2 - J_3)} { J_1}  & 0\\
     0 & 0 & 0 & 0 & 0 & \frac{x_{a4} (J_3 - J_1)} { J_2} \\
     0 & 0 & 0 & \frac{x_{a5} (J_1 - J_2)} { J_3} & 0 & 0
   \end{array}\right],\
   B_a ^{\mathrm{c}} = \left[\begin{array}{ccc}
     0 & 0 & 0\\
     0 & 0 & 0\\
     0 & 0 & 0\\
    \frac  {\left( \tilde{J}_1^2 - J_1 \tilde{J}_1 \right)}{{J_1}^2} & 0 & 0\\
     0 &  \frac{\left( \tilde{J}_2^2 - J_2 \tilde{J}_2 \right)} {{J_2}^2} & 0\\
     0 & 0 &  \frac{\left(\tilde{J}_3^2 - J_3 \tilde{J}_3\right)} {{J_3}^2}
   \end{array}\right].
\end{equation*}

\subsection{Re-construction of RVD constraints}
\label{Construction_of_constraints}
To meet the requirements of safe docking, observing the noncooperative target in real-time, and other practical needs, it is necessary to construct the RVD constraints.
As introduced in Section \ref{INTRO}, the control accuracy and the efficiency benefit from simplifying the transformation between the states and the constraints since the QP solver requires linear constraints.
Hence the re-constructed RVD constraints are directly imposed on the states in the LOS-Euler framework.
The advantages are especially obvious once the constraints are extended to time-varying and coupled cases.

\subsubsection{Control input constraints}
Consider the limited torque generated by the actuators, the constraints of the thrusters and the reaction wheels are described by
\begin{equation}
 |  {u}_{pi}  |  \leqslant {u}_{pi}^{\max},\ i=1,\ 2,\ 3
\label{input_p}
\end{equation}
and
\begin{equation}
 | {u}_{ai} | \leqslant {u}_{ai}^{\max},\ i=1,\ 2,\ 3
\label{input_a}
\end{equation}
where ${u}_{pi}$ and ${u}_{ai}$ are described in (\ref{up}) and (\ref{ua}), $\pmb{u}_p^{\max}$ = $ \left[u_{p1}^{\max},\ u_{p2}^{\max},\ u_{p3}^{\max} \right]^{\mathrm{T}}$ and $\pmb{u}_a^{\max}$ = $\left[ u_{a1}^{\max},\ u_{a2}^{\max},\  u_{a3}^{\max} \right]^{\mathrm{T}}$ denote the maximum control force of the thrusters and reaction wheels, respectively.

\subsubsection{Collision avoidance constraint}
For safe docking, collision avoidance is realized by maintaining the chaser outside a keep-out sphere around the target, with the fixed safe radius described by $r_{\mathrm{safe}}$. For the relative translation formulated by C-W equations, the keep-out zone is nonlinear ($\sqrt{x^2+y^2+z^2}\geqslant r_{\mathrm{safe}}$, where $x,y,z$ are relative coordinates), which reduces the control accuracy for the linearization. In the LOS-Euler framework, the constraint is imposed on the state of the LOS range directly, yields
\begin{equation}
\rho \geqslant r_{\mathrm{safe}}
\label{r_safe}.
\end{equation}


\subsubsection{Entry cone constraint}
\label{subsubsectionentry}
\begin{figure}[tbh]
\centering
\subfigure{\label{entry111}}\addtocounter{subfigure}{-2}
\subfigure{\subfigure[Entry cone]{\includegraphics[width=0.23\textwidth]{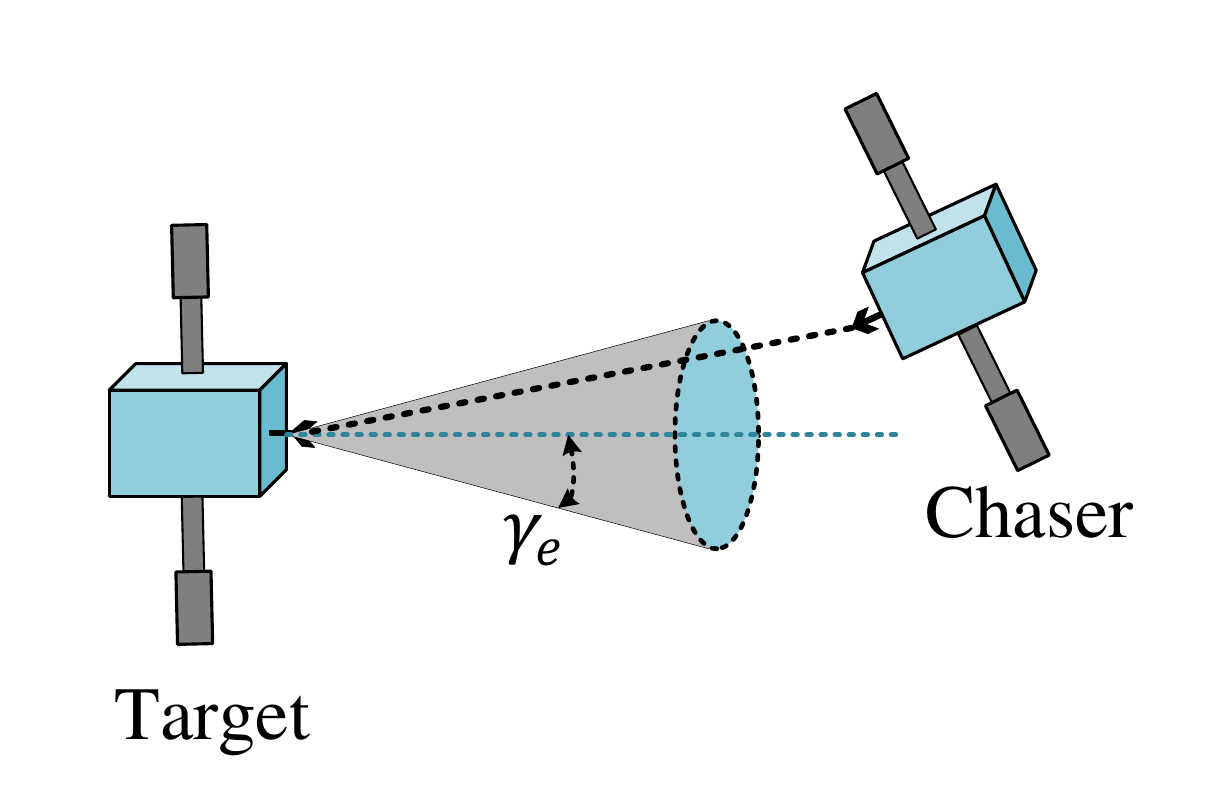}}}
\subfigure{\label{field111}}\addtocounter{subfigure}{-2}
\subfigure{\subfigure[Field of view]{\includegraphics[width=0.23\textwidth]{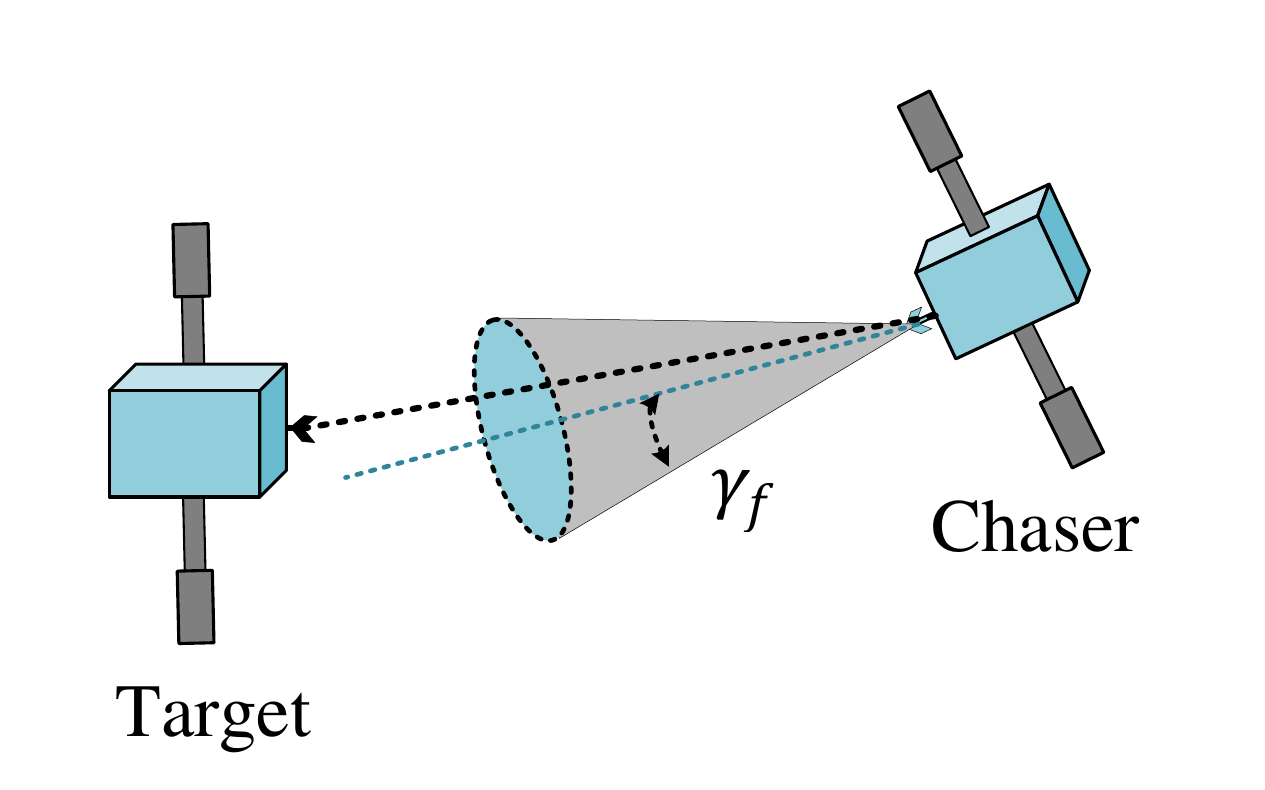}}}
\caption{Entry cone and the field of view constraints}\label{xingzuotu}\vspace{-1em}
\end{figure}

As shown in Fig. \ref{entry111}, the chaser should be kept in an approaching cone around the target's docking port during the final phase of docking. 
Such constraint is approximated by a four (six)-sided pyramid in most existing papers \cite{gavilan2012chance-constrained,weiss2015model}, which increases the computation cost and reduce the control accuracy. 
In this paper, the modeling is improved by imposing the constraint on the states directly to avoid linearization and transformation.
Additionally, a more challenging problem is extended and settled that the entry cone is time-varying and the position of the target's docking port is coupled with its rotation if the target is tumbling.

Assume: (i) the target's docking port is fixed at the $\hat{x}_{bt}$-axis in $\mathcal{F}_{bt}$; (ii) the attitude of the target is described by Euler rotation from $\mathcal{F}_{l}$ to $\mathcal{F}_{bt}$ with $2$-$3$-$1$ sequence; (iii) half of the entry cone angle is $\gamma_{\mathrm{e}}$. By defining the same rotation sequence from $\mathcal{F}_{l}$ to $\mathcal{F}_s$ and $\mathcal{F}_{l}$ to $\mathcal{F}_{bt}$, the desired elevation is equivalent to the target's pitch $\theta_t(t)$, and the desired azimuth is equivalent to the target's yaw $ \psi_t(t) $. Then the constraint can be imposed on the states directly, yields
\begin{equation}
\left\{
\begin{tabular}{l}
$ \theta_t (t)- \gamma_{\mathrm{e}} \leqslant \varepsilon\leqslant \theta_t(t) + \gamma_{\mathrm{e}}$,\\
$ - {\pi}/{2} < \varepsilon< \pi/{2}$,\\
$\psi_t(t) - \gamma_{\mathrm{e}} \leqslant \beta  \leqslant  \psi_t(t) + \gamma_{\mathrm{e}}$,\\
$ -\pi  \leqslant\beta  \leqslant  \pi$,\\
\end{tabular}
\right.  \label{entry cone}
\end{equation}
where $\varepsilon$ and $\beta$ are adopted states. The above relations can be converted to 
\begin{equation}
\left\{
\begin{tabular}{l}
$ \varepsilon^{\min}(t) < \varepsilon < \varepsilon^{\max}(t)$,\\
$ \beta^{\min}(t) \leqslant  \beta \leqslant   \beta^{\max}(t)$,
\label{entry_1}
\end{tabular}
\right.
\end{equation}
where $\varepsilon^{\min}(t) = \max\{- {\pi}/{2}, \ \theta_t (t)- \gamma_{\mathrm{e}} \}$, $\varepsilon^{\max}(t) = \min \{ \pi/{2}$, $ \theta_t(t) + \gamma_{\mathrm{e}}\}$, $\beta^{\min}(t) = \max\{-\pi,\  \psi_t(t) - \gamma_{\mathrm{e}} \}$, and $\beta^{\max}(t) = \min \{ \pi,\ \psi_t(t) + \gamma_{\mathrm{e}} \}$ are time-varying functions. Thus the coupling between the time-varying position of the target's docking port and its rotation is also reflected in (\ref{entry_1}).

\subsubsection{Field of view constraint}
\begin{figure}[tbh]
\centering
\includegraphics[width=0.35\textwidth]{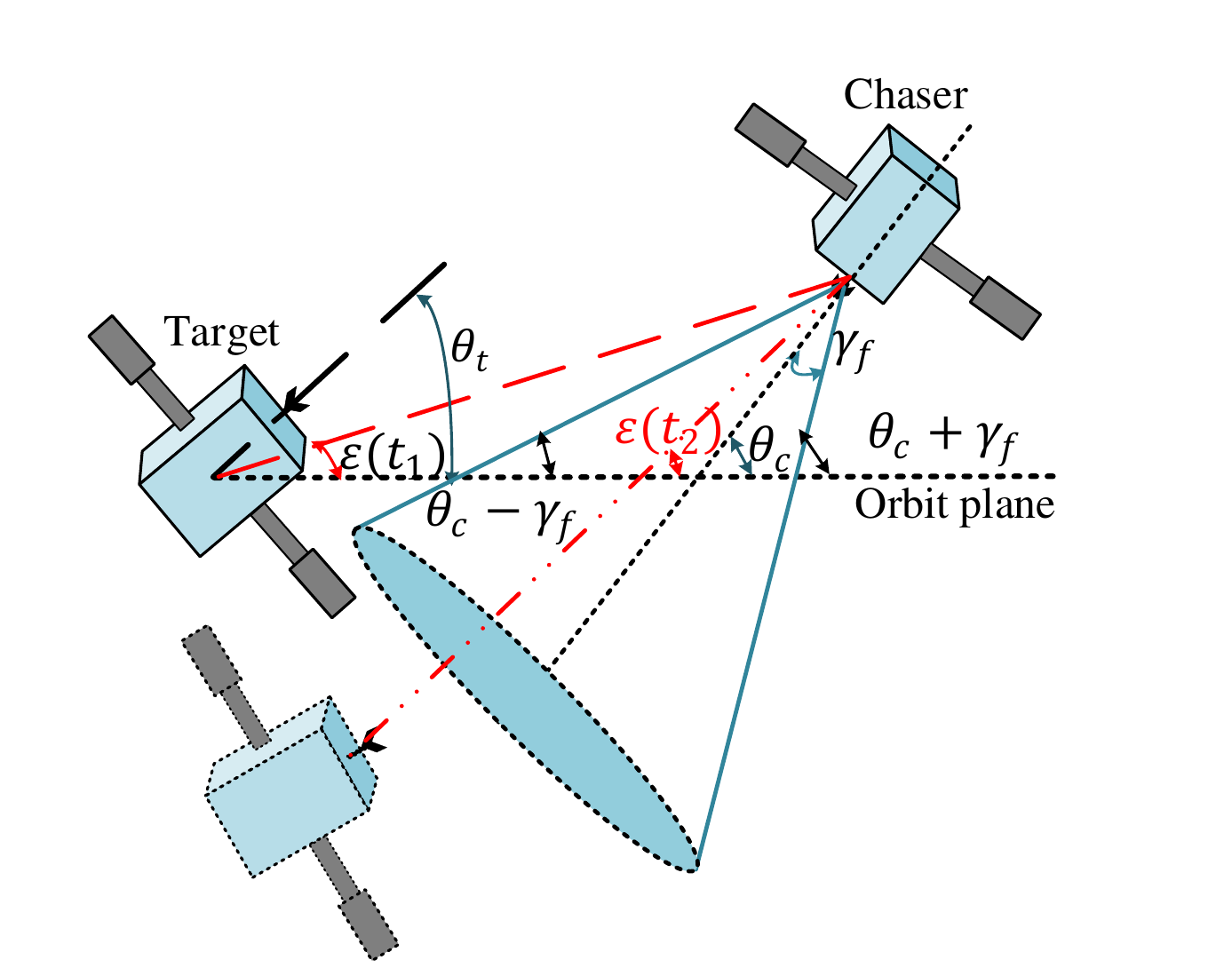}
\caption{Field of view constraint on the pitch angle}
\label{field1}
\end{figure}

As shown in Fig. \ref{field111}, for an active chaser equipped with a camera, it should be guaranteed that the vision sensors obtain the target's information, which is achieved by imposing the constraints on the chaser's attitude angles. For the rotation formulated by unit quaternion \cite{GOLZARI2020105677, 8267197}, the transformation between the quaternion and the attitude angles are complex and highly nonlinear, due to the complex triangular transformations and linearization. Besides, the constraint is position-attitude coupled, since it is achieved by keeping the target's pitch $ \theta_c$ and yaw $\psi_c $ in the cone around the elevation and azimuth, respectively, which yields
\begin{equation}
\left\{
\begin{tabular}{l}
$ \theta_c^{\min}(t) < \theta_c< \theta_c^{\max}(t)$,\\
$ \psi_c^{\min}(t) \leqslant \psi_c  \leqslant  \psi_c^{\max}(t)$,
\end{tabular}
\right.
\label{field_1}
\end{equation}
where $\theta_c^{\min}(t)$ $= \max\{- {\pi}/{2},\  \varepsilon(t) - \gamma_{\mathrm{f}} \}$, $\theta_c^{\max}(t) = \min \{ \pi/{2}, \ \varepsilon(t) + \gamma_{\mathrm{f}}\}$, $\psi_c^{\min}(t) = \max\{-\pi,\   \beta(t) - \gamma_{\mathrm{f}} \}$, $\psi_c^{\max}(t) = \min \{ \pi,\  \beta(t) + \gamma_{\mathrm{f}} \}$ are time-varying functions, and $\gamma_{\mathrm{f}}$ denotes half of the cone angle. 
For understanding convenience, the constraint of the pitch angle $\theta_c$ is taken as an example to illustrate detailed. As shown in Fig. \ref{field1}, the following relation should be guaranteed, i.e.,
  \[ \theta_c  - \gamma_{\mathrm{f}} \leqslant \varepsilon(t)  \leqslant \theta_c  + \gamma_{\mathrm{f}}. \]
The above relation can be rewritten as
 \[ \varepsilon(t)  - \gamma_{\mathrm{f}} \leqslant \theta_c  \leqslant \varepsilon(t) + \gamma_{\mathrm{f}}. \]
Thus the position-attitude coupling in the field of view control is reflected in (\ref{field_1}). 

\subsection{Control objective}
\label{Advantages of LOS-Euler RVD framework}
In summary, the proposed LOS-Euler RVD framework contains: (i) the LOS dynamics and the Euler dynamics described by the rotation from $\mathcal{F}_{l}$ to $\mathcal{F}_{bc}$ with $2$-$3$-$1$ sequence; (ii) re-construction of RVD constraints; (iii) using $2$-$3$-$1$ rotation sequence from $\mathcal{F}_{l}$ to $\mathcal{F}_{bt}$ to represent the tumbling target's attitude. The framework can achieve: (i) better matching with navigation since no transformation exists between the navigation information and the states; (ii) better matching with MPC since no transformation exists between the constraints and the states; (iii) linearly reflect the position-attitude coupling between the target's docking port position and its rotation; (iv) linearly reflect the position-attitude coupling between the field of view and the LOS range.  
However, there exist drawbacks of the framework but could be solved in this paper: (i) the dynamics are nonlinear, which is converted to the PWA model to employ the PWA MPC framework; (ii) the interaction between the position-attitude actuators, which is considered as disturbance; (iii) there exists singularities in the attitude, azimuth, and elevation angles which are settled in Section \ref{RVD strategy}.

In this paper, the control objective is to drive the chaser to track the time-varying position and attitude of the target's docking port by employing MPC.
In the relative position control loop, the relative distance, elevation, and azimuth are driven to track the time-varying desired distance, target's pitch, and target's yaw in real-time.
For the relative attitude control loop, the field of view of the chaser is controlled to cover the LOS range in real-time during the long-distance rendezvous phase. The chaser's attitude is driven to track the tumbling target's attitude in the short-distance docking phase.
Meanwhile, all the RVD constraints should be satisfied.

\section{Sampling-based PWA model predictive control}

MPC is widely employed to achieve optimal performance under various complicated constraints. 
 \emph{Prediction}: At each sampling instant, the ahead predictions over a finite prediction horizon are generated by the dynamical model. 
 \emph{Optimization}: Take the predictions into a constrained optimization index, converting the index to a convex QP problem, then minimizing it online to get a control input sequence.
 \emph{Control}:  According to the receding horizon mechanism, only the input signal related to the current instant is adopted.

\subsection{Problem statement}

Consider the dynamics in the LOS-Euler framework are nonlinear, nonlinear MPC (NMPC) is unsuitable for real-time RVD for solving a non-convex optimization, which requires far more computation resources and relies on local solutions.  Since PWA model can approximate the nonlinear dynamics using a set of piecewise linear models, PWA MPC scheme is employed due to linear MPC (LMPC) is still applicable to the PWA model, and the computation cost is far lower than NMPC  on the premise that the optimization is convex. 

Note that the state matrix and the control input matrix are constant in LMPC but switched in PWA MPC. 
During the prediction process, once the varying state and control input matrices are obtained by recursive linearization at each prediction state,
the predictive error will result in deviations between the prediction models and the actual control models, which finally affect the control performance of PWA MPC.
To this end, the calculated control sequence by MPC affected by the accumulated prediction error could result in the practical controlled states not satisfying the original constraints, then the MPC may be unsolvable.
    
\begin{figure}[tbh]
\centering
\includegraphics[width=0.7\textwidth]{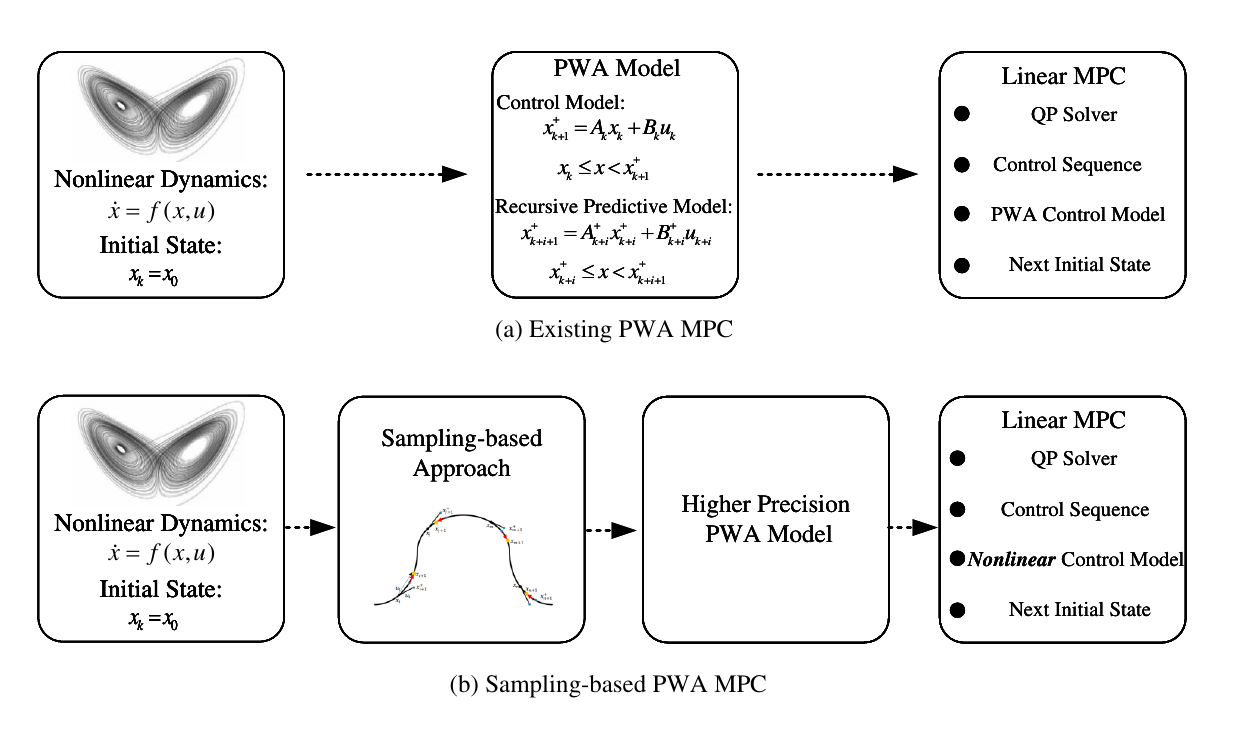}
\caption{Problem statement}
\label{S3_framework}
\end{figure}

As shown in Fig. \ref{S3_framework}, during each prediction horizon $[k, k+N_p]$ in PWA MPC, the control model is linearized at each current actual state $x_k$, the prediction models used to predict the evolution are linearized at the predictions $ x^{+}_{k+i} (i\in [1, N_p-1])$. 
Unlike with LMPC that the prediction models are equivalent to the control models at the corresponding actual instants,
the prediction models have deviation with the corresponding actual control models, because of the error that the linear predictions approximating their corresponding actual nonlinear states. Note that the model deviation usually increases with the multi-step recursion.
Hence,  a sampling-based approach that provides a higher-precision PWA model is presented by reducing the accumulated prediction error. 
Besides, the nonlinear dynamics rather than the PWA model is adopted as the control model.




\subsection{Sampling-based approach: reducing predictive error}
\label{continuous approach}
The basic idea of reducing the prediction error is to judge the relation between the ahead prediction and the corresponding nonlinear state by the convexity information of the nonlinear dynamics. Then constructing a directional sampling interval to drive the prediction to a closer neighborhood of the actual state.
\subsubsection{SISO system}
\label{SISO}
\begin{figure}[tbh]
\centering
\includegraphics[width=0.4\textwidth]{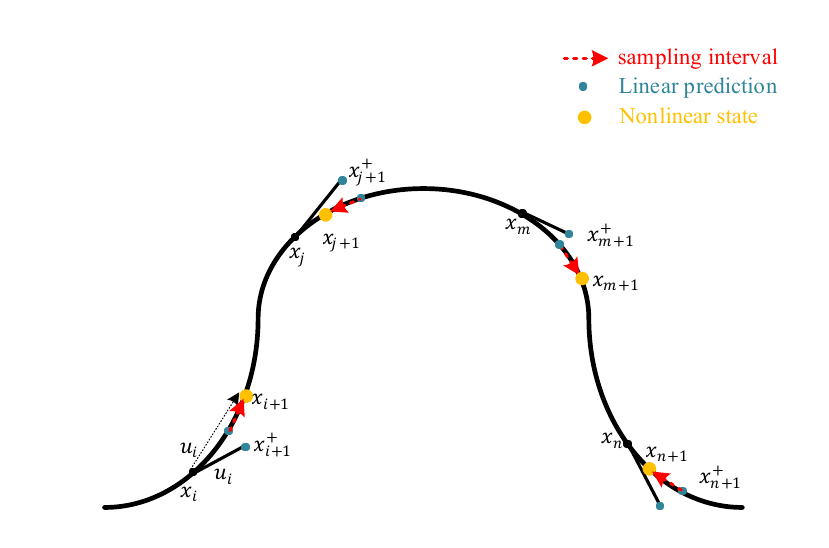}
\caption{Basic idea of sampling}
\label{sampling}
\end{figure}

As shown in Fig. \ref{sampling}, consider the continuous SISO nonlinear system $\dot{x} = f(x,u)$, where $x$ is the state and $u$ represents the control input, linearize the system at $(x_0, u_0)$ to obtain the first-order approximate model corresponding to an ahead state interval, i.e.,
\begin{equation}
\delta \dot{x}_l = f^{'}_{x|(x_0,u_0)} \delta x_l + f^{'}_{u|(x_0,u_0)} \delta u.
\label{SISO_dynamics}
\end{equation}
To study the predictive error, assume a bounded control signal $\delta u = u - u_0$ imposes on the linear and nonlinear model separately to obtain the linear prediction and the nonlinear state. Consider the following four cases at the sampling state: (i) the nonlinear dynamics is concavity and increasing; (ii) the nonlinear dynamics is convexity and increasing; (iii) the nonlinear dynamics is convexity and decreasing; (iv) the nonlinear dynamics is concavity and decreasing. Conclude that the modulus of the input signal should be appropriately adjusted larger in (i) and (iii), and lower in (ii) and (iv).

Introduce the following sampling control input, yields
\begin{equation}
  u_{\mathrm{s}} = w_{\mathrm{s}} \mathrm{sign}\left[\ddot{f}(x_0,u_0)\right] \mathrm{sign} \left[ \dot{f}(x_0,u_0) \right] \left|  \delta u \right|, 
\label{sampling_input_1}
\end{equation}
where $w_{\mathrm{s}}\in [0,\ 1]$ is the sampling factor to control the value of sampling input, $ \mathrm{sign} \left[ \ddot{f}(x_0,u_0) \right] = \pm 1$ and $\mathrm{sign} \left[ \dot{f}(x_0,u_0) \right]$ are the auxiliary items of the nonlinear model to control the sampling direction. Note that (\ref{sampling_input_1}) assumes $f^{'}_{u|(x_0,u_0)}>0$ considering $f^{'}_{u|(x_0,u_0)}=1$ in most systems, (\ref{sampling_input_1}) should multiply by $-1$ if $f^{'}_{u|(x_0,u_0)}<0$. Thus there exists a sampling interval from the prediction to the actual state by appropriately choosing $w_{\mathrm{s}}$, then the prediction is driven to the closer neighborhood of the actual state. (\ref{sampling_input_1}) can be converted to
\begin{equation}
  u_{\mathrm{s}} =w_{\mathrm{s}} \mathrm{sign} \left[ \ddot{f}(x_0,u_0) \right] \mathrm{sign} \left[ \dot{f}(x_0,u_0) \right] \mathrm{sign}( \delta u) \delta u. 
\label{sampling_input}
\end{equation}
Note that the nonlinearity and $\delta u$ jointly affect the prediction error, thus the sampling input $u_s$ is designed based on $\delta u$ to avoid over compensation, and dynamically adjust $\delta u$ to reduce the prediction error. Besides, the proposed method is a kind of pre-compensation, since it uses the current state information to compensate the future nonlinearity.

\begin{theorem}
Imposing a bounded control input $\delta u$ on the  nonlinear SISO system $\dot{x} = f(x,u)$ and its linearized model (\ref{SISO_dynamics}) at $(x_0, u_0)$, the sampling control input (\ref{sampling_input}) can reduce the prediction error between the linear prediction $\delta x_l$ and the actual nonlinear state $\delta x_a$, by driving the linear prediction to a closer neighborhood of the actual nonlinear state, if choosing an appropriate $w_{\mathrm{s}} \in [0,\ 1]$.
\end{theorem}

\begin{proof}
Consider the SISO nonlinear system which is described by
\begin{equation}
\delta \dot{x}_a = f^{'}_{x|(x_0,u_0)} \delta x_a + \frac{f^{''}_{x|(x_0,u_0)}}{2} \delta x^2_a+ \cdots + \frac{ f^{(n)}_{x|(x_0,u_0)}}{n !} \delta x^n_a + R_n(x_a)+f^{'}_{u|(x_0,u_0)} \delta u,
\label{Nonlinear dynamics}
\end{equation}
where $\delta x_a$ denotes the nonlinear state, $R_n(x_a)$ is the bounded high-order remainder. The objective is to reduce the error between the linear prediction and the actual nonlinear state driven by the same $\delta u$. Define the prediction error $e_l = \delta x_a - \delta x_l$, then
\begin{equation}
\dot{e}_l = f^{'}_{x|(x_0,u_0)} e_l + \frac{f^{''}_{x|(x_0,u_0)} }{2} \delta x^2_a + \sum^{n}_{i = 3}\frac{ f^{(i)}_{x|(x_0,u_0)}}{i !} \delta x^i_a + R_n(x).
\end{equation}

By adding the proposed sampling input (\ref{sampling_input_1}), the linear controlled system converts to
\begin{equation}
\delta \dot{x}_s = f^{'}_{x|(x_0,u_0)} \delta x_l + f^{'}_{u|(x_0,u_0)} \delta u+ f^{'}_{u|(x_0,u_0)}\mathrm{sign}[\ddot{f}(x_0,u_0)] \mathrm{sign}[\dot{f}(x_0,u_0)]  w_{\mathrm{s}}  | \delta u |.
\end{equation}
Note that the above equation assumes $f^{'}_{u|(x_0,u_0)}>0$, it should multiply by $-1$ if $f^{'}_{u|(x_0,u_0)}<0$. Define another prediction error $e_s = \delta x_a - \delta x_s$, then
\begin{equation}
\dot{e}_s = f^{'}_{x|(x_0,u_0)} e_l + \frac{f^{''}_{x|(x_0,u_0)} }{2} \delta x^2_a + \sum^{n}_{i = 3}\frac{ f^{(i)}_{x|(x_0,u_0)}}{i !} \delta x^i_a + R_n(x_a) - \mathrm{sign} \left[ \ddot{f}(x_0,u_0) \right] \mathrm{sign}\left[ \dot{f}(x_0,u_0) \right] w_{\mathrm{s}} f^{'}_{u|(x_0,u_0)} |\delta u|.
\end{equation}

Consider the above four cases, it can be found that $e_l$ is positive in (i) and (iv), and negative in (ii) and (iii). For case (i), $f^{'}_{x|(x_0,u_0)}>0$ and $f^{''}_{x|(x_0,u_0)}>0$, the model error $\dot{e}_s$ by adopting the sampling strategy is formulated as follows,
\begin{equation}
\dot{e}_s = f^{'}_{x|(x_0,u_0)} e_l + \frac{f^{''}_{x|(x_0,u_0)} }{2} \delta x^2_a  -  w_{\mathrm{s}} f^{'}_{u|(x_0,u_0)} |\delta u_l| + \sum^{n}_{i = 3}\frac{ f^{(i)}_{x|(x_0,u_0)}}{i !} \delta x^i_a + R_n(x_a) .
\end{equation} 
If neglecting the terms higher than third-order, it can be concluded that $\left| f^{'}_{x|(x_0,u_0)} e_l + \frac{f^{''}_{x|(x_0,u_0)} }{2} \delta x^2_a  -  w_{\mathrm{s}} f^{'}_{u|(x_0,u_0)} |\delta u_l| \right| < \left| f^{'}_{x|(x_0,u_0)} e_l + \frac{f^{''}_{x|(x_0,u_0)} }{2} \delta x^2_a \right|$, hence $|\dot{e}_s|< |\dot{e}_l| $ if $w_{\mathrm{s}} \in [0,1]$ is selected appropriately. 
Note that one can add other sampling inputs formulated by $w_s   \mathrm{sign} \left[ f^{(i)}(x_0,u_0) \right]  \mathrm{sign} \left[ \dot{f}(x_0,u_0) \right]  | \delta u|$ to pre-compensating the high-order terms. Consider the computational cost in real-time RVD, this paper neglects the high-order remainder.

For case (ii), $f^{'}_{x|(x_0,u_0)}>0$, $f^{''}_{x|(x_0,u_0)}<0$ and $e_l<0$, $\dot{e}_s$ is formulated as follows,
\begin{equation}
\dot{e}_s = f^{'}_{x|(x_0,u_0)} e_l + \frac{f^{''}_{x|(x_0,u_0)} }{2} \delta x^2_a  +  w_{\mathrm{s}} f^{'}_{u|(x_0,u_0)} |\delta u_l| + \sum^{n}_{i = 3}\frac{ f^{(i)}_{x|(x_0,u_0)}}{i !} \delta x^i_a + R_n(x_a).
\end{equation} 
It can be concluded that $\left| f^{'}_{x|(x_0,u_0)} e_l + \frac{f^{''}_{x|(x_0,u_0)} }{2} \delta x^2_a  +  w_{\mathrm{s}} f^{'}_{u|(x_0,u_0)} |\delta u_l| \right| < \left| f^{'}_{x|(x_0,u_0)} e_l + \frac{f^{''}_{x|(x_0,u_0)} }{2} \delta x^2_a \right|$, hence $\left| \dot{e}_s|< |\dot{e}_l \right| $ if $w_{\mathrm{s}} \in [0,1]$ is selected appropriately. Similarly, $|\dot{e}_s|< |\dot{e}_l| $ is valid in case (iii) and case (iv). Thus the linear prediction is driven to a closer neighborhood of the actual nonlinear state by imposing the sampling control input.  
\end{proof}

\subsubsection{MIMO system}
\label{MIMO}
There exist couplings between the states in a MIMO system. Consider the nonlinear system $\pmb{\dot{x}} = \pmb{f}(\pmb{x}, \pmb{u})$,
where $\pmb{x} = [x_1,\ x_2,\ \cdots,\ x_n]^{\mathrm{T}} \in \mathbb{R}^{n}$ is the state vector, $\pmb{u} = [u_1,\ u_2,\ \cdots,\ u_n]^{\mathrm{T}} \in \mathbb{R}^{n}$ denotes the input vector. For $i=1$ to $n$, $\dot{x}_i = f_i (\pmb{x},u_i)$. Through using Taylor expansion at $(\pmb{x}_0,\pmb{u}(0))$ and imposing a bounded control $\delta \pmb{u}= [u_1-u_1(0),\  \cdots,\ u_n-u_n(0)]^{\mathrm{T}} \in \mathbb{R}^{n}$, the first-order approximation model is formulated by

\begin{equation}
\delta \dot{\pmb{x}}_l = \left[ \begin{array}{cccc}
     \frac{\partial f_1}{\partial x_1} & \frac{\partial f_1}{\partial x_2} & \cdots &\frac{\partial f_1}{\partial x_n}\\
     \frac{\partial f_2}{\partial x_1} & \frac{\partial f_2}{\partial x_2} & \cdots &\frac{\partial f_2}{\partial x_n}\\
     \vdots  & \vdots  & \ddots & \vdots\\
     \frac{\partial f_n}{\partial x_1} & \frac{\partial f_n}{\partial x_2} & \cdots &\frac{\partial f_n}{\partial x_n}
   \end{array} \right]_{(\pmb{x}_0, \pmb{u}(0))} \delta \pmb{x}_l+ \left[ \begin{array}{cccc}
     \frac{\partial f_1}{\partial u_1} & 0 & \cdots &0\\
    0 & \frac{\partial f_2}{\partial u_2} & \cdots &0\\
     \vdots  & \vdots  & \ddots & \vdots\\
     0 & 0 & \cdots &\frac{\partial f_n}{\partial u_n}
   \end{array} \right]_{(\pmb{x}_0, \pmb{u}(0))} \delta \pmb{u} \\
   =  \left[ \begin{array}{c}
     \frac{\partial f_1}{\partial \pmb{x}}^{\mathrm{T}} \\
      \frac{\partial f_2}{\partial \pmb{x}}^{\mathrm{T}} \\
     \vdots  \\
      \frac{\partial f_n}{\partial \pmb{x}}^{\mathrm{T}}    \end{array} \right]_{(\pmb{x}_0, \pmb{u}(0))} \delta \pmb{x}_l  + {\frac{\partial \pmb{f}}{\partial  \pmb{u}}}_{|(\pmb{x}_0, \pmb{u}(0))}  \delta \pmb{u}.
\label{MIMO_linearized_dynamics}
\end{equation}
Introduce the following sampling control input, yields
\begin{equation}
  \pmb{u}_{\mathrm{s}} = \left[ \begin{array}{cccc}
                        w_{\mathrm{s1}} s_{11} s_{12} \mathrm{sign}(\delta u_1)           & 0 & \cdots &0\\
                       0 & w_{\mathrm{s2}}  s_{21} s_{22} \mathrm{sign}(\delta u_2) & \cdots &0\\
                        \vdots  & \vdots  & \ddots & \vdots\\
                        0 & 0 & \cdots &w_{\mathrm{sn}}  s_{n1} s_{n2} \mathrm{sign}(\delta u_n)
                        \end{array} \right] \delta \pmb{u},
\label{MIMO_sampling_input}
\end{equation}
where $w_{\mathrm{si}}$ denote the sampling factor to control the value of sampling input, 
$s_{ij}$ ($i\in[1,n]$, $j\in[1,2]$) are the auxiliary items of the nonlinear model to control the sampling direction.
$s_{i1}= \pm 1$ represents the direction information (first-order) of the total increment,
$s_{i2}=  \pm 1$ contains the convexity information (second-order). 

Compared with the proposed method in SISO system that the increment is determined by a single state, the increment in the MIMO system is determined by the state vector.
Besides, compared with using the signal of the second-order derivative as the auxiliary term, the auxiliary term of the convexity information in MIMO system is represented by the matrix $\bigtriangledown^2 f_i$. $s_{i2}=  1$ if $\bigtriangledown^2 f_{i| \pmb{x}_0} $ is semi-positive definite and $s_{i2}=  -1$ if $\bigtriangledown^2 f_{i| \pmb{x}_0} $ is negative definite. 
Note that $\bigtriangledown^2 f_i$ is a Hessian matrix, thus $s_{i2}$ can be judged by the signal of the diagonal entries, which means $s_{i2}=1$ for all $\frac{\partial^2 f_i}{\partial x^2_j} $ ($j=1 \cdots n$) are semi-positive.

\begin{theorem}
Imposing a bounded control input $\delta \pmb{u}_l$ on the nonlinear MIMO system $\pmb{\dot{x}} = \pmb{f}(\pmb{x}, \pmb{u})$ and its linearized model (\ref{MIMO_linearized_dynamics}) at $(\pmb{x}_0,\pmb{u}(0))$, the sampling control input (\ref{MIMO_sampling_input}) can reduce the prediction error between the linear prediction $\delta \pmb{x}_l$ and the actual nonlinear state $\delta \pmb{x}_a$, by driving the linear prediction to a closer neighborhood of the actual nonlinear state, if choosing a set of appropriate $w_{\mathrm{si}} \in [0,\ 1]$, $i\in [1,n]$.
\label{theorem_2}
\end{theorem}

\begin{proof}
Consider the MIMO nonlinear system which is described by
\begin{equation}
\delta \dot{\pmb{x}}_a 
   =  \left[ \begin{array}{c}
     \frac{\partial f_1}{\partial \pmb{x}}^{\mathrm{T}} \\
      \frac{\partial f_2}{\partial \pmb{x}}^{\mathrm{T}} \\
     \vdots  \\
      \frac{\partial f_n}{\partial \pmb{x}}^{\mathrm{T}}    \end{array} \right]_{(\pmb{x}_0, \pmb{u}(0))} \delta \pmb{x}_a +
      \left[ \begin{array}{c}
     \delta \pmb{x}_a^{\mathrm{T}} H_1  \delta \pmb{x}_a\\
     \delta \pmb{x}_a^{\mathrm{T}} H_2  \delta \pmb{x}_a \\
     \vdots  \\
      \delta \pmb{x}_a^{\mathrm{T}} H_n  \delta \pmb{x}_a   \end{array} \right]_{(\pmb{x}_0, \pmb{u}(0))} + R_n(\pmb{x}_a) +{\frac{\partial \pmb{f}}{\partial  \pmb{u}}}_{|(\pmb{x}_0, \pmb{u}(0))}  \delta \pmb{u},
\end{equation}
where $\delta {\pmb{x}}_a$ denotes the nonlinear state, $R_n(\pmb{x}_a)$ is the bounded high-order remainder, $H_i$ represents the Hessian matrix $\bigtriangledown^2 f_i$. Define the predictive error $\pmb{e}_l = \delta \pmb{x}_a - \delta \pmb{x}_l$, then
\begin{equation}
\dot{\pmb{e}}_l =  \left[ \begin{array}{c}
     \frac{\partial f_1}{\partial \pmb{x}}^{\mathrm{T}} \\
      \frac{\partial f_2}{\partial \pmb{x}}^{\mathrm{T}} \\
     \vdots  \\
      \frac{\partial f_n}{\partial \pmb{x}}^{\mathrm{T}}    \end{array} \right]_{(\pmb{x}_0, \pmb{u}(0))} \pmb{e}_l + \left[ \begin{array}{c}
     \delta \pmb{x}_a^{\mathrm{T}} H_1  \delta \pmb{x}_a\\
     \delta \pmb{x}_a^{\mathrm{T}} H_2  \delta \pmb{x}_a \\
     \vdots  \\
      \delta \pmb{x}_a^{\mathrm{T}} H_n  \delta \pmb{x}_a   \end{array} \right]_{(\pmb{x}_0, \pmb{u}(0))} + R_n(\pmb{x}_a).\end{equation}

By imposing the proposed sampling input (\ref{MIMO_sampling_input}), the linear model of the sampling state $\delta \pmb{x}_s$ can be obtained. Define the prediction error $\pmb{e}_s = \delta \pmb{x}_a - \delta \pmb{x}_s$, then \begin{equation}
\dot{\pmb{e}}_s =\left[ \begin{array}{c}
     \frac{\partial f_1}{\partial \pmb{x}}^{\mathrm{T}} \\
      \frac{\partial f_2}{\partial \pmb{x}}^{\mathrm{T}} \\
     \vdots  \\
      \frac{\partial f_n}{\partial \pmb{x}}^{\mathrm{T}}    \end{array} \right]_{(\pmb{x}_0, \pmb{u}_0)} \pmb{e}_l  + \left[ \begin{array}{c}
     \delta \pmb{x}_a^{\mathrm{T}} H_1  \delta \pmb{x}_a - w_{s1} s_{11}s_{12} | \delta u_1 |\\
     \delta \pmb{x}_a^{\mathrm{T}} H_2  \delta \pmb{x}_a - w_{s2} s_{21}s_{22}| \delta u_2 |\\
     \vdots  \\
      \delta \pmb{x}_a^{\mathrm{T}} H_n  \delta \pmb{x}_a   - w_{sn} s_{n1}s_{n2} | \delta u_n |\end{array} \right]_{(\pmb{x}_0, \pmb{u}_0)} + R_n(\pmb{x}_a).
\end{equation}
If neglecting the terms higher than the third-order, consider the above four cases for each $\dot{e}_{si} = {\frac{\partial f_i}{\partial \pmb{x}}}_{|(\pmb{x}_0, \pmb{u}_0)}^{\mathrm{T}}  \pmb{e}_l +  \delta \pmb{x}_a^{\mathrm{T}} H_i  \delta \pmb{x}_a - w_{si} s_{i1}s_{i2} | \delta u_i | $. If $\dot{e}_{si}$ is concavity and increasing, ${\frac{\partial f_i}{\partial \pmb{x}}}_{|(\pmb{x}_0, \pmb{u}_0)}^{\mathrm{T}}  \pmb{e}_l  $ and $\delta \pmb{x}_a^{\mathrm{T}} H_i  \delta \pmb{x}_a $ are positive, 
then $\left| {\frac{\partial f_i}{\partial \pmb{x}}}_{|(\pmb{x}_0, \pmb{u}_0)}^{\mathrm{T}}  \pmb{e}_l +  \delta \pmb{x}_a^{\mathrm{T}} H_i  \delta \pmb{x}_a - w_{si}  | \delta u_i   | \right| < \left| {\frac{\partial f_i}{\partial \pmb{x}}}_{|(\pmb{x}_0, \pmb{u}_0)}^{\mathrm{T}}  \pmb{e}_l +  \delta \pmb{x}_a^{\mathrm{T}} H_i  \delta \pmb{x}_a \right|$.
If $\dot{e}_{si}$ is convexity and increasing, ${\frac{\partial f_i}{\partial \pmb{x}}}_{|(\pmb{x}_0, \pmb{u}_0)}^{\mathrm{T}}  \pmb{e}_l  $ and $\delta \pmb{x}_a^{\mathrm{T}} H_i  \delta \pmb{x}_a $ are negative, then $\left| {\frac{\partial f_i}{\partial \pmb{x}}}_{|(\pmb{x}_0, \pmb{u}_0)}^{\mathrm{T}}  \pmb{e}_l +  \delta \pmb{x}_a^{\mathrm{T}} H_i  \delta \pmb{x}_a + w_{si}  | \delta u_i |  \right| < \left| {\frac{\partial f_i}{\partial \pmb{x}}}_{|(\pmb{x}_0, \pmb{u}_0)}^{\mathrm{T}}  \pmb{e}_l +  \delta \pmb{x}_a^{\mathrm{T}} H_i  \delta \pmb{x}_a \right|$.
The other cases are similar.
Thus the linear prediction is driven to a closer neighborhood of the actual nonlinear state by imposing the sampling control input.  
\end{proof}

Theorem \ref{theorem_2} demonstrates the sampling input can reduce the one-step ahead prediction error in an MIMO system. Then using the method to improve the model deviation in the prediction process of the PWA MPC is considered herein.
Assuming a bounded control input sequence $ \left[\delta \pmb{u}^{\mathrm{T}}_k, \cdots, \delta \pmb{u}^{\mathrm{T}}_{k+N_p-1}\right]^{\mathrm{T}}$ is given, where $N_p$ is the prediction horizon. The actual PWA model approximating the nonlinear system is formulated by $\pmb{x} _{k+i+1} = A_{k+i} \pmb{x}_{k+i} + B_{k+i} \pmb{u}_{k+i}$, $i\in \left[0, N_p-1\right]$. Note that each $A_{k+i}$ and $B_{k+i}$ are $\pmb{x}_{k+i-1}$ dependent (linearized at each actual nonlinear state $\pmb{x}_{k+i-1}$). In standard PWA MPC, the model used to predict the evolution is described by $\pmb{x}^{+}_{k+i+1} = A^{+}_{k+i} \pmb{x}_{k+i} + B^{+}_{k+i} \pmb{u}_{k+i}$, $i\in \left[1, N_p-1\right]$ (the model is equivalent with the actual PWA model if $i=0$). Each $A^{+}_{k+i}$ and $B^{+}_{k+i}$ are the linear prediction $\pmb{x}^{+}_{k+i-1}$ dependent, and the error between each $A^{+}_{k+i}$/$B^{+}_{k+i}$ and $A_{k+i}$/$B_{k+i}$ increases with $i$, which is called the model deviation. However, according to Theorem \ref{theorem_2}, the model deviation could be improved since the accumulated prediction error is reduced.

\subsection{Controller design}
\label{controller design}
In this subsection, the proposed sampling-based method is applied in the double-loop control. 
There are several practical issues in the application.
Consider MPC is based on the discrete-time system, the proposed sampling-based method is extended to the discrete-time system firstly. The main problem using the sampling-based method is the convexity information in a discrete-time system is not equivalent with the continuous-time system. 
However, it is difficult to directly discrete a highly nonlinear system to get the convexity information in real-time. 
Thus two methods are provided to approximate the discrete-time convexity information. The first method uses the convexity in the continuous system at the sampling state to approximate on the premise that the sampling interval approaches to zero enough, since the nonlinearity in the continuous system can approximate the nonlinearity of its discrete-time system in a local interval. The second method is based on the difference $f(x_{k+2})-2f(x_{k+1})+f(x_{k})$. However, obtain the actual nonlinear $f(x_{k+1})$ and $f(x_{k+2})$ are difficult since there is a lack of the direct discretize method of the complex nonlinear system. Thus the approximation of $f(x_{k+1})$ and $f(x_{k+2})$ is obtained by the recursion of taking $x_{k}$ into the linear uncontrolled model to generate the approximate $x_{k+1}$, $f(x_{k+1})$, $x_{k+2}$ and $f(x_{k+2})$. 
In addition, since discrete-time MPC is sensitive to noise and bias, to the best knowledge of the authors, most papers adopted state estimation methods (Kalman filter\cite{Lee1994}, moving horizon estimation\cite{Tenny2002}, set-member estimation\cite{Scholte2008}, etc.) to obtain the high-precision measurement data. Consider the state estimator and MPC can be decoupled in this paper, hence it is reasonably assumed that the data in MPC have been pre-processed.

Another issue is the first-order auxiliary term $s_{i1}$ depends on the signal of the total increment $ {\frac{\partial f_i}{\partial \pmb{x}}}_{|(\pmb{x}_0, \pmb{u}_0)}^{\mathrm{T}}  \pmb{e}_l$ in the MIMO system, which is determined both by the first-order derivative vector and the predictive error. However, $\pmb{e}_l$ can not be obtained, which takes difficulty in determing $s_{i1}$. The solution is through the convexity information to adjust the signal of the elements in $\pmb{e}_l$, e.g., the elements are positive if the system is convexity (similar with Fig. \ref{sampling}). Besides, consider $|\pmb{e}_l |$ is bounded since the control input is bounded. Thus the signal of the total increment mainly depends on the signal of the summation of the elements in the first-order derivative vector.
Besides, the judgement of the signal of the control input in (\ref{MIMO_sampling_input}) in the following controller design is obtained by the relation between the current state and the desired state.

To reduce the computation cost of linearization and guarantee the real-time performance, the discrete-time PWA model is obtained by embedding, which converts the nonlinear model to a pseudo-linear form (the state matrix and input matrix are state-dependent), take the state matrix and input matrix as constant during each state interval. Note that the proposed strategy still applies to the Taylor expansion linearization.

\subsubsection{Optimization index of the position loop}

Consider the continous LOS dynamics (\ref{position state}), and discrete (\ref{position state}) at each sampling instant.
Denote $\pmb{x}_p^{\mathrm{+}}(k+1)$ as the ahead prediction generated by the linear model, $\pmb{x}_p (k+1)$ as the actual nonlinear state, the sampling approach is to decrease the error between $\pmb{x}_p^{\mathrm{+}}(k+1)$ and $\pmb{x}_p (k+1)$. By introducing the sampling control input $\pmb{u}_{sp}$, $\pmb{x}_p^{\mathrm{+}}(k+1)$ is expressed by
\begin{equation}
  \pmb{x}_p^{\mathrm{+}}(k+1) = A_{p}(k)  \pmb{x}_p(k) + B_{p}(k)  \pmb{u}_p(k) +  B_{p }(k) \pmb{u}_{sp}(k), 
\label{discrete position state}
\end{equation}
where $A_{p}$ = $e^{A_p^{\mathrm{c}} T_s}$, $B_{p}$ = $\left(\int_0^{T_s} e^{A_p^{\mathrm{c}} \tau} d \tau \right)B_p^{\mathrm{c}} $, and $T_s$ is the sampling interval. (\ref{discrete position state}) is equivalent to
\begin{equation}
  \pmb{x}_p^{\mathrm{+}}(k+1) = A_{p}(k)  \pmb{x}_p(k) + B_{p}(k)  \pmb{u}_p(k) +  \left[ \begin{array}{cccc}
                        S_1     & 0 & \cdots &0\\
                       0 & S_2 & \cdots &0\\
                        \vdots  & \vdots  & \ddots & \vdots\\
                        0 & 0 & \cdots &S_6
                        \end{array} \right] B_{p }(k) \left|\pmb{u}_{p}(k) \right|, 
\label{discrete position state_1}
\end{equation}
where $S_i =  w_{\mathrm{si}} s_{i1} s_{i2} $ denotes the sampling parameter ($i=1\cdots6$), 
where $s_{i1} =1$ if the discrete dynamics is increasing at the sampling state $x_{pi}(k)$, $s_{i2} =1$ if the discrete dynamics is concave at $x_{pi}(k)$.
According to (\ref{position state}), then (\ref{discrete position state_1}) is further converted to 
\begin{equation}
  \pmb{x}_p^{\mathrm{+}}(k+1) = A_{p}(k)  \pmb{x}_p(k) + B_{p}(k)  \pmb{u}_p(k) +  B_{p }(k) W_p(k)  \pmb{u}_{p}(k), 
\label{discrete position state_2}
\end{equation}
where
\begin{equation*}
W_p(k) = \left[ \begin{array}{ccc}
                        S_4 \mathrm{sign}\left[u_{p1}(k)\right]     & 0 & 0\\
                       0 & S_5 \mathrm{sign}\left[u_{p2}(k)\right] & 0\\
                        0 & 0 & S_6 \mathrm{sign}\left[u_{p3}(k)\right]
                        \end{array} \right] .
\end{equation*}


According to (\ref{position state}), define $f_4(\pmb{x}_p) = \dot{x}_{p4}$, then
\begin{equation}
\bigtriangledown f_4 = \left[\begin{array}{cccccc}
    \omega^2 \cos^2 x_{p 2} - \frac{\mu}{R^3} + 3\frac{\mu}{R^3}  \cos^2 x_{p 2} \sin^2 x_{p 3} -  \frac{x_{p 5}^2} {x_{p 1}^2} -  \frac{x_{p 6}^2} {x_{p 1}^2} \cos^2 x_{p 2}\\
- \omega^2 x_{p 1}\sin 2x_{p 2}  -3\frac{\mu}{R^3} x_{p 1} \sin 2x_{p 2}  \sin^2 x_{p 3} +2 \omega x_{p 6} \sin 2x_{p 2}  - \frac{x_{p 6}^2} {x_{p 1}} \sin 2x_{p 2}  \\
3\frac{\mu}{R^3} x_{p 1}  \cos^2 x_{p 2} \sin 2x_{p 3}\\
0\\
2 \frac{ x_{p 5}} {x_{p 1}}\\
- 2 \omega \cos^2 x_{p 2}+ 2 \frac{x_{p 6}} {x_{p 1}} \cos^2 x_{p 2}
   \end{array}\right],
\end{equation}
and 
\begin{equation}
\bigtriangledown^2 f_4= \left[\begin{array}{cccccc}
    2 \frac{x_{p5}^2+x_{p6}^2 \cos^2 x_{p2}}{x_{p1}^3} & F_1\sin 2x_{p 2}  & \frac{3 \mu \cos^2 x_{p 2} \sin 2x_{p 3}}{R^3}   & 0 & \frac{-2x_{p 5}} {x_{p 1}^2} & \frac{- 2x_{p 6} \cos^2 x_{p 2}} {x_{p 1}^2}  \\

\sin 2x_{p 2} F_1 &  
F_2 \cos 2x_{p 2} &  \frac{-3 \mu  x_{p 1} \sin 2x_{p 2}  \sin 2x_{p 3}}{R^3} &0 & 0 & \left(2 \omega   - \frac{2x_{p 6}}{x_{p 1}} \right)  \sin 2x_{p 2}\\

\frac{3 \mu \cos^2 x_{p 2} \sin 2x_{p 3}}{R^3}  & \frac{-3 \mu x_{p 1}  \sin 2 x_{p 2} \sin 2x_{p 3}    }{R^3}  & \frac{6 \mu x_{p 1}  \cos^2 x_{p 2} \cos 2x_{p 3}}{R^3}  & 0 & 0 &0\\
0 & 0 & 0 & 0 & 0 & 0\\
\frac{-2x_{p 5}} {x_{p 1}^2} &0 &0 & 0 & \frac{2} {x_{p 1}} &0 \\
 \frac{-2 x_{p 6} \cos^2 x_{p 2}} {x_{p 1}^2}  & 2\left( \omega -  \frac{x_{p 6}} {x_{p 1}}\right)\sin 2x_{p2} &0 & 0 &0 &  \frac{2 \cos^2 x_{p 2}} {x_{p 1}} 

   \end{array}\right],
\end{equation}
where
\begin{equation*}
 F_1 = \left(  - \omega^2  -3\frac{\mu}{R^3}\sin^2 x_{p 3} +  \frac{x_{p 6}^2} {x_{p 1}^2} \right),\ 
 F_2 = -2 \omega^2 x_{p 1} -6\frac{\mu}{R^3} x_{p 1}  \sin^2 x_{p 3} +4 \omega x_{p 6} -2 \frac{x_{p 6}^2} {x_{p 1}}.
\end{equation*}

Define $f_5(\pmb{x}_p) = \dot{x}_{p5}$, then
\begin{equation}
\bigtriangledown f_5 = \left[\begin{array}{cccccc}
 \left( - \omega^2 - \frac{ 3\mu }{R^3} \sin^2 x_{p 3} \right)
  \frac{ \sin 2x_{p 2}}{2} +  \frac{x_{p 4} x_{p 5} } {x_{p 1}^2}+ \frac{x_{p 6}^2} {x_{p 1}^2} \frac{ \sin 2x_{p 2}}{2} \\
 \left( - \omega^2 x_{p 1}- \frac{3 \mu x_{p 1}}{R^3} \sin^2 x_{p 3}  + 2 \omega x_{p 6} - \frac{x_{p 6}^2} {x_{p 1}} \right)
   \cos 2x_{p 2}\\
 - 3\frac{ \mu x_{p 1}}{R^3} \sin2 x_{p 3}\frac{ \sin 2x_{p 2}}{2} \\
-\frac{x_{p 5}}{x_{p 1}} \\
 -\frac{x_{p 4}}{x_{p 1}} \\
 \sin 2x_{p 2}\left( \omega-\frac{x_{p 6}}{x_{p 1}}\right)
   \end{array}\right],
\end{equation}
and
\begin{equation}
\bigtriangledown^2 f_5 =
 \left[\begin{array}{cccccc}
    \frac{ -2 x_{p 4} x_{p 5} - x_{p 6}^2 \sin 2x_{p 2}}{x_{p 1}^3}  &  F_1  \cos 2x_{p 2}  &
  \frac{ -3\mu \sin 2 x_{p 3}\sin 2x_{p 2}}{2R^3}  &
 \frac{x_{p 5}} {x_{p 1}^2} &   \frac{x_{p 4}} {x_{p 1}^2} & \frac{x_{p 6}\sin 2x_{p 2}} {x_{p 1}^2}  \\

  F_1  \cos 2x_{p 2} &  
 -2F_3 \sin 2x_{p 2}    &  
 \frac{-3 \mu x_{p1}\sin 2 x_{p 3}\cos 2x_{p 2}}{R^3}   &
 0  &
 0 &
 2 \left( \omega - \frac{x_{p 6}} {x_{p 1}} \right)\cos 2x_{p 2}  \\

  \frac{-3 \mu \sin2 x_{p 3} \sin 2x_{p 2}}{2 R^3}  &
  \frac{ -3 \mu x_{p1} \sin2 x_{p 3} \cos 2x_{p 2}}{R^3}  &
  \frac{-6 \mu x_{p1} \cos2 x_{p 3} \sin 2x_{p 2}}{2R^3}  &
 0 &
 0 &
0\\

\frac{x_{p 5}} {x_{p 1}^2}  & 0 & 0 & 0 & \frac{-1} {x_{p 1}}  & 0\\

\frac{x_{p 4}} {x_{p 1}^2} &0 &0 &  \frac{-1} {x_{p 1}} & 0 &0 \\

 \frac{x_{p 6}\sin 2x_{p 2}} {x_{p 1}^2} & 2\cos 2x_{p 2}\left( \omega-\frac{x_{p 6}}{x_{p 1}}\right) &0 & 0 &0 & -\frac{\sin 2x_{p 2}} {x_{p 1}}

   \end{array}\right],
\end{equation}
where
\begin{equation*}
F_3 = \left( - \omega^2 x_{p 1}- \frac{3 \mu x_{p 1}}{R^3} \sin^2 x_{p 3}  + 2 \omega x_{p 6} - \frac{x_{p 6}^2} {x_{p 1}} \right).
\end{equation*}

Define $f_6(\pmb{x}_p) = \dot{x}_{p6}$, then
\begin{equation}
\bigtriangledown f_6 = \left[\begin{array}{cccccc}
\dot{\omega} +  \frac{3 \mu } { 2R^3} \sin 2x_{p 3} +  \frac{x_{p 4} x_{p 6}} {x_{p 1}^2}  -  \frac{2 x_{p 5} x_{p 6} \tan x_{p 2}} {x_{p 1}^2}\\
 
 2\left( -\omega +  \frac{x_{p 6}}{x_{p 1}}  \right)x_{p 5} \sec^2 x_{p 2} - \frac{\sin x_{p 2} }{\cos^2 x_{p 2}}u_{p3} \\

\frac{3 \mu } { R^3} x_{p 1} \cos 2x_{p 3}  \\

 2 \omega - \frac{x_{p 6}}{x_{p 1}}  \\
  2\left( -\omega +  \frac{x_{p 6}}{x_{p 1}}  \right) \tan x_{p 2} \\
 - \frac{x_{p 4}}{x_{p 1}}  + \frac{ 2x_{p 5} \tan x_{p 2} }{x_{p 1}}
 
   \end{array}\right],
\end{equation}
and
\begin{equation}
\bigtriangledown^2 f_6 = \left[\begin{array}{cccccc}
  \frac{2 x_{p 6}\left( -x_{p 4} + 2 x_{p 5}\tan x_{p 2}    \right)} {x_{p 1}^3}   &
  \frac{-2 x_{p 5}x_{p 6} \sec^2 x_{p 2}}{x_{p 1}^2}  &
  \frac{3 \mu \cos 2x_{p 3} } { R^3}   &
 \frac{ x_{p 6} } {x_{p 1}^2} &
\frac{ -2 x_{p 6} \tan x_{p 2}} {x_{p 1}^2} &
\frac{  x_{p 4}- 2 x_{p 5} \tan x_{p 2}} {x_{p 1}^2} \\

  \frac{-2 x_{p 5}x_{p 6} \sec^2 x_{p 2}}{x_{p 1}^2}  &  
 F_4   &  
 0  &
 0  &
F_5 &
  \frac{ 2x_{p 5} \sec^2 x_{p 2}}{x_{p 1}}   \\

 \frac{3 \mu } { R^3} \cos 2x_{p 3}  &
  0 &
  \frac{-6 \mu } { R^3} x_{p 1} \sin 2x_{p 3}  &
 0 &
 0 &
0\\

\frac{ x_{p 6} } {x_{p 1}^2} & 0 & 0 & 0 & 0 & \frac{ -1 } {x_{p 1}}\\

\frac{ -2x_{p 6} \tan x_{p 2} } {x_{p 1}^2} &
F_5 &0 & 0 & 0 & \frac{2 \tan x_{p 2}} {x_{p 1}}\\

 \frac{x_{p 4}- 2 x_{p 5} \tan x_{p 2}} {x_{p 1}^2} & \frac{2 x_{p 5} \sec^2 x_{p 2}}{x_{p 1}} &0 & \frac{-1} {x_{p 1}} &\frac{2 \tan x_{p 2} } {x_{p 1}} & 0

   \end{array}\right],
\end{equation}
where
\begin{equation*}
 F_4 =  4\left( -\omega +  \frac{x_{p 6}}{x_{p 1}}  \right)x_{p 5} \sec^2 x_{p 2} \tan x_{p 2}  -\left(   \frac{1+2\tan^2 x_{p 2} }{\cos x_{p 2}}         \right) u_{p3},\
 F_5 = 2\left( -\omega +  \frac{x_{p 6}}{x_{p 1}}  \right) \sec^2 x_{p 2}.
\end{equation*}

Note that the states in the LOS dynamics are coupled ($x_{p5}=\rho \dot{\varepsilon}$, $x_{p6}=\rho \dot{\beta}$), thus let $S_4 = w_{\mathrm{s4}} s_{11} s_{42}$, $S_5 = w_{\mathrm{s5}} s_{21} s_{52}$, $S_6 = w_{\mathrm{s6}} s_{31} s_{62}$ to decouple. 
By introducing an augmented vector $ \pmb{x}^{\ast}_p(k) \in \mathbb{R}^{6N_p}$, the $N_p$ ahead predictions of $\pmb{x}_p(k)$ can be described as
\begin{equation}
 \pmb{x}^{\ast}_p(k) = \left[{\pmb{{x}}_p^{+}}^{\mathrm{T}}(k+1|k),\ {\pmb{{x}}_p^{+}}^{\mathrm{T}}(k+2|k),\ \cdots ,\ {\pmb{{x}}_p^{+}}^{\mathrm{T}}(k+N_p|k)\right]^{\mathrm{T}}.
\end{equation}
Define the control horizon as $N_c$ ($N_c<N_p$), then $\pmb{x}^{\ast}_p(k)$ is rewritten to a compact form by iteration, i.e.,
\begin{equation}
   \pmb{x}^{\ast}_p(k) = A_p^{\ast} \pmb{x}_{p}(k) + B_p^{\ast} \pmb{u}^{\ast}_p +B_p^{\ast} W_p^{\ast} \pmb{u}^{\ast}_{p},
\label{AX}
\end{equation}
where $W_p^{\ast} =  \oplus ^{N_c-1}_{i = 0} W_p(k+i)\in \mathbb{R}^{{3N_c}\times {3N_c}}$ ($\oplus$ is the direct sum notation), and
\begin{equation*}
   A_p^{\ast}(k) = \left[\begin{array}{c}
     A_p(k)\\
     A_p(k + 1) A_p(k)\\
     \vdots\\
     \prod^{N_p - 1}_{i = 0} A_p(k + i)
   \end{array}\right] \in \mathbb{R}^{{6N_p}\times {6}},
\end{equation*}
and
\begin{align*}
 {B}_p^{\ast}(k) =  \left[ \begin{array}{cccc}
     B_p (k) \! & \!\cdots \!&\! 0\\
     A_p (k \!+\! 1) B_p (k) \! & \! \ddots\! &\!\vdots\\
     \vdots  \!& \!\ddots \!&\! \vdots\\
     \prod^{N_p \!-\! 1}_{i \!=\! 1} \! A_p (k \!+\! i) B_p (k) \!&\! \cdots\! &\! \prod^{N_p \!-\! N_c}_{i \!=\! 1} \! A_p (k \!+\! i) B_p (k)
   \end{array} \right]
 \in \mathbb{R}^{{6N_p}\times {3N_c}},
\end{align*}
and
\begin{equation*}
 \pmb{u}^{\ast}_{p} = \left[ \pmb{u}_p(k)^{\mathrm{T}},\ \pmb{u}_p(k+1)^{\mathrm{T}},\ \cdots, \ \pmb{u}_p(k+N_c-1)^{\mathrm{T}} \right]^{\mathrm{T}} \in \mathbb{R}^{3N_c}.
\end{equation*}

Consider the response of the system is smoother if controlling the increment of the input rather than the total input \cite{Li2017}. Define the recursive equation of $N_c$ control inputs $\pmb{u}^{\ast}_p \in \mathbb{R}^{{3N_c}}$ as
\begin{equation}
   \begin{cases} 
      \pmb{u}_p(k) = \Delta\pmb{u}_p (k) + \pmb{u}_p(k-1),\\
      \pmb{u}_p(k+1) =  \Delta \pmb{u}_p(k+1) + \Delta\pmb{u}_p(k) + \pmb{u}_p(k-1), \\
      \qquad \qquad \qquad \qquad   \vdots\\
      \pmb{u}_p(k+N_c-1) =  \sum^{N_c-1}_{i = 0} \Delta \pmb{u}_p(k+i) + \pmb{u}_p(k-1). \\
   \end{cases}
\label{BBB}
\end{equation}
Rewriting (\ref{BBB}) as follows,
\begin{equation}
   \pmb{u}^{\ast}_p(k) = \Lambda \pmb{u}_p(k-1) + \Gamma \Delta \tilde{\pmb{u}}_p(k),
\label{DDD}
\end{equation}
where $\Delta \tilde{\pmb{u}}_p(k) = \sum^{N_c-1}_{i = 0} \Delta {\pmb{u}}_p(k+i) | i\rangle
\in \mathbb{R}^{{3N_c}}$, $\Lambda = \sum^{N_c}_{i = 1} I_3 | i\rangle $ $ \in \mathbb{R}^{{3N_c}\times {3}}$, and $\Gamma(i,j)=I_3 \ (i>j) \in \mathbb{R}^{{3N_c}\times {3N_c}} $ denotes a lower triangular matrix. 
Substituting (\ref{DDD}) into (\ref{AX}) yields
\begin{equation} 
    \pmb{x}^{\ast}_p(k) = A_p^{\ast} \pmb{x}_{p}(k) + \left(B_{p}^{\ast}  + B_{p}^{\ast} W_p^{\ast}\right)\Lambda \pmb{u}_p(k-1) + \left(B_{p}^{\ast}  + B_{p}^{\ast} W_p^{\ast}\right) \Gamma \Delta \tilde{\pmb{u}}_p.
\label{EEE}
\end{equation}

Define the optimization index as
\begin{equation}
{\min J}_p(k) =\sum^{N_p}_{i = 1} \left\| \pmb{x}_p (k + i) - \pmb{x}_{d,p} (k + i) \right\|_Q^2 + \sum^{N_c - 1}_{i =
   0} \left\| \Delta{\pmb{u}}_p(k+i) \right\|_P^2,
\end{equation}
where $\pmb{x}_{d,p}(k+i)$ denotes the desired state corresponding to $\pmb{x}_{p}(k+i)$, $P$ is a semi-positive-definite matrix and $Q$ is a positive definite weight matrix. Converting above optimization index to a compact form, i.e.,
\begin{equation}
   {\min J}_p(k) = \left[  \pmb{x}^{\ast}_p(k) - \pmb{x}^{\ast}_{d,p} (k) \right]^{\mathrm{T}} \tilde{Q} \left[  \pmb{x}^{\ast}_p(k) - \pmb{x}^{\ast}_{d,p} (k) \right] + \Delta
   \tilde{\pmb{u}}_p^T \tilde{P} \Delta \tilde{\pmb{u}}_p,
\label{FFF}
\end{equation}
where $\pmb{x}^{\ast}_{d,p} (k) = \sum^{N_p}_{i = 1} \pmb{x}_{d,p}(k+i) | i\rangle$ denotes the augmented desired states, $\tilde{Q} = \oplus ^{N_p}_{i = 1} Q$ and $\tilde{P} = \oplus ^{N_c}_{i = 1} P$.

Define 
\begin{equation} 
\pmb{E}_p = \pmb{x}^{\ast}_{d,p} (k) - A_p^{\ast} \pmb{x}_{p}(k) - \left(B_{p}^{\ast}  + B_{p}^{\ast} W_p^{\ast}\right)\Lambda \pmb{u}_p(k-1),
\label{E_p}
\end{equation}
and substitute (\ref{E_p}) into (\ref{FFF}) yields
\begin{equation}
\begin{array}{l}
    {\min}J_p(k) = \left[ (B_{p}^{\ast}  +  B_{p}^{\ast} W_p^{\ast}) \Gamma \Delta \tilde{\pmb{u}}_p - \pmb{E}_p \right]^{\mathrm{T}} \tilde{Q} \left[ (B_{p}^{\ast}  + B_{p}^{\ast} W_p^{\ast}) \Gamma \Delta \tilde{\pmb{u}}_p - \pmb{E}_p\right]+ \Delta \tilde{\pmb{u}}_p^{\mathrm{T}} \tilde{P} \Delta \tilde{\pmb{u}}_p\\ 
          \qquad\quad\quad   = \frac{1}{2} \Delta
   \tilde{\pmb{u}}_p^{\mathrm{T}} H_p \Delta \tilde{\pmb{u}}_p + \pmb{f}_p^{\mathrm{T}} \Delta
   \tilde{\pmb{u}}_p^{\mathrm{T}} + \pmb{E}_p^{\mathrm{T}} \tilde{Q} \pmb{E}_p,
\label{JJ_p}
\end{array}
\end{equation}
where $H_p = 2 \left[ \Gamma^{\mathrm{T}} \left(B_{p}^{\ast}  + B_{p}^{\ast} W_p^{\ast}\right)^{\mathrm{T}} \tilde{Q} \left(B_{p}^{\ast}  + B_{p}^{\ast} W_p^{\ast}\right) \Gamma + \tilde{P} \right]$ and $\pmb{f}_p = -2 \Gamma^{\mathrm{T}} \left( B_{p}^{\ast}  + B_{p}^{\ast} W_p^{\ast} \right)^{\mathrm{T}} \tilde{Q} \pmb{E}_p $.

\subsubsection{Optimization index of the attitude loop}

Consider the proposed sampling-based method requires more computation cost, and the standard PWA MPC can achieve ideal performance under the simulations in several cases. Thus the attitude control in this paper is paid more attention on the singularity-free problem, which is discussed in the next section.
Consider the attitude dynamics (\ref{attitude state}),
introduce an augmented state vector $\pmb{x}^{\ast}_a(k) \in \mathbb{R}^{6N_p}$, which represent $N_p$ ahead prediction states of $\pmb{x}_a(k)$  in (\ref{attitude state}). According to the iterative relation, rewrite the prediction states $\pmb{x}^{\ast}_a(k)$ in a compact form such that,
\begin{equation}
   \pmb{x}^{\ast}_a(k) = A_a^{\ast} \pmb{x}_{a}(k) +  
B_{a}^{\ast}\pmb{u}^{\ast}_{a},
\label{attitude dynamics_1}
\end{equation}
where the form of $A_a^{\ast}$ and ${B}_a^{\ast}$ are similar to $A_p^{\ast}$ and ${B}_p^{\ast}$ in (\ref{AX}).
Consider the following optimization index,
\begin{equation}
   {\min J}_a(k) = \left[ \pmb{x}^{\ast}_a(k) - \pmb{x}^{\ast}_{d,a} (k) \right]^{\mathrm{T}} \tilde{Q} \left[ \pmb{x}^{\ast}_a(k) - \pmb{x}^{\ast}_{d,a} (k) \right] + \Delta
   \tilde{\pmb{u}}_a^{\mathrm{T}} \tilde{P} \Delta \tilde{\pmb{u}}_a
\label{J_a},
\end{equation}
where $\pmb{x}^{\ast}_{d,a} (k) = \sum^{N_p}_{i = 1} \pmb{x}_{d,a} (k+i)| i\rangle$ denotes the augmented desired states.
Define 
\begin{equation} 
\pmb{E}_a = \pmb{x}^{\ast}_{d,a} (k) - A_a^{\ast} \pmb{x}_{a}(k) - B_{a}^{\ast}\Lambda \pmb{u}_a(k-1),
\label{E_a}
\end{equation}
and substitute (\ref{E_a}) into (\ref{J_a}), yields
\begin{align}
    {\min}J_a(k)
             = \frac{1}{2} \Delta
   \tilde{\pmb{u}}_a^{\mathrm{T}} H_a \Delta \tilde{\pmb{u}}_a + \pmb{f}_a^{\mathrm{T}} \Delta
   \tilde{\pmb{u}}_a^{\mathrm{T}} + \pmb{E}_a^{\mathrm{T}} \tilde{Q} \pmb{E}_a,
\label{JJ_a}
\end{align}
where $H_a = 2 \left( {\Gamma^{\mathrm{T}} B_a^{\ast}}^{\mathrm{T}} \tilde{Q} B_{a}^{\ast}  \Gamma + \tilde{P} \right)$, $\pmb{f}_a = -2 \Gamma^{\mathrm{T}}{B_{a}^{\ast} } ^{\mathrm{T}} \tilde{Q} \pmb{E}_a $.

\subsection{Constraints reconfiguration}
Based on Section 2.3, the constraints related with the augmented vectors $\pmb{x}^{\ast}_p(k)$ and $\pmb{x}^{\ast}_a(k)$ are reconfigured in this subsection.
   
\subsubsection{Control input constraints}
The control input constraint of thrusters is described as,
\begin{equation} 
   - \tilde{\pmb{u}}_p^{\max} \leqslant \Lambda \pmb{u}_p(k - 1) + \Gamma \Delta \tilde{\pmb{u}}_p \leqslant \tilde{\pmb{u}}_p^{\max},
\label{control input constraints_position}
\end{equation}    
where $\tilde{\pmb{u}}_p^{\max} = \sum^{N_c}_{i = 1} \pmb{u}_p^{\max} | i \rangle \in \mathbb{R}^{3N_c}$ is the augmented vector of $\pmb{u}_p^{\max}$ defined in (\ref{input_p}), $\Lambda$ and $\Gamma$ are defined in (\ref{DDD}).

Converting (\ref{control input constraints_position}) to the following form,
\begin{equation}
G_{c} \Delta \tilde{\pmb{u}}_p \leqslant \pmb{g}_{c,p}
\label{control_p} 
\end{equation}
where
\begin{equation*} 
G_{c} =   \left[\begin{array}{c}
     \Gamma\\
     - \Gamma
   \end{array}\right],  \
\pmb{g}_{c,p} =\left[\begin{array}{c}
     \tilde{\pmb{u}}_p^{\max} - \Lambda \pmb{u}_p(k - 1)\\
     \tilde{\pmb{u}}_p^{\max} + \Lambda \pmb{u}_p(k - 1)
   \end{array}\right].
\end{equation*}

Similarly, the control input constraint on the reaction wheels is given by
\begin{equation}
G_{c} \Delta \tilde{\pmb{u}}_a \leqslant \pmb{g}_{c,a}
\label{control_a} 
\end{equation}
where
\begin{equation*} 
G_{c} =   \left[\begin{array}{c}
     \Gamma\\
     - \Gamma
   \end{array}\right],\
  \pmb{g}_{c,a} =\left[\begin{array}{c}
     \tilde{\pmb{u}}_a^{\max} - \Lambda \pmb{u}_a(k - 1)\\
     \tilde{\pmb{u}}_a^{\max} + \Lambda \pmb{u}_a(k - 1)
   \end{array}\right],
\end{equation*}    
where $\tilde{\pmb{u}}_a^{\max} = \sum^{N_c}_{i = 1} \pmb{u}_a^{\max} | i \rangle \in \mathbb{R}^{3N_c}$ is the augmented vector of $\pmb{u}_a^{\max}$ described in (\ref{input_a}). 
\subsubsection{Collision avoidance constraint}
The collision avoidance constraint is re-configurated as follows,
\begin{equation} 
  \tilde{\pmb{l}}_1 \pmb{x}^{\ast}_p (k) \geq \tilde {\pmb{r}}_{\mathrm{safe}},
\label{collision1}
\end{equation}    
where $\tilde{\pmb{l}}_1 = E_{N_p} \otimes \pmb{l}_1 \in \mathbb{R}^{{N_p}\times {6N_p}}$ with $\otimes$ being the Kronecker product of two matrices,  $E_{N_p}\in \mathbb{R}^{{N_p}\times {N_p}}$ is an identity matrix with $N_p$ dimension, $\pmb{l}_1 = [1,\ 0,\ 0,\ 0,\ 0,\ 0] \in \mathbb{R}^{{1}\times {6}} $,  and $\tilde{\pmb{r}}_{\mathrm{safe}} = \sum^{N_p}_{i = 1} \pmb{r}_{\mathrm{safe}} | i \rangle \in \mathbb{R}^{{N_p}}$ is the augmented vector of $\pmb{r}_{\mathrm{safe}}$ described in (\ref{r_safe}).

Substituting (\ref{EEE}) into (\ref{collision1}), yields
\begin{equation}
G_{a} \Delta \tilde{\pmb{u}}_p \leqslant \pmb{g}_{a},
\label{Constraint_coliision} 
\end{equation}
where
\begin{equation*}
\begin{array}{l}
G_{a} =-\tilde{\pmb{l}}_{1} \left(B_{p}^{\ast}  + B_{p}^{\ast} W_p^{\ast}\right) \Gamma ,\
\pmb{g}_{a}= - \tilde {\pmb{r}}_{\mathrm{safe}}  +  \tilde{\pmb{l}}_1 A_p^{\ast} \pmb{x}_{p}(k) + \tilde{\pmb{l}}_1 \left(B_{p}^{\ast}  \!+\! B_{p}^{\ast} W_p^{\ast} \right) \Lambda \pmb{u}_p(k\!-\!1).
\end{array}
\end{equation*}

\subsubsection{Entry Cone Constraint}
The entry cone constraint is described by
\begin{equation} 
 \tilde {\pmb{\varepsilon}}^{\min}(k)  \leqslant \tilde{\pmb{l}}_2 \pmb{x}^{\ast}_p (k) \leqslant \tilde {\pmb{\varepsilon}}^{\max}(k) ,
\label{cone1}
\end{equation}    
and
\begin{equation} 
 \tilde {\pmb{\beta}}^{\min}(k)  \leqslant \tilde{\pmb{l}}_3 \pmb{x}^{\ast}_p (k) \leqslant \tilde {\pmb{\beta}}^{\max}(k),
\label{cone2}
\end{equation}    
where $\tilde{\pmb{l}}_2 = E_{N_p} \otimes \pmb{l}_2$ with $\pmb{l}_2 = [0,\ 1,\ 0,\ 0,\ 0,\ 0]\in \mathbb{R}^{{1}\times {6}}$, $\tilde{\pmb{l}}_3 = E_{N_p} \otimes \pmb{l}_3$ with $\pmb{l}_3 = [0,\ 0,\ 1,\ 0,\ 0,\ 0]\in \mathbb{R}^{{1}\times {6}}$.
$\tilde {\pmb{\varepsilon}}^{\min}(k) = \sum^{N_p}_{i = 1} \varepsilon^{\min}(k+i) | i \rangle \in \mathbb{R}^{{N_p}}$  is the augmented vector of $\varepsilon^{\min}(k)$ shown in (\ref{entry_1}).
Similarly, $\tilde {\pmb{\varepsilon}}^{\max}(k) = \sum^{N_p}_{i = 1} \varepsilon^{\max}(k+i) | i \rangle \in \mathbb{R}^{{N_p}} $, $\tilde {\pmb{\beta}}^{\min}(k) = \sum^{N_p}_{i = 1} \beta^{\min}(k+i) | i \rangle \in \mathbb{R}^{{N_p}}$, $\tilde {\pmb{\beta}}^{\max}(k) = \sum^{N_p}_{i = 1} \beta^{\max}(k+i) | i \rangle \in \mathbb{R}^{{N_p}}$. 

Substituting (\ref{EEE}) into (\ref{cone1}) and (\ref{cone2}), i.e., 
\begin{equation}
G_{e1} \Delta \tilde{\pmb{u}}_p \leqslant \pmb{g}_{e1},
\label{Constraints_cone1}
\end{equation}
where
\begin{align*}
G_{e1} =  \left[\begin{array}{c}
     \tilde{\pmb{l}}_{2} \left(B_{p}^{\ast}  + B_{p}^{\ast} W_p^{\ast}\right) \Gamma\\
     - \tilde{\pmb{l}}_{2} \left(B_{p}^{\ast}  + B_{p}^{\ast} W_p^{\ast}\right) \Gamma
   \end{array}\right] ,\
\pmb{g}_{e1}= \left[\begin{array}{c}
   \tilde {\pmb{\varepsilon}}^{\max}(k) - \tilde{\pmb{l}}_2 \left[A_p^{\ast} \pmb{x}_{p}(k)  +  \left(B_{p}^{\ast}  + B_{p}^{\ast} W_p^{\ast}\right) \Lambda \pmb{u}_p(k\!-\!1)\right] \\
     - \tilde {\pmb{\varepsilon}}^{\min}(k)  +  \tilde{\pmb{l}}_2 \left[A_p^{\ast} \pmb{x}_{p}(k) + \left(B_{p}^{\ast}  + B_{p}^{\ast} W_p^{\ast}\right) \Lambda \pmb{u}_p(k\!-\!1)\right] 
   \end{array}\right].
\end{align*}
and

\begin{equation}
G_{e2} \Delta \tilde{\pmb{u}}_p \leqslant \pmb{g}_{e2},
\label{Constraints_cone2}
\end{equation}
where
\begin{align*}
G_{e2} =  \left[\begin{array}{c}
     \tilde{\pmb{l}}_{3} \left(B_{p}^{\ast}  + B_{p}^{\ast} W_p^{\ast}\right) \Gamma\\
     - \tilde{\pmb{l}}_{3} \left(B_{p}^{\ast}  + B_{p}^{\ast} W_p^{\ast}\right) \Gamma
   \end{array}\right] ,\
\pmb{g}_{e2} = \left[\begin{array}{c}
      \tilde {\pmb{\beta}}^{\max}(k) -  \tilde{\pmb{l}}_3 \left[A_p^{\ast} \pmb{x}_{p}(k) + \left(B_{p}^{\ast}  + B_{p}^{\ast} W_p^{\ast}\right) \Lambda \pmb{u}_p(k-1)\right] \\
     -\tilde {\pmb{\beta}}^{\min}(k) + \tilde{\pmb{l}}_3 \left[A_p^{\ast} \pmb{x}_{p}(k) +  \left(B_{p}^{\ast}  + B_{p}^{\ast} W_p^{\ast}\right) \Lambda \pmb{u}_p(k-1)\right]
   \end{array}\right].
\end{align*}


\subsubsection{Field of view constraint}
The constraint on the $N_p$ predictions of the roll angle is re-configurated as
\begin{equation}
  -\tilde {\pmb{\pi}} \leqslant \tilde{\pmb{l}}_1 \pmb{x}^{\ast}_a (k) \leqslant \tilde {\pmb{\pi}},
\label{Nfield1}
\end{equation}
where $\tilde {\pmb{\pi}} = \sum^{N_p}_{i = 1} {\pi} | i \rangle $.
Substituting (\ref{attitude dynamics_1}) into (\ref{Nfield1}), yields
\begin{equation}
G_{f1} \Delta \tilde{\pmb{u}}_p \leqslant \pmb{g}_{f1},
\label{Constraint_field_1}
\end{equation}
where
\begin{equation*}
\begin{aligned}
G_{f1} &=   \left[ \begin{array}{c}
    \tilde{\pmb{l}}_{1} \left(B_{a}^{\ast}  + B_{a}^{\ast} W_a^{\ast} \right) \Gamma \\
     - \tilde{\pmb{l}}_{1} \left(B_{a}^{\ast}  + B_{a}^{\ast} W_a^{\ast}\right) \Gamma
   \end{array}\right] ,\\
\pmb{g}_{f1}&= \left[\begin{array}{c}
    \tilde {\pmb{\pi}} - \tilde{\pmb{l}}_1 \left[A_a^{\ast} \pmb{x}_{a}(k) +  \left(B_{a}^{\ast}  +  B_{a}^{\ast} W_a^{\ast}\right) \Lambda \pmb{u}_a(k-1)\right]
   - \tilde {\pmb{\pi}} + \tilde{\pmb{l}}_1 \left[A_a^{\ast} \pmb{x}_{a}(k) + \left(B_{a}^{\ast}  + B_{a}^{\ast} W_a^{\ast}\right) \Lambda \pmb{u}_a(k-1)\right]
   \end{array}\right].
\end{aligned}
\end{equation*}

The constraint on the $N_p$ predictions of the pitch angle is re-configurated as
\begin{equation}
  \tilde {\pmb{\theta}}_c^{\min}(k) \leqslant \tilde{\pmb{l}}_2 \pmb{x}^{\ast}_a (k) \leqslant \tilde {\pmb{\theta}}_c^{\max}(k),
\label{Nfield2}
\end{equation}
where $\tilde {\pmb{\theta}}_c^{\min}(k) = \sum^{N_p}_{i = 1} \theta_c^{\min}(k+i) | i \rangle$ is the augmented vector of $\theta_c^{\min}(k)$ described in (\ref{field_1}). Similarly, $\tilde {\pmb{\theta}}_c^{\max}(k) = \sum^{N_p}_{i = 1} \theta_c^{\max}$ $(k+i) | i \rangle$.
Substituting (\ref{attitude dynamics_1}) into (\ref{Nfield2}), yields
\begin{equation}
G_{f2} \Delta \tilde{\pmb{u}}_p \leqslant \pmb{g}_{f2},
\label{Constraint_field_2}
\end{equation}
where
\begin{equation*}
\begin{aligned}
G_{f2} &= \left[ \begin{array}{c} \tilde{\pmb{l}}_{2} (B_{a}^{\ast}  + B_{a}^{\ast} W_a^{\ast}) \Gamma\\
     - \tilde{\pmb{l}}_{2} (B_{a}^{\ast}  + B_{a}^{\ast} W_a^{\ast}) \Gamma
   \end{array} \right],\\
\pmb{g}_{f2}&= \left[ \begin{array}{c}
  \tilde {\pmb{\theta}}_c^{\max}(k) - \tilde{\pmb{l}}_2 [A_a^{\ast} \pmb{x}_{a}(k) +  (B_{a}^{\ast}  \!+ \! B_{a}^{\ast} W_a^{\ast}) \Lambda \pmb{u}_a(k\!-\!1)] \\
- \tilde {\pmb{\theta}}_c^{\min}(k) + \tilde{\pmb{l}}_2 [A_a^{\ast} \pmb{x}_{a}(k) + (B_{a}^{\ast}  \!+\! B_{a}^{\ast} W_a^{\ast}) \Lambda \pmb{u}_a(k\!-\!1)]
   \end{array}\right].
\end{aligned}
\end{equation*}

The constraint on the $N_p$ predictions of the yaw angle is re-configurated as follows,
\begin{equation}
  \tilde {\pmb{\psi}}_c^{\min}(k) \leqslant \tilde{\pmb{l}}_3 \pmb{x}^{\ast}_a (k) \leqslant \tilde {\pmb{\psi}}_c^{\max}(k),
\label{Nfield3}
\end{equation}
where $\tilde {\pmb{\psi}}_c^{\min}(k) = \sum^{N_p}_{i = 1} \psi_c^{\min}(k+i) | i \rangle$, and $\tilde {\pmb{\psi}}_c^{\max}(k) = \sum^{N_p}_{i = 1}$ $ \psi_c^{\max}$ $(k+i) | i \rangle$.
Substituting (\ref{attitude dynamics_1}) into (\ref{Nfield3}), yields
\begin{equation}
G_{f3} \Delta \tilde{\pmb{u}}_p \leqslant \pmb{g}_{f3},
\label{Constraint_field_3}
\end{equation}
where
\begin{align*}
G_{f3} =   \left[ \begin{array}{c}
    \tilde{\pmb{l}}_{3} \left(B_{a}^{\ast}  + B_{a}^{\ast} W_a^{\ast} \right) \Gamma\\
    - \tilde{\pmb{l}}_{3} \left(B_{a}^{\ast}  + B_{a}^{\ast} W_a^{\ast} \right) \Gamma
   \end{array}\right], \
\pmb{g}_{f3}= \!\left[\begin{array}{c}
    \tilde {\pmb{\psi}}_c^{\max}(k) - \tilde{\pmb{l}}_3\left[ A_a^{\ast} \pmb{x}_{a}(k) + \left( B_{a}^{\ast}  +  B_{a}^{\ast} W_a^{\ast} \right) \Lambda \pmb{u}_a(k-1)\right] \\
     - \tilde {\pmb{\psi}}_c^{\min}(k) + \tilde{\pmb{l}}_3\left[ A_a^{\ast} \pmb{x}_{a}(k) + \left( B_{a}^{\ast}  + B_{a}^{\ast} W_a^{\ast} \right) \Lambda \pmb{u}_a(k-1)\right]
   \end{array}\right].
\end{align*}

\subsection{Implement of sampling-based  PWA MPC }
Consider the optimization index (\ref{JJ_p}), and the reconfiguration of constraints (\ref{control_p}), (\ref{Constraint_coliision}), (\ref{Constraints_cone1}), and (\ref{Constraints_cone2}), the optimal control problem over the predictive horizon is converted to the following convex QP problems, i.e.,
\begin{align}
&\min\,\, J_p=\frac{1}{2} \Delta
   \tilde{\pmb{u}}_p^{\mathrm{T}} H_p \Delta \tilde{\pmb{u}}_p + \pmb{f}_p^{\mathrm{T}} \Delta
   \tilde{\pmb{u}}_p^{\mathrm{T}} + \pmb{E}_p^{\mathrm{T}} \tilde{Q} \pmb{E}_p\\
&s.t.\quad
 \begin{cases} 
      \pmb{x}_p(k|k)=\pmb{x}_p(k)\\
     \pmb{x}^{\ast}_p(k) = A_p^{\ast} \pmb{x}_{p}(k) + {B}_p^{\ast}W_p^{\ast}\pmb{u}^{\ast}_{p}  + 
B_{p}^{\ast}\pmb{u}^{\ast}_{p}\\
     G^{\ast}_P \Delta \tilde{\pmb{u}}_p \leqslant \pmb{g}^{\ast}_P
   \end{cases}
\label{OP_p}
\end{align}
where $G^{\ast}_{P} = [G_{c}^{\mathrm{T}},\ G_a^{\mathrm{T}},\ G_{e1}^{\mathrm{T}},\ G_{e2}^{\mathrm{T}}]^{\mathrm{T}}$, $\pmb{g}^{\ast}_P = [\pmb{g}_{c,p}^{\mathrm{T}},\ \pmb{g}_a^{\mathrm{T}},\ \pmb{g}_{e1}^{\mathrm{T}},\ $ $\pmb{g}_{e2}^{\mathrm{T}}]^{\mathrm{T}}$.

Similarly,  consider the optimization index (\ref{JJ_a}), and the reconfiguration of constraints (\ref{control_a}),  (\ref{Constraint_field_1}),  (\ref{Constraint_field_2}), and (\ref{Constraint_field_3}), the optimal control problem over the predictive horizon is converted to the following convex QP problems, i.e.,
\begin{align}
&\min\,\, J_a=\frac{1}{2} \Delta
   \tilde{\pmb{u}}_a^{\mathrm{T}} H_a \Delta \tilde{\pmb{u}}_a + \pmb{f}_a^{\mathrm{T}} \Delta
   \tilde{\pmb{u}}_a^{\mathrm{T}} + \pmb{E}_a^{\mathrm{T}} \tilde{Q} \pmb{E}_a\\
&s.t.\quad
 \begin{cases} 
      \pmb{x}_a(k|k)=\pmb{x}_a(k)\\
     \pmb{x}^{\ast}_a(k) = A_a^{\ast} \pmb{x}_{a}(k) +  
B_{a}^{\ast}\pmb{u}^{\ast}_{a}\\
     G^{\ast}_A \Delta \tilde{\pmb{u}}_a \leqslant \pmb{g}^{\ast}_A
   \end{cases}
\label{OP_a}
\end{align}
where $G^{\ast}_A = \left[G_{c}^{\mathrm{T}},\  G_{f1}^{\mathrm{T}},\ G_{f2}^{\mathrm{T}},\ G_{f3}^{\mathrm{T}} \right]^{\mathrm{T}}$, $\pmb{g}^{\ast}_A = \left[\pmb{g}_{c,a}^{\mathrm{T}},\pmb{g}_{f1}^{\mathrm{T}},\ \pmb{g}_{f2}^{\mathrm{T}},\ \pmb{g}_{f3}^{\mathrm{T}} \right]^{\mathrm{T}}$. 

Then the optimal control input sequence is solved by the QP solver. Compared with the existing paper, the input sequence is obtained by the PWA model, and the input signal related to the sampling instant is imposed on the actual nonlinear model rather than the PWA model, which is more coincident with the real case.

\section{Singularity free strategy}
\label{RVD strategy}
\begin{figure}[tbh]
\centering
\includegraphics[width=0.5\textwidth]{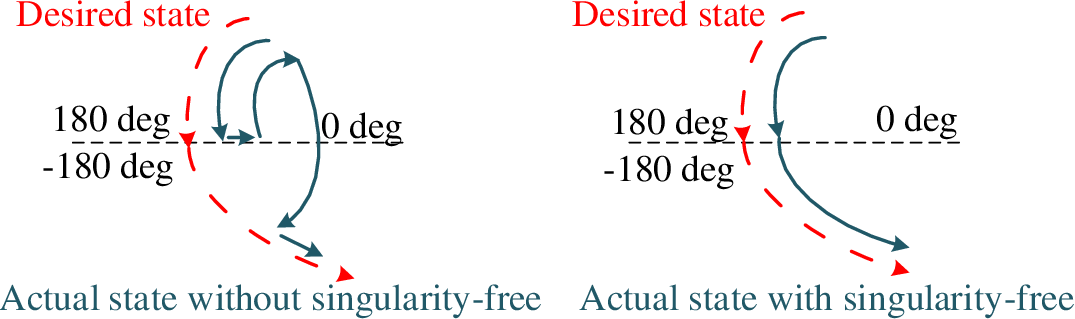}
\caption{Singularity}
\label{Singularity}
\end{figure}

In the LOS-Euler RVD framework, the LOS frame $\mathcal{F}_s$ is obtained by the Euler rotation of LVLH frame $\mathcal{F}_l$, the target's/chaser's attitude is described by the Euler rotation from LVLH frame $\mathcal{F}_l$ to the body frame $\mathcal{F}_{bt}$/$\mathcal{F}_{bc}$, which requires the second rotation angle is set within $[ - \pi/2,\ \pi/2]$, the first and the third rotation angle are set within $[ - \pi,\ \pi]$ to avoid the gimbal lock phenomenon. However, it should be noted that $-\pi$ and $\pi$ represent a same position. As shown in Fig. \ref{Singularity}, singularity means two different values represent a same position physically. Consider a tracking problem in which the desired state crosses the singularity, the mathematical expression of the desired state jumps once it reaches the singularity, however, it is difficult for an input constraint system to control the actual state to track the unnecessary jump desired state immediately (the error doesn't jump physically). Actually, the actual state tracks the desired state in the opposite direction, which results considerable tracking error and the system can not realize continuous tracking.  
In this subsection, a singularity free strategy is provided to avoid the above situation. 

The basic idea of the proposed strategy is to select the correct value from different singular values, according to the continuity of motion. Then design a mechanism to switch the mathematical expression both of the desired signal and the actual state without changing the physical position. Therefore, the actual state tracks along the direction of the desired state's movement continuously with crossing the singularities. 

As shown in the Algorithm \ref{alg1}, assuming that the motion of the tumbling target is continuous without continuous switching at the singularity. Consider the following cases: (i) the desired state reaches the singularity clockwise and the actual state tracks in a lag way; (ii) the desired state reaches the singularity clockwise and the actual state tracks in a leading way; (iii) the desired state reaches the singularity counterclockwise and the actual state tracks in a lag way; (iv) the desired state reaches the singularity counterclockwise and the actual state tracks in a leading way. 

For case 1, the motion of the desired state is described by $-\pi+\delta (t),\ \delta(t)>0$ once it reaches $\pi$, the actual state can be expressed by $\pi-\zeta(t),\ \zeta(t)>0$. According to the proposed strategy, the desired state is switched to $\pi+\delta (t)$ firstly (without changing the physical position) to ensure that the actual state still moves along the desired direction, until the actual signal reaches the singularity. Assume the actual state reaches the singularity at $t_1$, then the actual state is switched to $-\pi$, and the desired state is switched to $-\pi+\delta (t_1)$ to ensure that the tracking continues. 

In case 2, the actual state reaches the singularity but the desired state doesn't, firstly set the actual state to $-\pi$, assuming that the motion of the desired state is described by $\pi-\delta (t),\ \delta( t)>0$, then switch the desired state to $-\pi- \delta (t)$ to ensure that the actual state still moves along the desired direction, until the actual signal or the desired state reaches the singularity. The strategy in case 3 and case 4 are similar.

It should be noted that the discrete state can't always reach the singular point precisely for a discrete-time system. Our method is to set a neighborhood field near the singular point, once the state reaches the neighborhood, we consider the state reaches the singular point approximately.

\begin{algorithm}
\caption{Singularity-free strategy} 
\label{alg1}
\begin{algorithmic}[1]
\If {  \{$x_{\mathrm{d}}$ moves clockwise\}  }
     \If { \{  $x_{\mathrm{d}} (k) = n_x\pi$ \&\& $x(k) < n_x\pi$ \} }
     \State $i=1$\\
     \qquad\quad Repeat
     \State $x_{\mathrm{d}}(k+i) \Leftarrow x_{\mathrm{d}}(k+i) +2n_x\pi$ 
     \State $i \Leftarrow i+1$\\
     \qquad\quad Until\
     {$x (k+N) = n_x\pi$},\ 
      $x(k+N+1) \Leftarrow -n_x\pi$
     \Else \{ $x_{\mathrm{d}} (k) < n_x\pi$ \&\& $x(k) = n_x\pi$ \}
     \State $x (k+1) = -n_x\pi$
     \State $i=1$\\
     \qquad\quad Repeat
     \State $x_{\mathrm{d}}(k+i) \Leftarrow x_{\mathrm{d}}(k+i) - 2n_x\pi$
     \State $i \Leftarrow i+1$\\ 
     \qquad\quad Until\
       {$x_{\mathrm{d}} (k+N) = n_x\pi$} 
     \EndIf 
\Else   \{ $x_{\mathrm{d}}$ moves counter-clockwise \}  

     \If { \{  $x_{\mathrm{d}} (k) = -n_x\pi$ \&\& $x(k) > n_x\pi$  \} }
     \State $i=1$\\
     \qquad\quad Repeat
     \State $x_{\mathrm{d}}(k+i) \Leftarrow x_{\mathrm{d}}(k+i) - 2n_x\pi$ 
     \State $i \Leftarrow i+1$\\
     \qquad\quad Until\
      {$x (k+N) = -n_x\pi$},\  
      $x(k+N+1) \Leftarrow n_x\pi$
     \Else \{ $x_{\mathrm{d}} (k) > -n_x\pi$ \&\& $x(k) = -n_x\pi$ \}
     \State $x (k+1) = n_x\pi$
     \State $i=1$\\
     \qquad\quad Repeat
     \State $x_{\mathrm{d}}(k+i) \Leftarrow x_{\mathrm{d}}(k+i) + 2n_x\pi$
     \State $i \Leftarrow i+1$ \\ 
     \qquad\quad Until\  {$x_{\mathrm{d}} (k+N) = -n_x\pi$} 
     \EndIf 
 \EndIf 
\end{algorithmic}
\end{algorithm}

\section{Numerical simulations}
In this section, three different RVD scenarios are presented to illustrate the effectiveness of the proposed approaches. Compared with the existing papers, the control input designed by the PWA model is imposed on the nonlinear system to simulate the actual situation. Consider there exists unsolvable optimization in standard PWA MPC under strong RVD constraint, only the control input constraint is considered in case 1. The performance of both adopting the sampling-based strategy and without the strategy are compared. In case 2, all the RVD constraints are considered, and the performance of the singularity free strategy is shown. Case 3 presents the control performance of sampling-based PWA MPC under disturbance. 

Assuming the target spacecraft is orbiting in an elliptical orbit, with eccentricity $e$ is $0.3$, semi-major axis $a$ is $10000\ \mathrm{km}$, initial true anomaly $\nu$ is $0\ \mathrm{deg}$. The target's dimension is assumed as $2.5\ \mathrm{m}\times 2.5\ \mathrm{m}\times 2.5\ \mathrm{m} $. The coordinate of the target's docking port is assumed $[0.5,\ 0,\ 0]^{\mathrm{T}} \mathrm{m}$ in the body frame $\mathcal{F}_{bt}$. The angular velocity of the tumbling target is $[0.02,\ 0.015,\ 0.02]^{\mathrm{T}}\ \mathrm{rad/s}$. The chaser's inertia matrix is $\mathrm{diag}(3.0514,\ 2.6628,\ 2.1879)\ \mathrm{kg \cdot m^2}$ in the body frame $\mathcal{F}_{bt}$. The wheels' inertia matrix is $\mathrm{diag}(0.5,\ 0.5,\ 0.5)\ \mathrm{kg \cdot m^2}$ in the body frame $\mathcal{F}_{bt}$. The dimension of the chaser is $2.5\ \mathrm{m}\times 2.5\ \mathrm{m}\times 2.5\ \mathrm{m} $. The coordinate of the chaser's docking port is assumed $[-0.5,\ 0,\ 0]^{\mathrm{T}} \mathrm{m}$ in the body frame $\mathcal{F}_{ct}$.

For the parameters of MPC, the prediction horizon $N_p$ is set as 30, and the control horizon $N_c$ is 15. The simulation duration is $200\ \mathrm{s}$, the sampling interval $T_s$ is $0.1\ \mathrm{s}$. The weight matrices $P$ and $Q$ in the relative position controller is set as $\mathrm{diag}(200,\ 200,\ 200 )$  and  $\mathrm{diag}(500,\ 3500,\ 3500 ,\ $ $ 500$,
\noindent $\ 500,\ 500)$. The weight matrices $P$ and $Q$ in the relative attitude controller are set $\mathrm{diag}(100,\ 100,\ 100 )$ and $\mathrm{diag}$ $(5000,\ 5000,\ 5000,$ 
\noindent $500,\ 500,\ 500)$.  The RVD conditions are provided in Table \ref{conditions}.  
\begin{remark}
Both the methods of approximating the discrete convexity in Section \ref{controller design} are proved effective. The following simulations are based on the first method. Note that the effectiveness of this method depends on the selection of the sampling interval. If the sampling interval is too large, the continuous convexity at the sampling state can not approximate the discrete nonlinearity, the second method should be adopted in this case.
\end{remark}


\subsection{Case 1: Rendezvous under control input constraint}

\begin{figure*}
\centering
\subfigure{\label{c1_p1}}\addtocounter{subfigure}{-2}
\subfigure{\subfigure[$\rho$ tracking]{\includegraphics[width=0.325\textwidth]{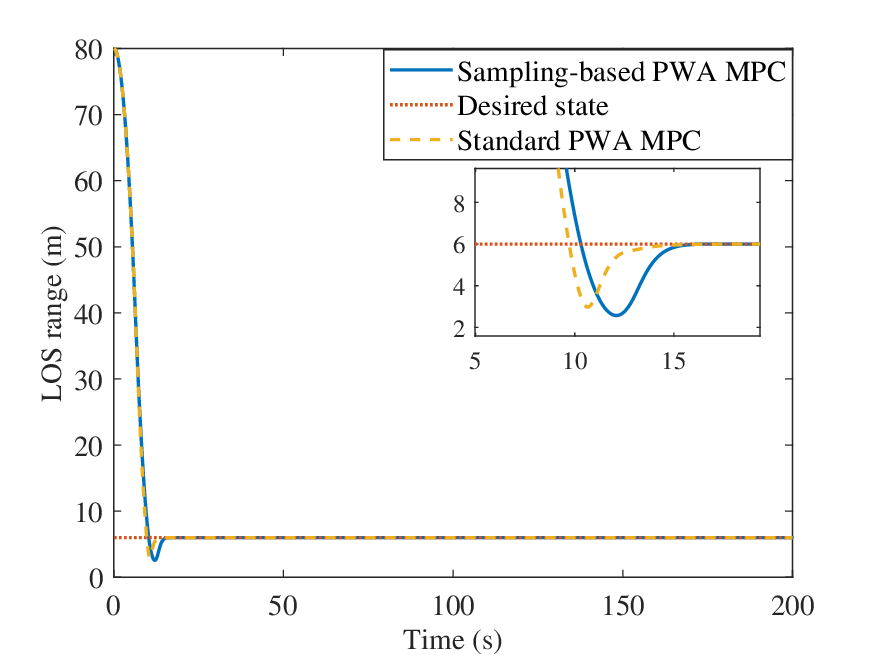}}}
\subfigure{\label{c1_p2}}\addtocounter{subfigure}{-2}
\subfigure{\subfigure[Elevation $\varepsilon$ tracking]{\includegraphics[width=0.325\textwidth]{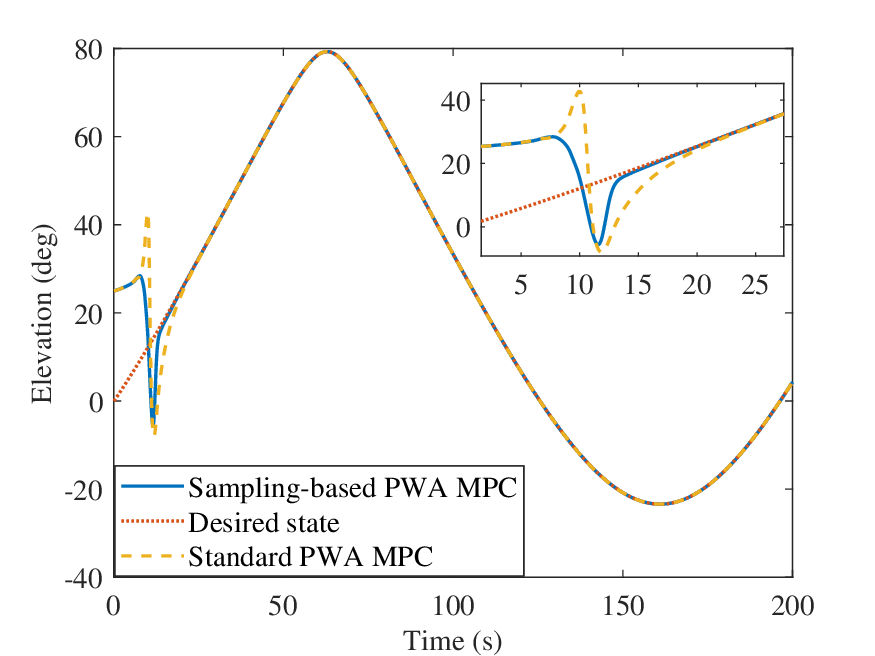}}}
\subfigure{\label{c1_p3}}\addtocounter{subfigure}{-2}
\subfigure{\subfigure[Azimuth $\beta$ tracking]{\includegraphics[width=0.325\textwidth]{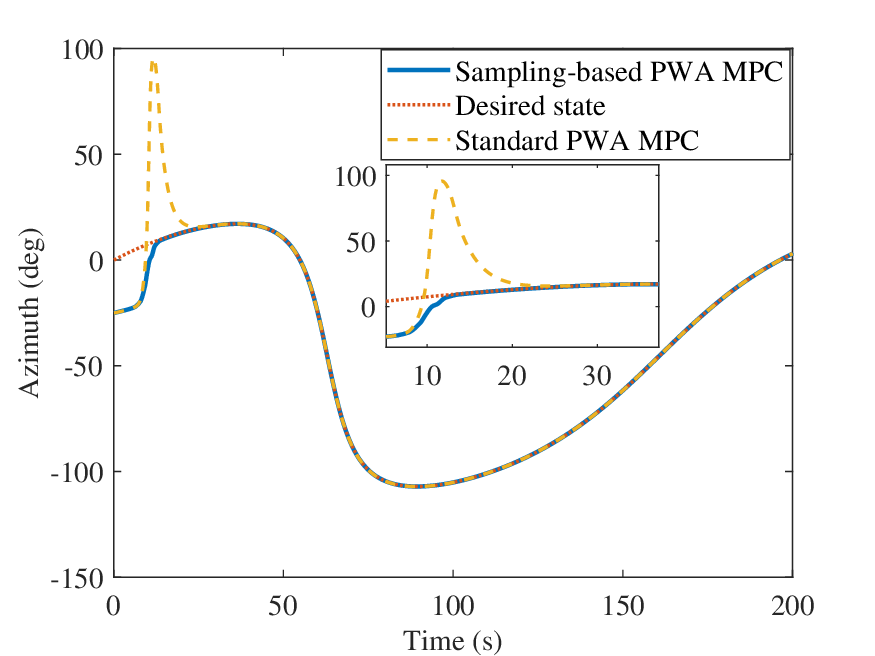}}}
\subfigure{\label{c1_p4}}\addtocounter{subfigure}{-2}
\subfigure{\subfigure[$\dot{\rho}$ tracking]{\includegraphics[width=0.325\textwidth]{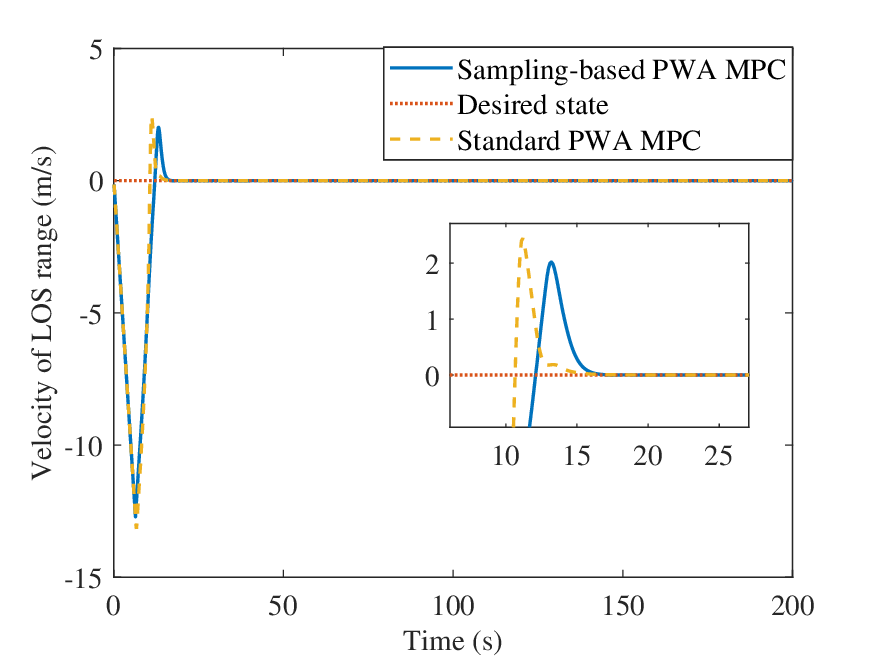}}}
\subfigure{\label{c1_p5}}\addtocounter{subfigure}{-2}
\subfigure{\subfigure[$\rho \dot{\varepsilon}$ tracking]{\includegraphics[width=0.325\textwidth]{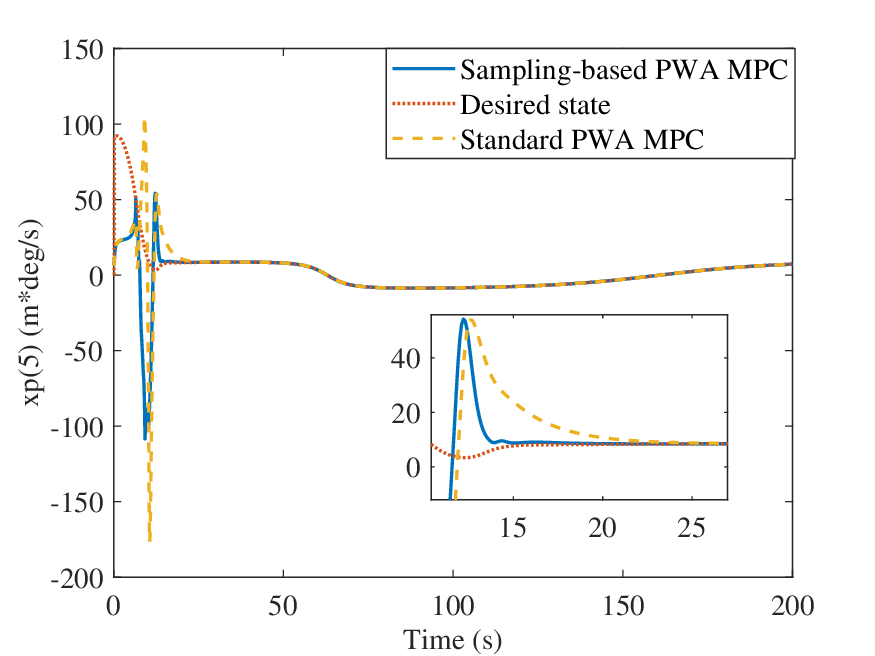}}}
\subfigure{\label{c1_p6}}\addtocounter{subfigure}{-2}
\subfigure{\subfigure[$\rho \dot{\beta}$ tracking]{\includegraphics[width=0.325\textwidth]{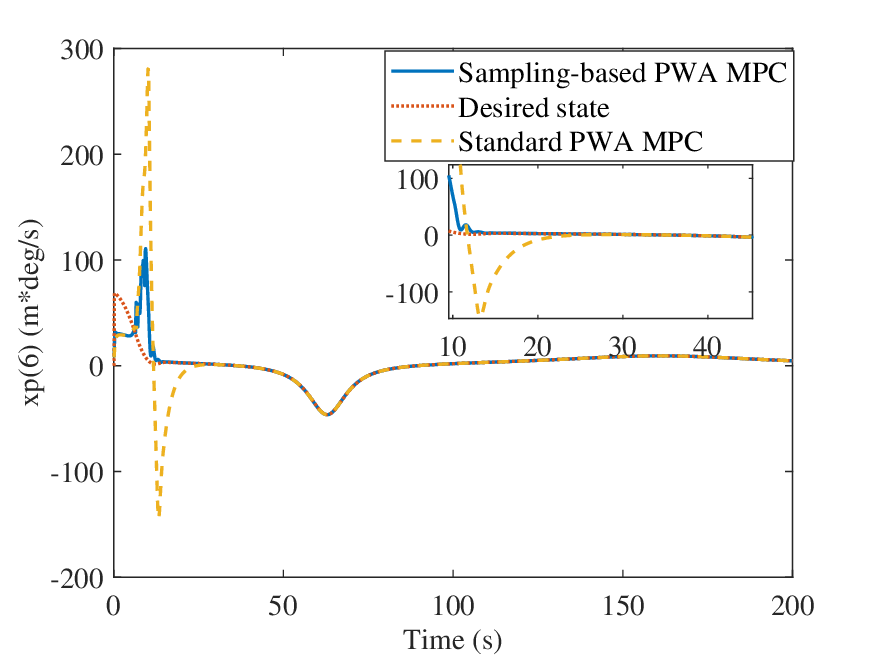}}}
\caption{Case 1: Performance of position tracking}
\label{Case1}
\end{figure*}

\begin{figure*}
\centering
\subfigure{\label{c1_xyz_0}}\addtocounter{subfigure}{-2}
\subfigure{\subfigure[3-D RVD by standard PWA MPC]{\includegraphics[width=0.33\textwidth]{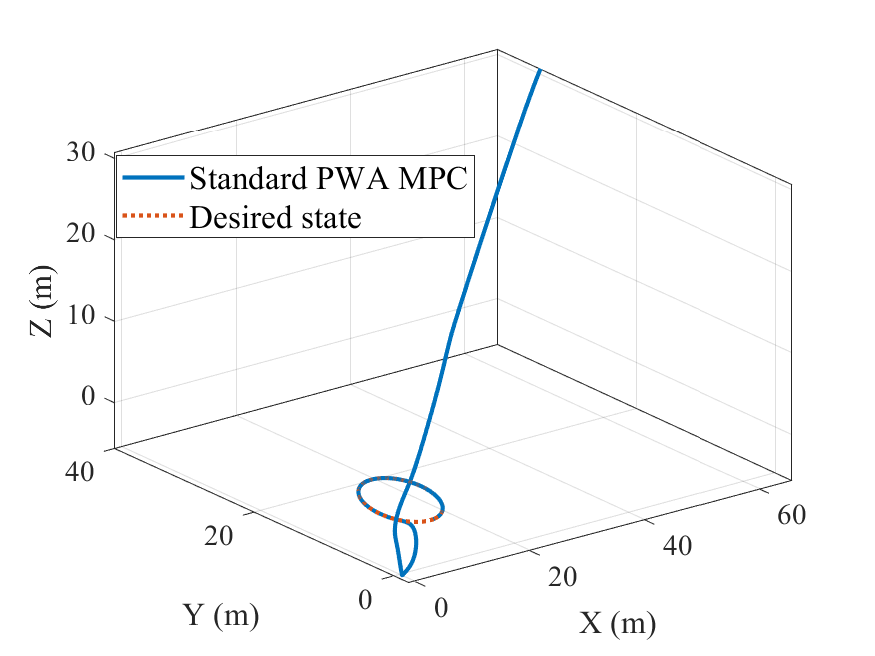}}}
\subfigure{\label{c1_xyz}}\addtocounter{subfigure}{-2}
\subfigure{\subfigure[3-D RVD by sampling based PWA MPC]{\includegraphics[width=0.33\textwidth]{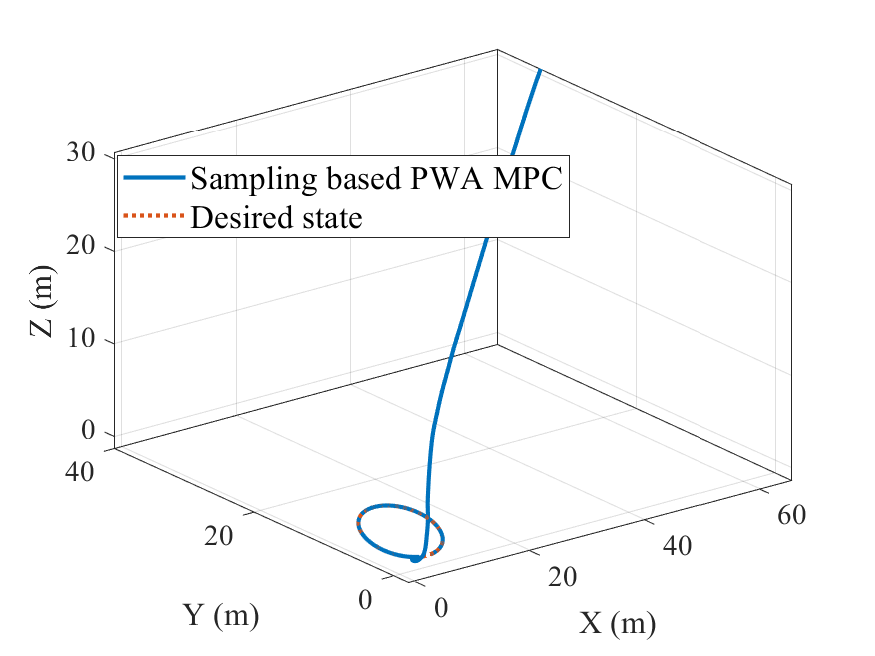}}}
\subfigure{\label{c1_e_xyz_0}}\addtocounter{subfigure}{-2}
\subfigure{\subfigure[Error described in LVLH by standard PWA MPC]{\includegraphics[width=0.33\textwidth]{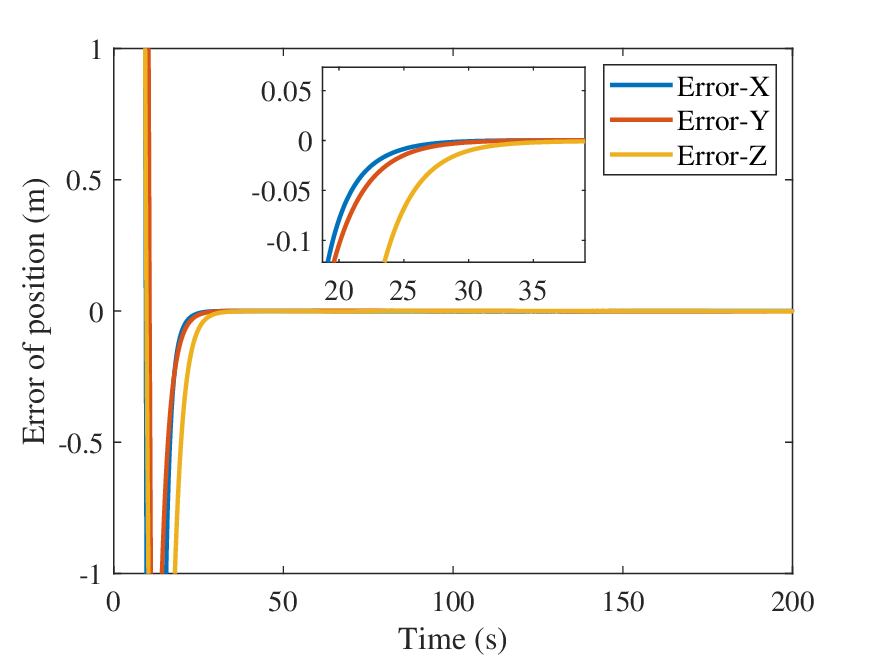}}}
\subfigure{\label{c1_e_xyz}}\addtocounter{subfigure}{-2}
\subfigure{\subfigure[Error described in LVLH by sampling based PWA MPC]{\includegraphics[width=0.33\textwidth]{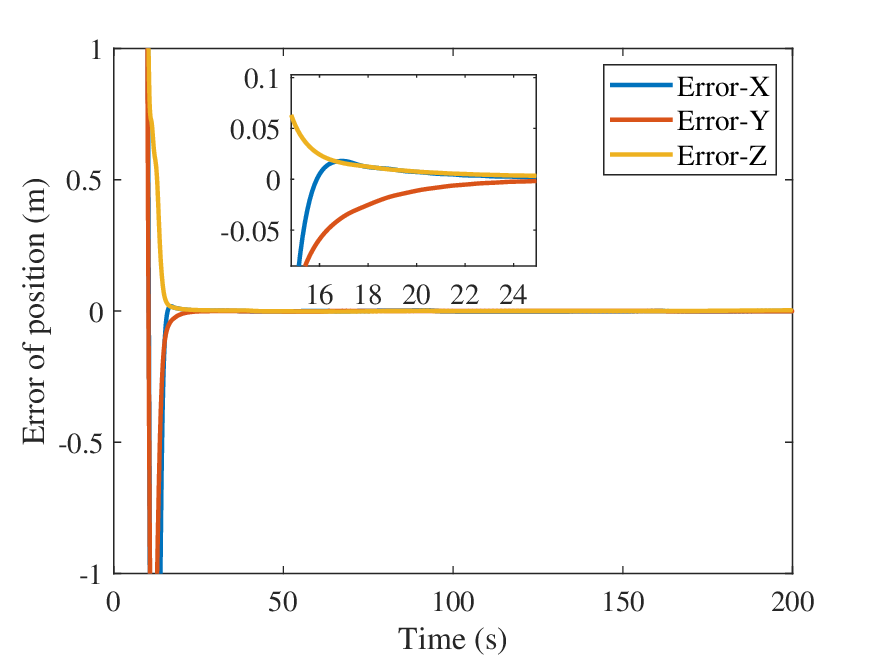}}}
\caption{Case 1: Position tracking in LVLH frame}
\label{Case1_3D}
\end{figure*}

\begin{figure*}
\centering
\subfigure{\label{c1_u0}}\addtocounter{subfigure}{-2}
\subfigure{\subfigure[standard PWA MPC]{\includegraphics[width=0.33\textwidth]{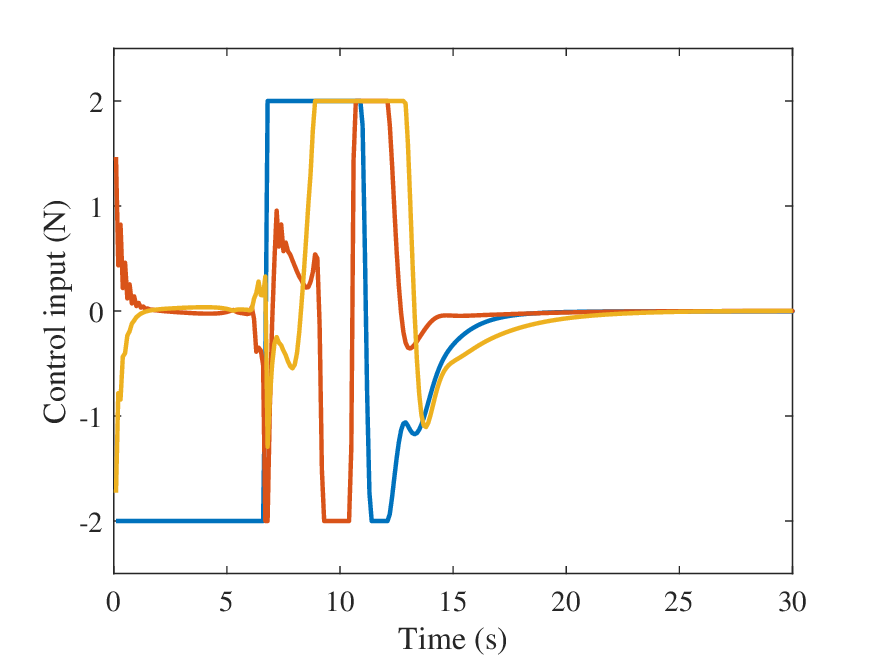}}}
\subfigure{\label{c1_u}}\addtocounter{subfigure}{-2}
\subfigure{\subfigure[sampling based PWA MPC]{\includegraphics[width=0.33\textwidth]{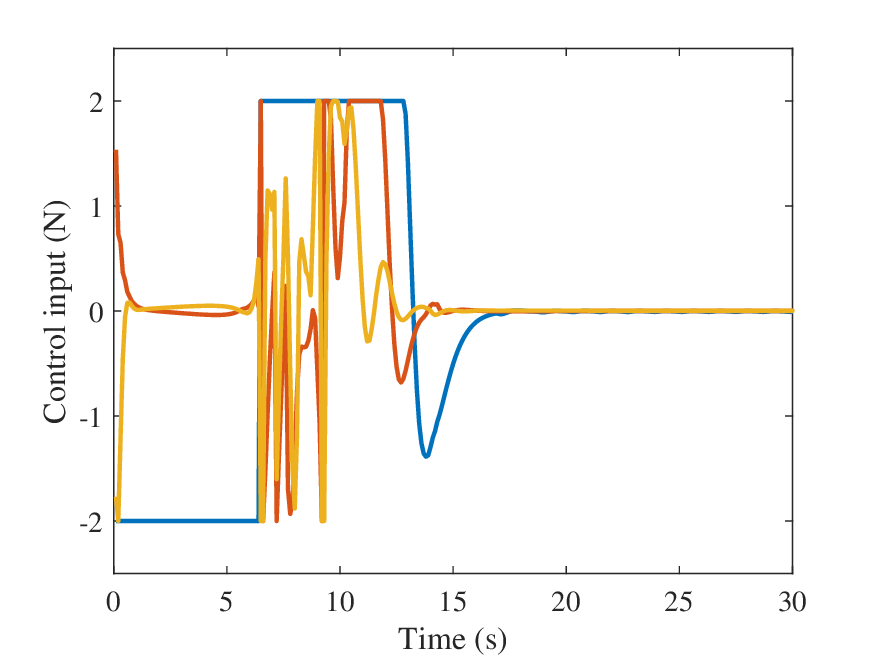}}}
\caption{Case 1: Control input}
\label{Case1_control}
\end{figure*}

In this case, the sampling parameter $w_{si}$ ($i=1,\ 2,\ 3$) is set $0.7,\ 0.3,\ 0.7$. The control performance both of the sampling based PWA MPC and standard PWA MPC under the control input constraint are shown in Fig. \ref{Case1}. Fig. \ref{c1_p1} and Fig. \ref{c1_p4} present the tracking performance of the LOS range $\rho$ and its velocity $\dot{\rho}$. Fig. \ref{c1_p2} and Fig. \ref{c1_p5} show the control performance of the elevation angle $\varepsilon$ and  $\rho \dot{\varepsilon}$, it can be seen that the proposed method can achieve lower overshoot and shorter convergence time. Fig. \ref{c1_p3} and Fig. \ref{c1_p6} show the tracking performance of the azimuth angle $\beta$ and  $\rho \dot{\beta}$, it can be seen that the proposed method can achieve no overshoot and shorter convergence time. Fig. \ref{Case1_3D} shows the RVD process described in LVLH frame, the proposed approach shown in Fig. \ref{c1_xyz} can achieve better docking performance compared with Fig. \ref{c1_xyz_0}. Besides, the sampling-based approach has faster convergence of the tracking error described in LVLH frame (Fig. \ref{c1_e_xyz}). According to Fig. \ref{c1_u0} and Fig. \ref{c1_u}, the control input constraint is satisfied.
Through reducing the predictive error using the sampling input, then the model deviation between the predictive model and the actual control model is improved.
The above figures also show that the proposed methods can achieve lower overshoot and faster convergence, since less control signal is needed to compensate the model deviation. 

Table \ref{comparison} provides the detailed index of the tracking performance. In this case, the convergence time is defined by the time when the tracking error of $\rho$ and the coordinates in LVLH frame are lower than $0.1 \ \mathrm{m}$; the error of $\varepsilon$ and $\beta$ are lower than $0.1 \ \mathrm{deg}$; the tracking error of $\dot{\rho}$ is lower than $0.1 \mathrm{m/s}$; the error of $\dot{\rho}\varepsilon$ and $\dot{\rho} \beta$ are lower than $0.1 \ \mathrm{m\cdot deg/s}$. The convergence accuracy is defined by the average of the error's modulus after the convergence time.

\subsection{Case 2: Docking under all RVD constraints}

\begin{figure*}
\centering
\subfigure{\label{c2_p1}}\addtocounter{subfigure}{-2}
\subfigure{\subfigure[$\rho$ tracking]{\includegraphics[width=0.325\textwidth]{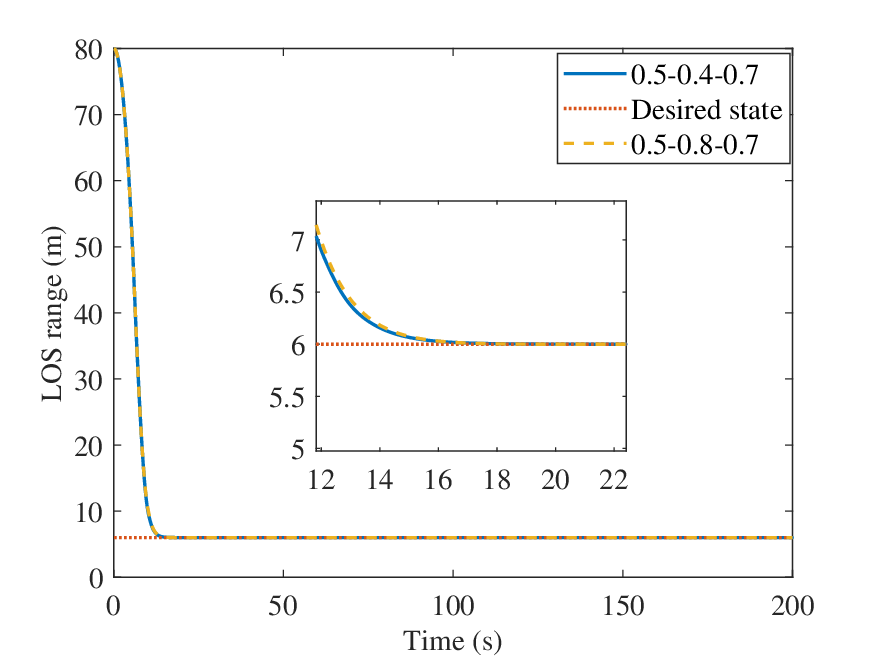}}}
\subfigure{\label{c2_p2}}\addtocounter{subfigure}{-2}
\subfigure{\subfigure[Elevation $\varepsilon$ tracking]{\includegraphics[width=0.325\textwidth]{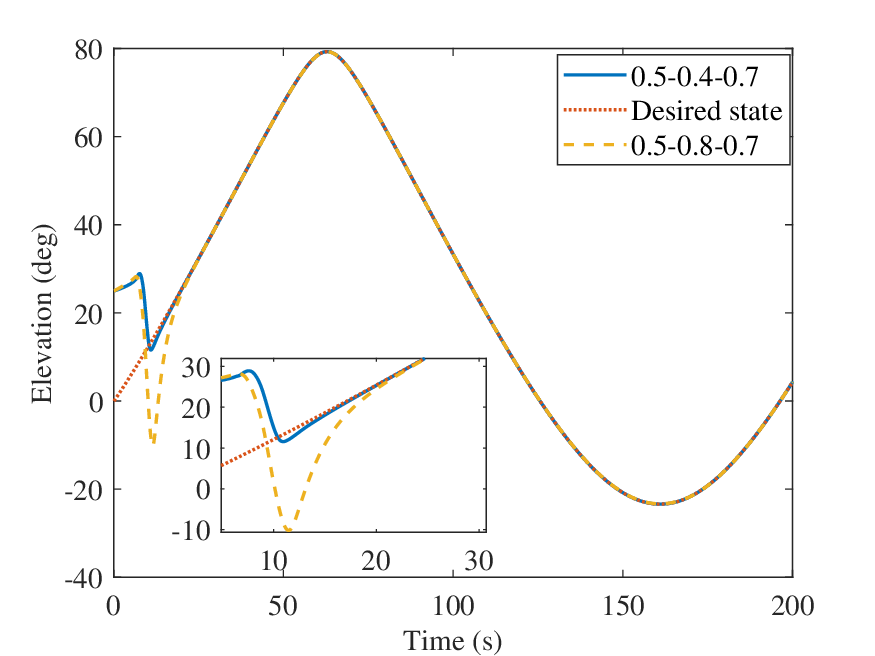}}}
\subfigure{\label{c2_p3}}\addtocounter{subfigure}{-2}
\subfigure{\subfigure[Azimuth $\beta$ tracking]{\includegraphics[width=0.325\textwidth]{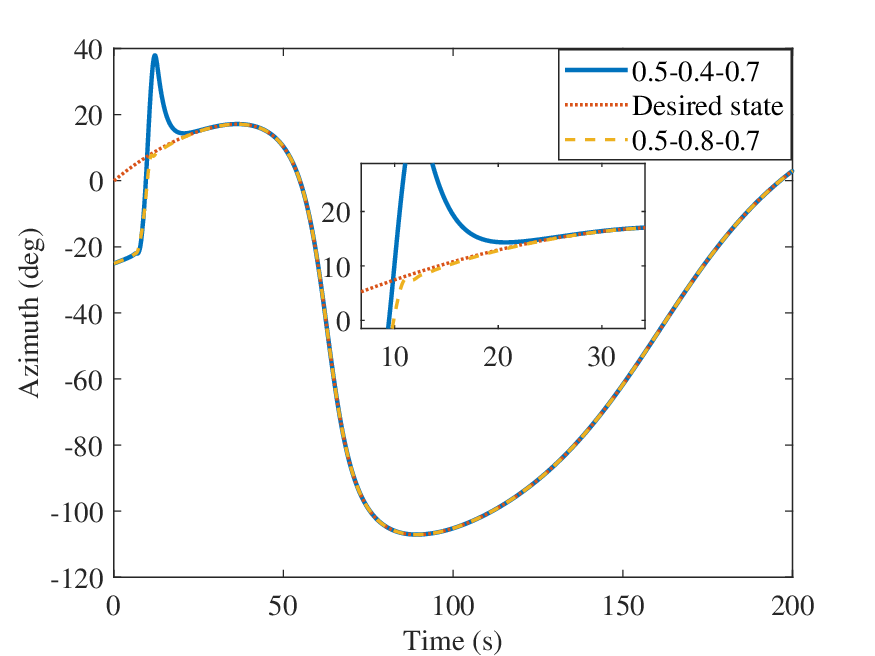}}}
\subfigure{\label{c2_p4}}\addtocounter{subfigure}{-2}
\subfigure{\subfigure[$\dot{\rho}$ tracking]{\includegraphics[width=0.325\textwidth]{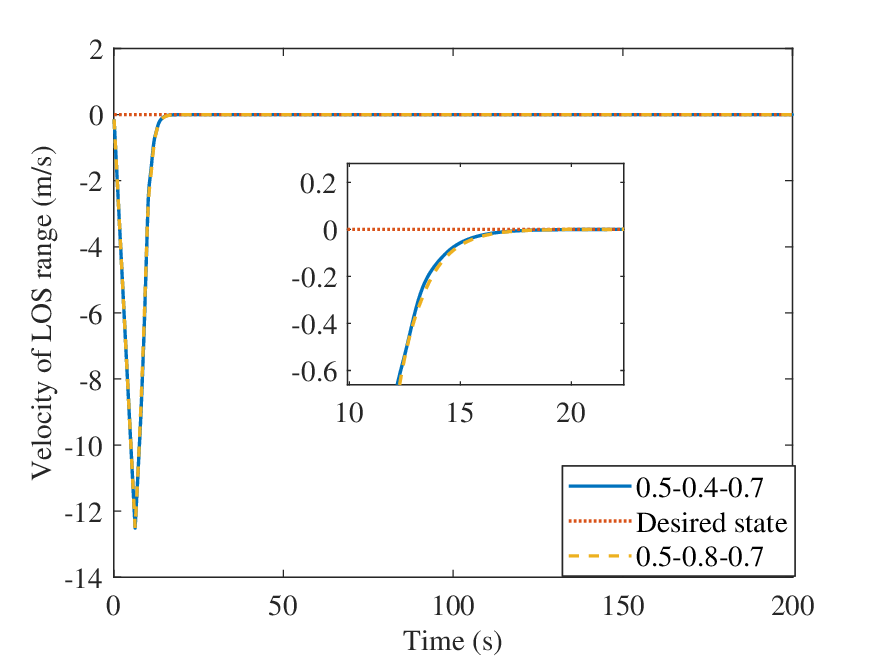}}}
\subfigure{\label{c2_p5}}\addtocounter{subfigure}{-2}
\subfigure{\subfigure[$\rho \dot{\varepsilon}$ tracking]{\includegraphics[width=0.325\textwidth]{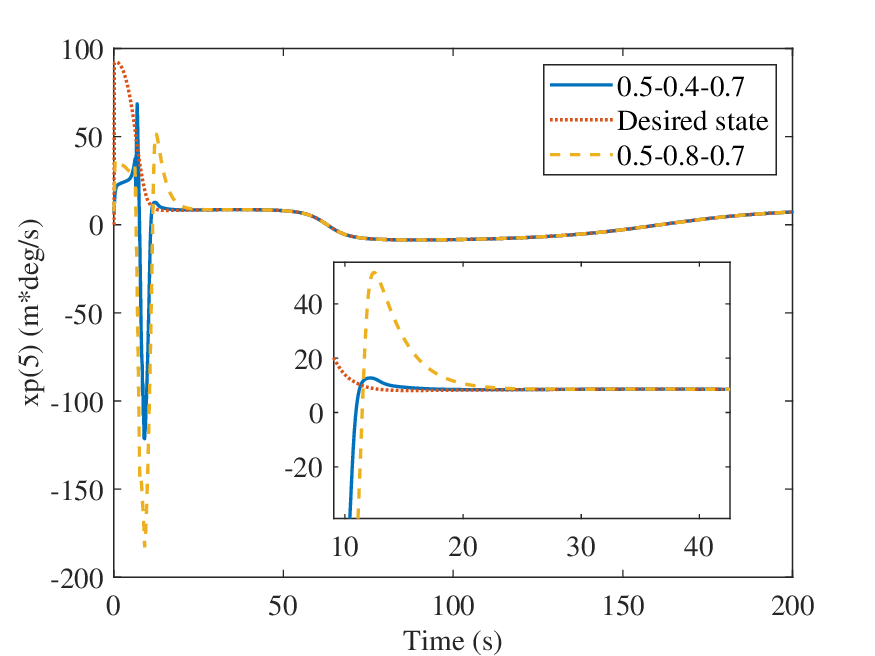}}}
\subfigure{\label{c2_p6}}\addtocounter{subfigure}{-2}
\subfigure{\subfigure[$\rho \dot{\beta}$ tracking]{\includegraphics[width=0.325\textwidth]{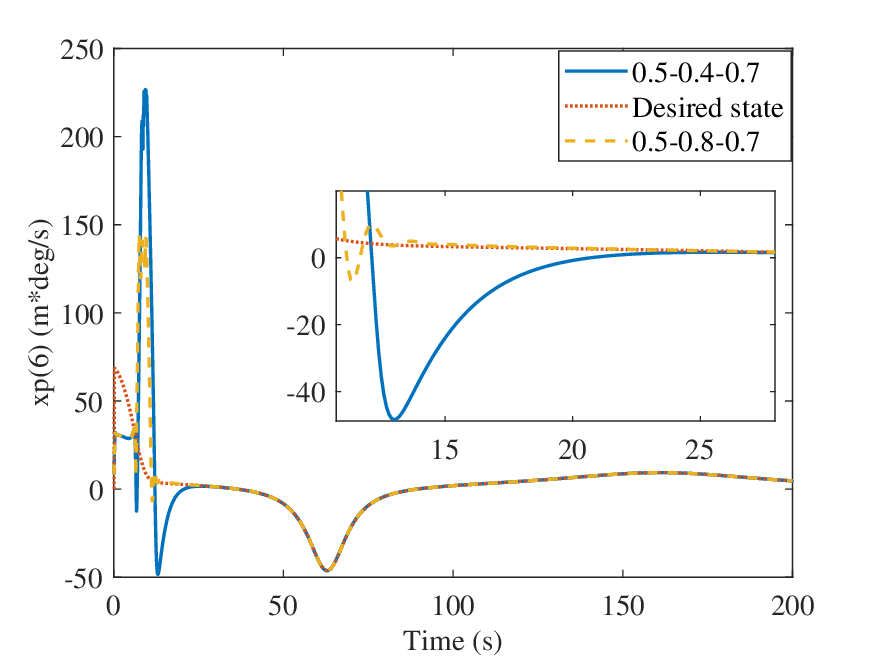}}}
\caption{Case 2: Performance of position tracking under RVD constraints}
\label{Case2}
\end{figure*}

\begin{figure*}
\centering
\subfigure{\label{c2_e1}}\addtocounter{subfigure}{-2}
\subfigure{\subfigure[Error of $\varepsilon$]{\includegraphics[width=0.33\textwidth]{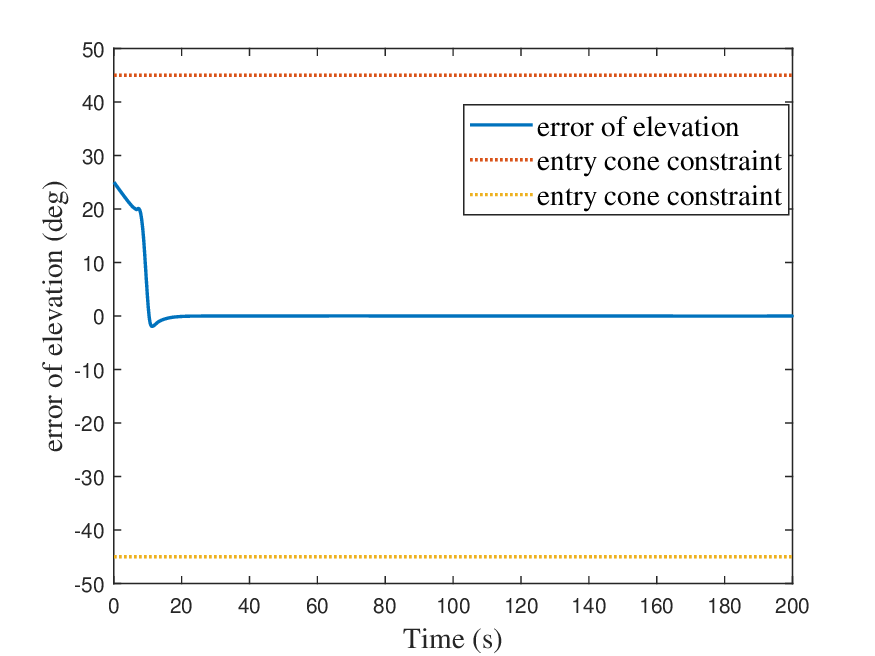}}}
\subfigure{\label{c1_e2}}\addtocounter{subfigure}{-2}
\subfigure{\subfigure[Error of $\beta$]{\includegraphics[width=0.33\textwidth]{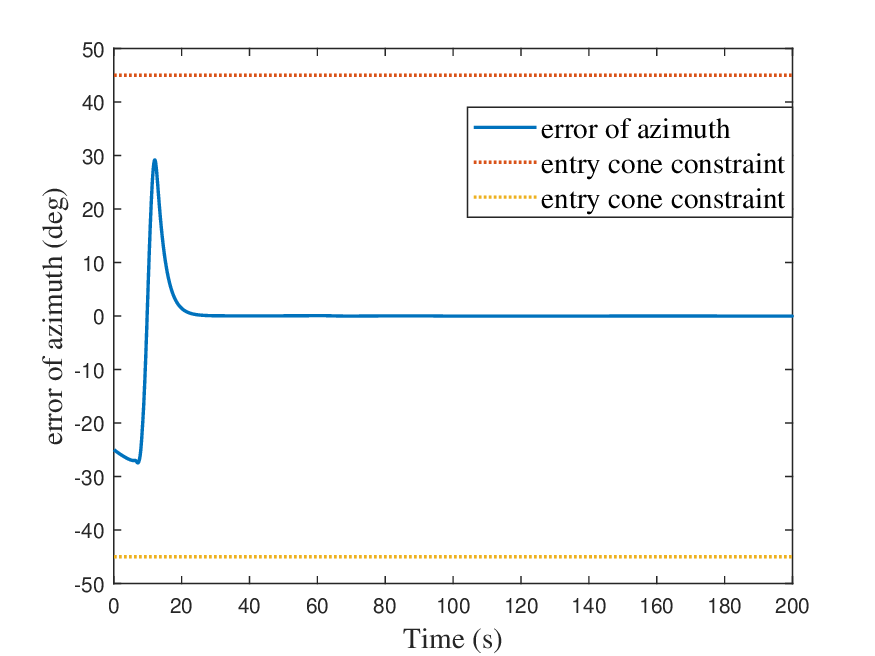}}}
\caption{Case 2: Position tracking error}
\label{Case2_error}
\end{figure*}

\begin{figure*}
\centering
\subfigure{\label{c2_xyz}}\addtocounter{subfigure}{-2}
\subfigure{\subfigure[3-D RVD by sampling based PWA MPC]{\includegraphics[width=0.325\textwidth]{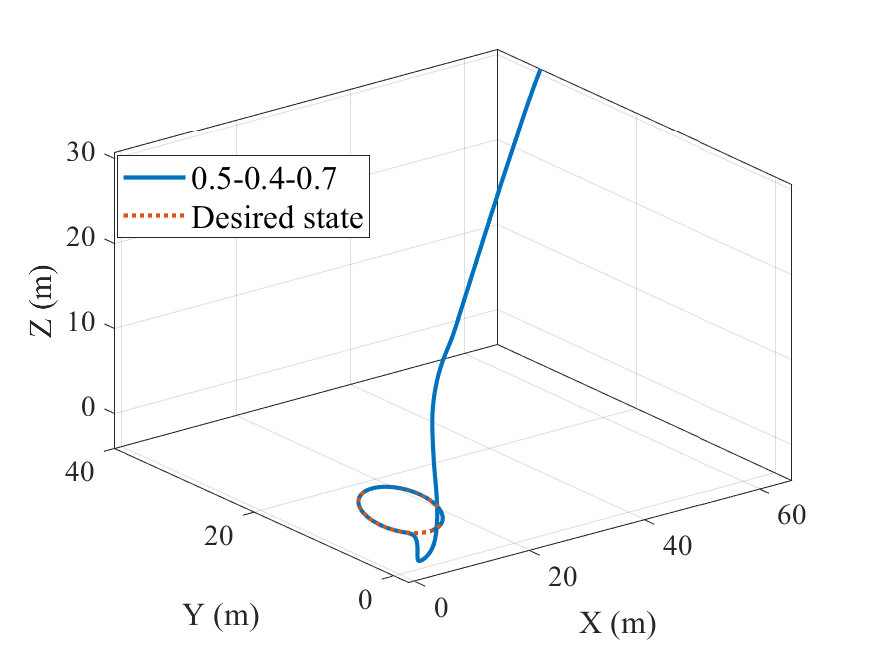}}}
\subfigure{\label{c2_e_xyz}}\addtocounter{subfigure}{-2}
\subfigure{\subfigure[Error described in LVLH by sampling based PWA MPC]{\includegraphics[width=0.325\textwidth]{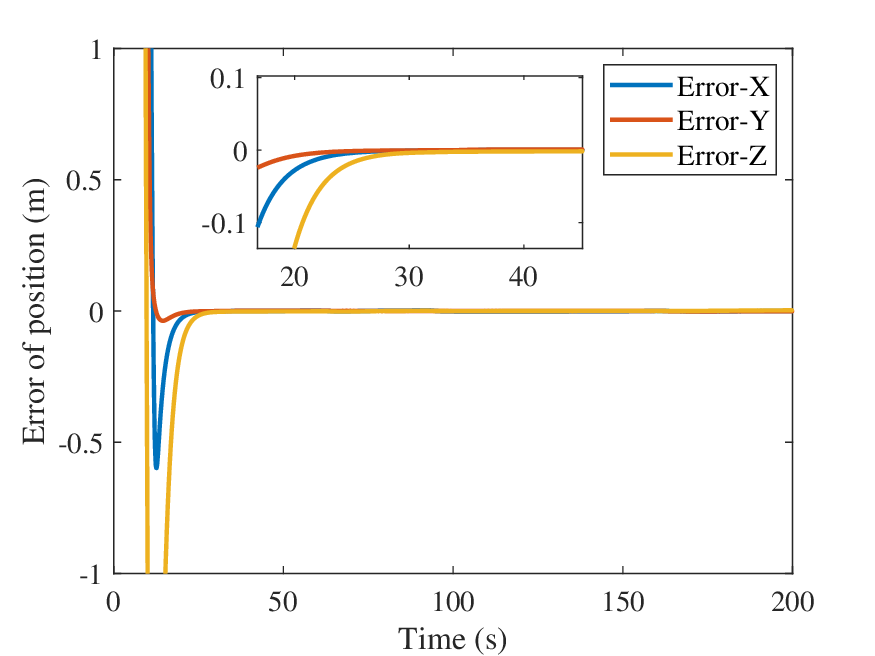}}}
\subfigure{\label{c2_u}}\addtocounter{subfigure}{-2}
\subfigure{\subfigure[sampling based PWA MPC]{\includegraphics[width=0.325\textwidth]{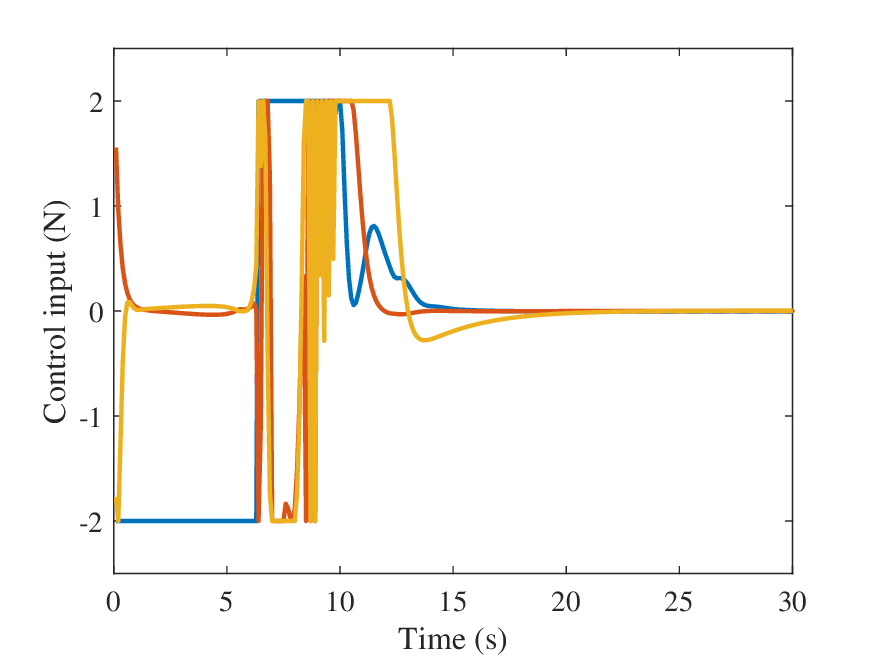}}}
\caption{Case 2: Position tracking in LVLH frame}
\label{Case2_LVLH}
\end{figure*}

In this case, all the RVD constraints are considered. 
According to the simulation, the standard PWA MPC in this case is unsolvable.
The reason is that the control sequence obtained by the standard PWA MPC is used to control the predictive model to track the desired states.
Once the control signal is used to compensate the model deviation caused by the accumulated predictive error, the actual control will conflict with strong RVD constraints.
Then the optimization will be unsolvable.
In this subsection, the performance under two different sets of the sampling parameter are provided, which are $0.5,\ 0.4,\ 0.7$ and $0.5,\ 0.8,\ 0.7$ are shown in Fig. \ref{Case2}. Fig. \ref{Case2_error} shows the tracking error which demonstrates the state constraints are satisfied. Fig.\ref{c2_xyz} and Fig.\ref{c2_e_xyz} shows the RVD process and the error in the LVLH frame, Fig.\ref{c2_u} demonstrates the input constraint is satisfied. Similar to the definition of the convergence time in case 1, the convergence time under all RVD constraints is shown in Table \ref{comparison}. The convergence accuracy is similar to case 1.

\begin{figure*}
\centering
\subfigure{\label{c2_a1}}\addtocounter{subfigure}{-2}
\subfigure{\subfigure[Roll tracking with singularity free]{\includegraphics[width=0.325\textwidth]{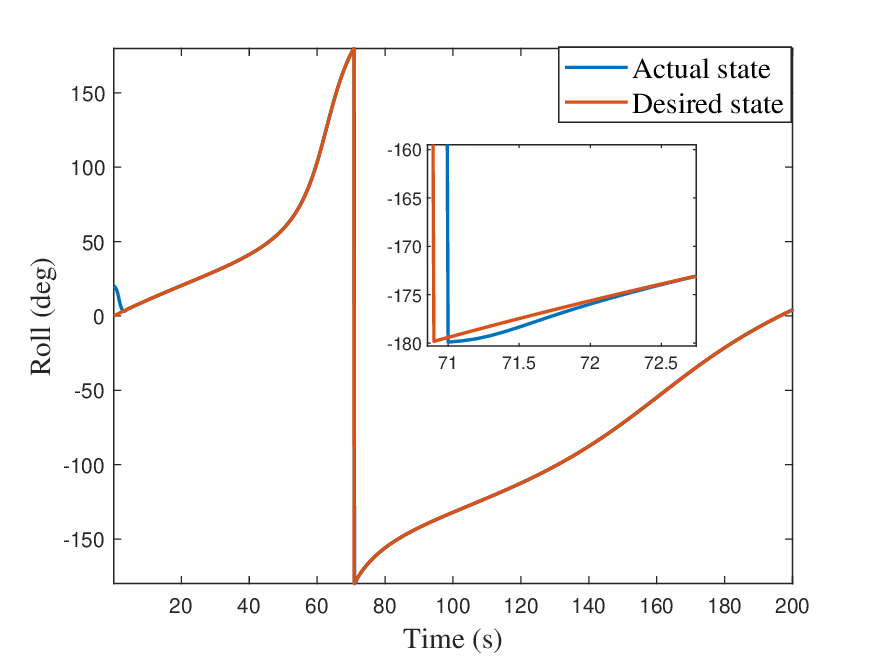}}}
\subfigure{\label{c2_a3}}\addtocounter{subfigure}{-2}
\subfigure{\subfigure[Pitch tracking with singularity free]{\includegraphics[width=0.325\textwidth]{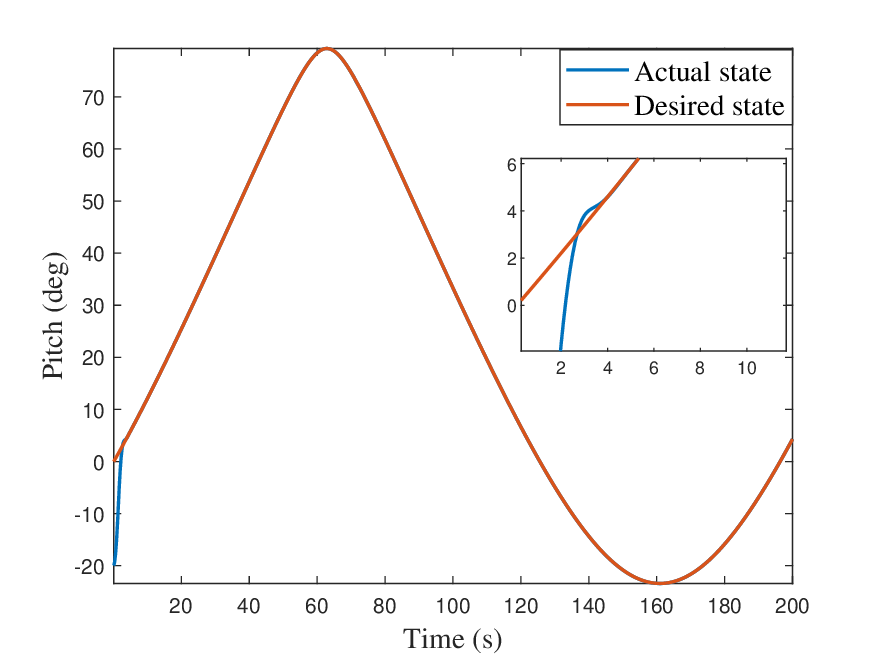}}}
\subfigure{\label{c2_a5}}\addtocounter{subfigure}{-2}
\subfigure{\subfigure[Yaw tracking with singularity free]{\includegraphics[width=0.325\textwidth]{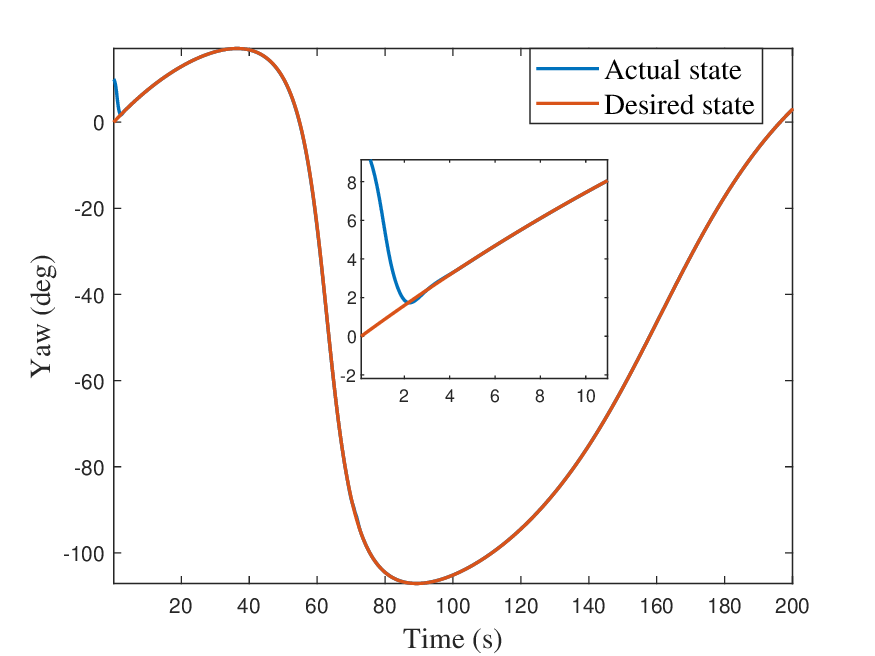}}}
\subfigure{\label{c2_a2}}\addtocounter{subfigure}{-2}
\subfigure{\subfigure[Roll tracking without singularity free]{\includegraphics[width=0.325\textwidth]{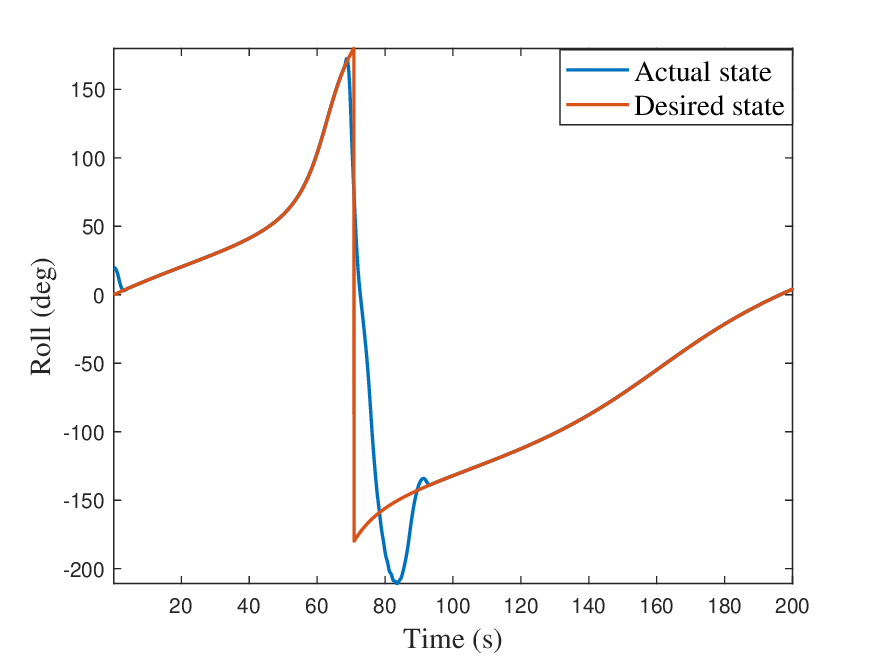}}}
\subfigure{\label{c2_a4}}\addtocounter{subfigure}{-2}
\subfigure{\subfigure[Pitch tracking without singularity free]{\includegraphics[width=0.325\textwidth]{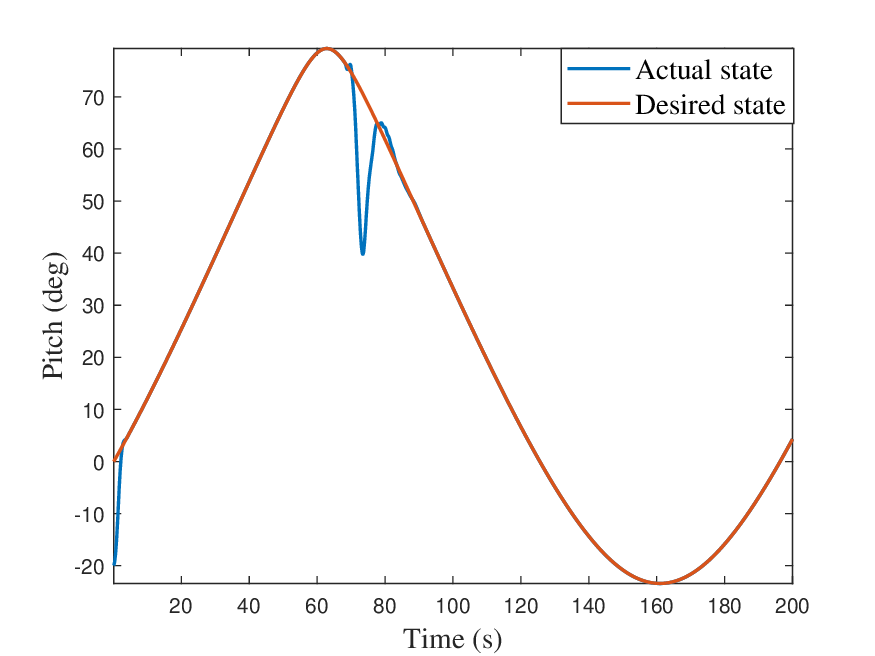}}}
\subfigure{\label{c2_a6}}\addtocounter{subfigure}{-2}
\subfigure{\subfigure[Yaw tracking without singularity free]{\includegraphics[width=0.325\textwidth]{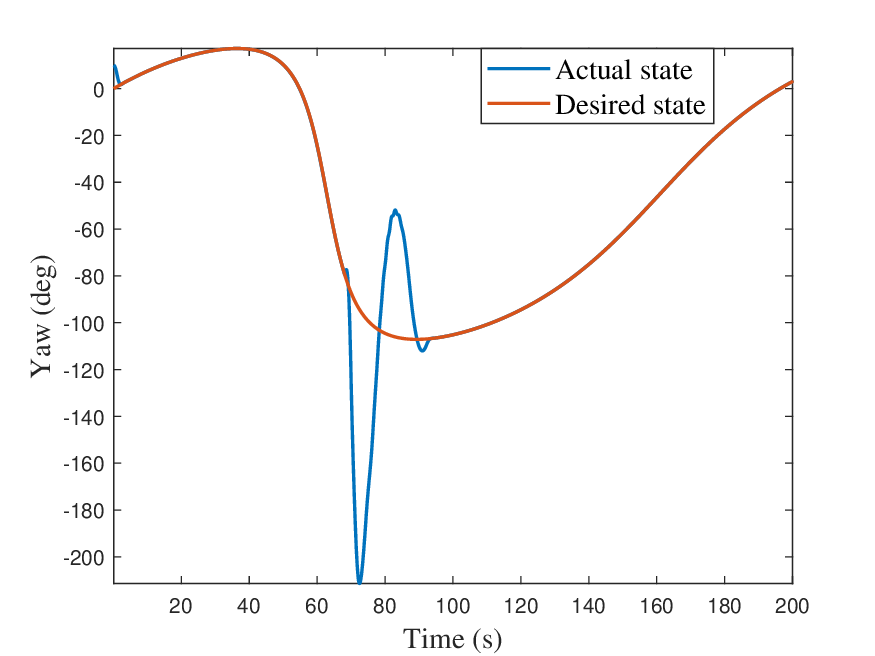}}}
\caption{Case 2: Performance of attitude tracking}
\label{case2_attitude}
\end{figure*}

\begin{figure*}
\centering
\subfigure{\label{c2_a_e}}\addtocounter{subfigure}{-2}
\subfigure{\subfigure[Error of $\varepsilon$]{\includegraphics[width=0.33\textwidth]{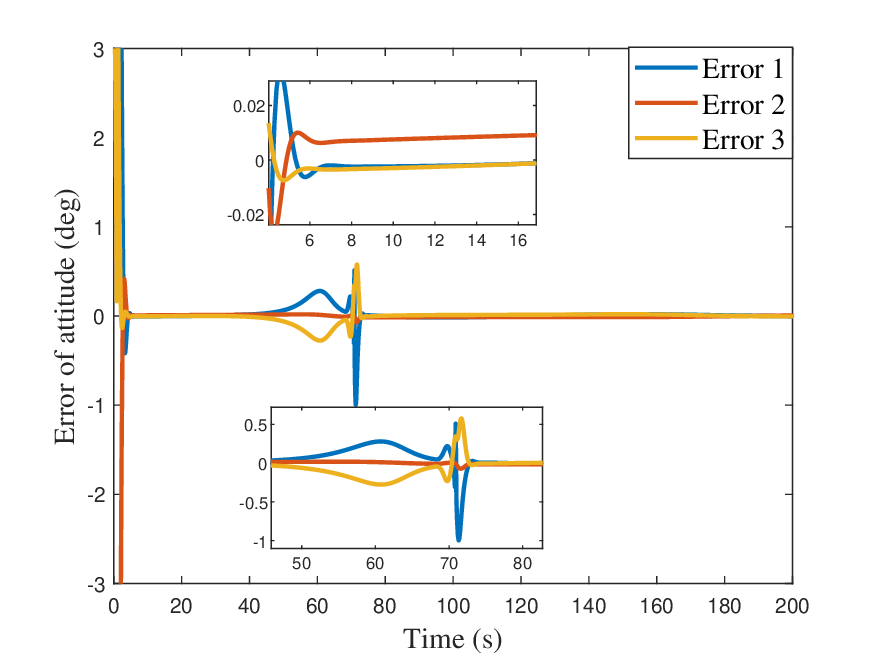}}}
\subfigure{\label{c2_au}}\addtocounter{subfigure}{-2}
\subfigure{\subfigure[Error of $\beta$]{\includegraphics[width=0.33\textwidth]{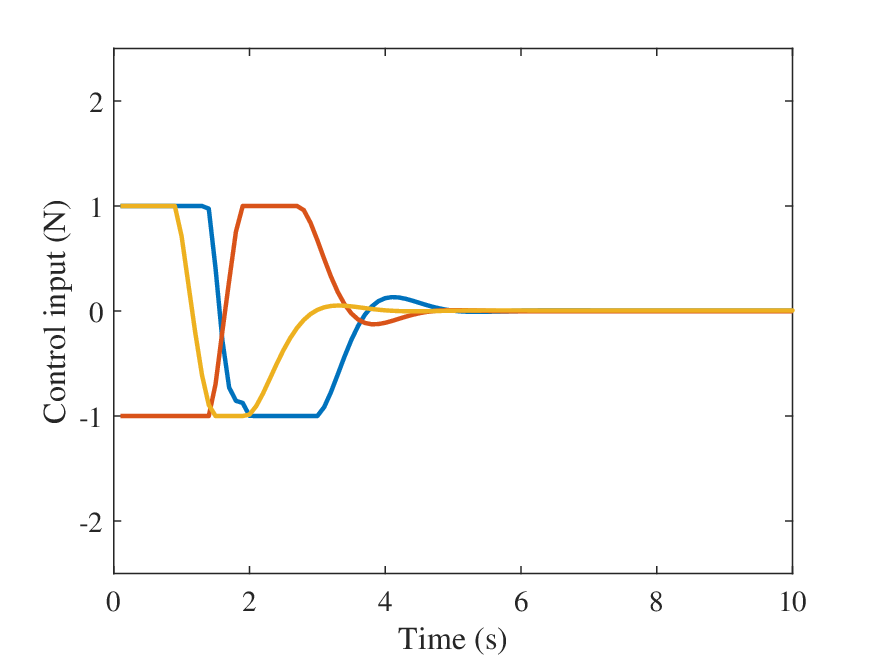}}}
\caption{Case 2: Error and control input in attitude tracking}
\label{case2_attitude_error}
\end{figure*}

Fig. \ref{case2_attitude} shows the attitude tracking performance with and without the proposed singularity free strategy, it can be concluded that the continuous attitude tracking can not be realized without the proposed method. Fig. \ref{case2_attitude_error} shows the tracking error and the control input. Since there exist switching of the mathematical expression of roll, the error which is lower than $1\ \mathrm{deg}$ is inevitable. Besides, the control input constraint is satisfied.

\begin{figure*}
\centering
\subfigure{\label{c3_p1}}\addtocounter{subfigure}{-2}
\subfigure{\subfigure[$\rho$ tracking]{\includegraphics[width=0.33\textwidth]{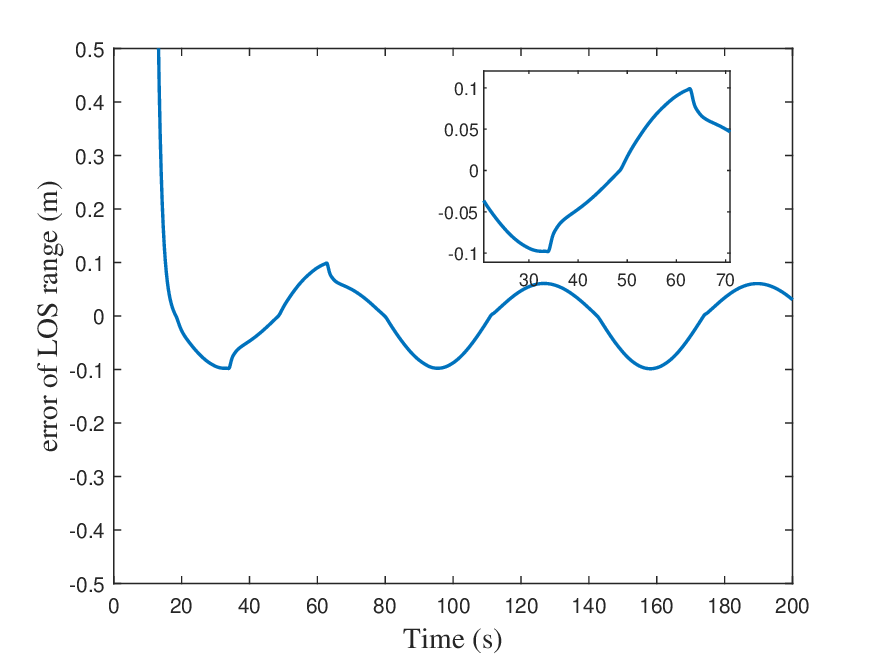}}}
\subfigure{\label{c3_p2}}\addtocounter{subfigure}{-2}
\subfigure{\subfigure[$\varepsilon$ tracking]{\includegraphics[width=0.33\textwidth]{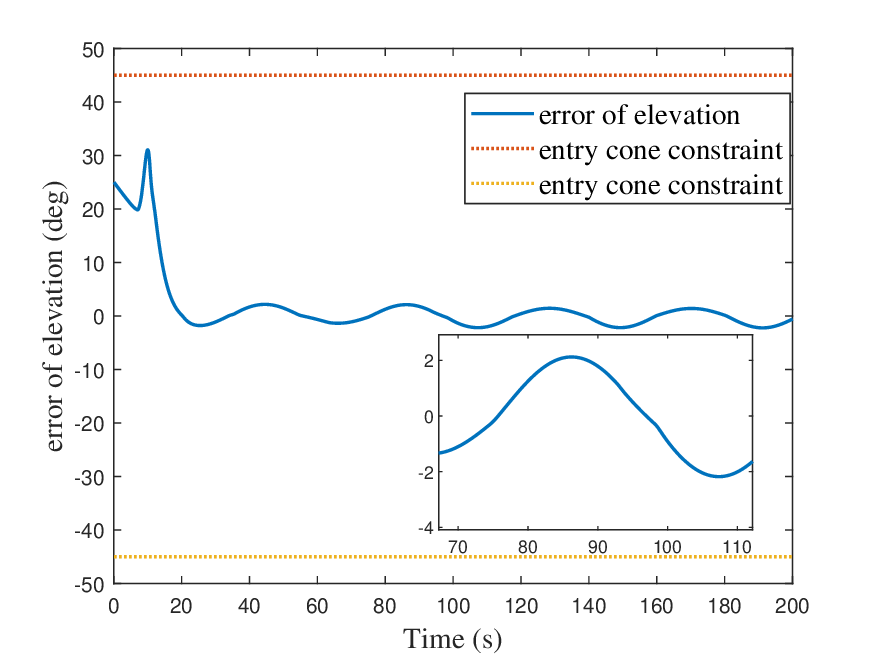}}}
\subfigure{\label{c3_p3}}\addtocounter{subfigure}{-2}
\subfigure{\subfigure[Elevation $\beta$ tracking]{\includegraphics[width=0.33\textwidth]{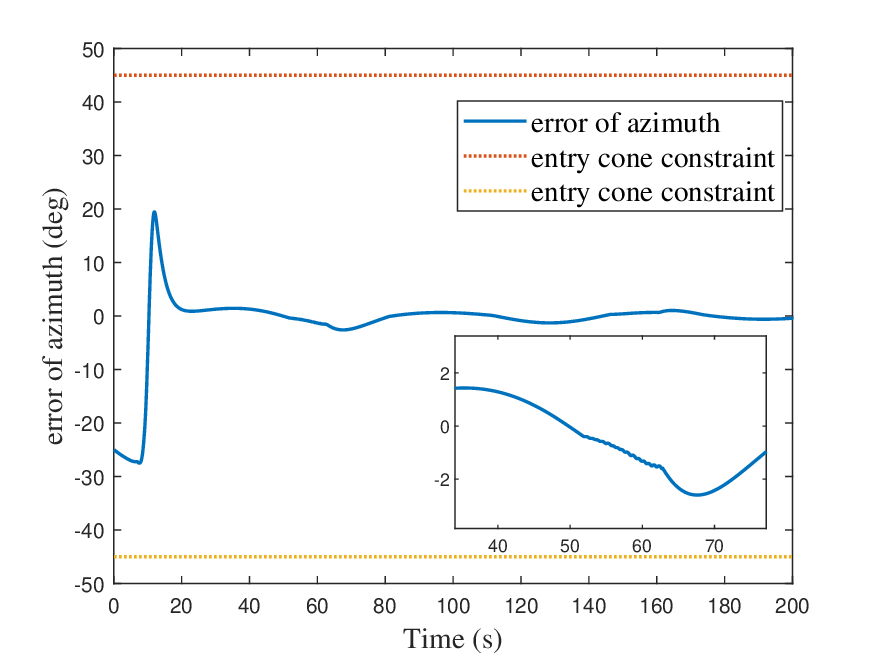}}}
\subfigure{\label{c3_p4}}\addtocounter{subfigure}{-2}
\subfigure{\subfigure[Error in LVLH frame]{\includegraphics[width=0.33\textwidth]{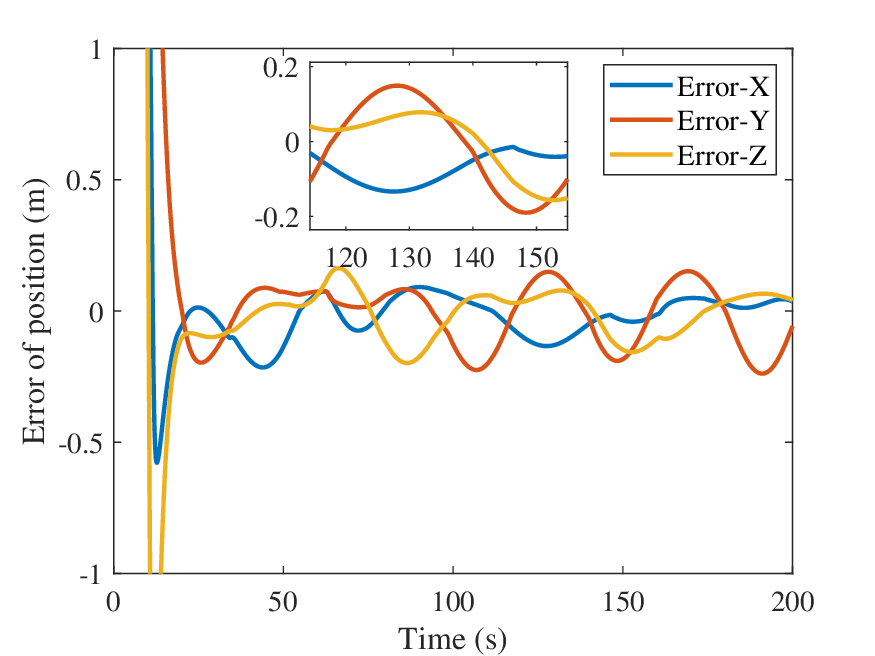}}}
\caption{Case 3: Performance of position tracking under disturbance}
\label{case3}
\end{figure*}

\subsection{Case 3: Docking under disturbance}

This case considers the disturbance. It should be noted that the purpose of the proposed strategy is to reduce the predictive error rather than improving robustness. There exist robust MPC just as tube-based MPC, this case is to present the convergence error is bounded under disturbance. In this case, the sampling parameter is set $(0.5,\ 0.4,\ 0.7)$. The disturbance is $[0.2 \cos(0.1t),\ 0.2 \sin(0.15t),\ 0.1 \sin(0.1t)]^{\mathrm{T}}$. From Fig. \ref{case3}, the tracking error is bounded. The detailed index is provided in Table \ref{comparison}.

Consider the real-time performance of the proposed sampling-based strategy, the simulation time is set $200 \ \mathrm{s}$ in each case. In case 1, the running time of the standard PWA MPC is $46 \ \mathrm{s}$, and the running time of the proposed approach is $45 \ \mathrm{s}$, including using ode45 to simulate the control process of nonlinear system. Therefore, the real-time is satisfied under the control input constraint. In case 2, the optimization in the standard PWA MPC is unsolvable, and the running time of the proposed approach is $100 \ \mathrm{s}$, including using ode45 to simulate the control process of nonlinear system. Therefore, the real-time is satisfied under all RVD constraints. Besides, the matrix $Q$ is used to adjust the convergence accuracy of each state. Within a specific range, the larger the value of the component, the higher the convergence accuracy. However, the dynamic performance of the system is affected if the value in $Q$ exceeds the range. Matrix $P$ represents the emphasis on hoping to complete the tracking task with a smaller input cost.

In summary, under meeting the requirements of real-time control, the proposed controller achieves lower overshoot and faster convergence than the standard PWA MPC, considering the input constraint.
The proposed method realizes optimal control under all position-attitude coupled constraints, and the standard PWA MPC is unsolvable.
The advantage of the proposed method origin from the sampling input compensates the model deviation between the predictive model and the control model.
Besides, consider the bounded disturbance, the robustness of the proposed controller is satisfied.
In addition, the singularity free strategy is  proved to cross the singularities of the Euler attitude angles.



\section{Conclusion}
This paper proposes a novel LOS-Euler RVD framework to improve the matching with the relative navigation system and MPC.
The 6 DoF information interactions among the sensor measurements, controlled states, and RVD constraints need no transformation and linearization.
The position-attitude couplings in the target observation and field of view control are linearly descirbed while docking with a tumbling target.
Besides, a singularity free strategy is provided to achieve continuous attitude tracking by crossing the singularity of angle states. 
The proposed sampling-based PWA MPC improves the dynamical performance and settles unsolvable optimization caused by the accumulated predictive error. 
The numerical simulations illustrate the effectiveness of the above approaches.

Future work based on this paper may include: (i) improvement of the robustness of the sampling-based PWA MPC; (ii) since the sampling parameters are manually selected to verify the validity of the proposed sapling-based idea, an adaptive mechanism to adjust the sampling parameter automatically is needed in future studies.


\bibliography{sbMPC}%

\clearpage

\begin{center}
\begin{table}[b]%
\centering
\caption{AR\&D conditions}
\label{conditions}%
\begin{tabular*}{300pt}{@{\extracolsep\fill}lcc@{\extracolsep\fill}}%
\toprule
\textbf{Parameters} & \textbf{Value}  \\
\midrule
Initial position state &  {$\rho=80\ \mathrm{m}$, $\varepsilon =25\ \mathrm{deg}$, $\beta = -25\ \mathrm{deg}$
                          }\\
Initial velocity state     &          $\dot{\rho}=0\ \mathrm{m/s}$, $\dot{\varepsilon} =\dot{\beta}=0\ \mathrm{deg/s}$ \\          
Control inputs constraints & {$\pmb{u}_p^{\max}=[2,\ 2,\ 2]^\mathrm{T}$, $\pmb{u}_a^{\max}=[1,\ 1,\ 1]^\mathrm{T}$}  \\
Keep-out zone & {$r_{\mathrm{safe}}=5\ \mathrm{m}$ } \\
Entry cone constraint  & {$\gamma_e = 45 \ \mathrm{deg}$}\\
Field of view constraint & {$\gamma_f = 30 \ \mathrm{deg}$}\\
\bottomrule
\end{tabular*}
\end{table}
\end{center}

\begin{center}
\begin{table}[b]%
\centering
\caption{RVD performance comparison}
(Note: c.t. denotes the convergence time; c.a. denotes the convergence accuracy.)
\label{comparison}
\begin{tabular*}{500pt}{@{\extracolsep\fill}lcc@{\extracolsep\fill}}%
\toprule
\textbf{Parameters}  & \textbf{Sampling-based PWA MPC} & \textbf{Standard PWA MPC}\\
\midrule
{c.t. of ($\rho$, $\varepsilon$, $\beta$)} &  {(15.3, 20.5, 19.8) $\mathrm{s}$ } & {(12, 17.8, 21.8) $\mathrm{s}$}\\
{c.t. of  ($\dot{\rho}$, $\rho \dot{\varepsilon}$, $\rho \dot{\beta}$)} &  {(12.6, 13.5, 12.2) $\mathrm{s}$ } & {(12, 17.8, 21.8) $\mathrm{s}$}\\
{c.a. of  ($\rho$, $\varepsilon$, $\beta$)} &  {(5.79 $\times 10^{-4} \ \mathrm{m}$, 0.0062 \ $ \mathrm{deg}$, 0.0206 \ $ \mathrm{deg}$) } & {(5$\times 10^{-4} \ \mathrm{m}$,  $0.0193\ \mathrm{deg}$, 0.0127\ $ \mathrm{deg}$)}\\
{Modulus of overshoot ($\rho$, $\varepsilon$, $\beta$)} &  {(3.43 $\mathrm{m}$, 19.78 $ \mathrm{deg}$, 0 $ \mathrm{deg}$) } & {(3.02 $ \mathrm{m}$, 30.69 $ \mathrm{deg}$, 87.1 $ \mathrm{deg}$)}\\
{c.t. of  ($x$, $y$, $z$)$<0.1 \mathrm{m}$} &  {(15.1, 15.2, 14.5) $\mathrm{s}$ } &{(19.6, 20.2, 24.1) $\mathrm{s}$ }\\
{c.a. of  ($x$, $y$, $z$)} & {(0.0015, 0.0017, 0.0017) $\mathrm{m}$}  & {(0.0016, 0.0020, 0.0019) $\mathrm{m}$} \\
{c.t. of ($\rho$, $\varepsilon$, $\beta$) under RVD constraints} &  {(14.6, 19.7, 27) $\mathrm{s}$ } & {unsolvable}\\
{c.t. of ($\dot{\rho}$, $\rho \dot{\varepsilon}$, $\rho \dot{\beta}$) under RVD constraints }& {(14.4, 21.8, 28.7) $\mathrm{s}$ } & {unsolvable}\\
{c.t. of ($x$, $y$, $z$) under RVD constraints }& {(16.9, 11.5, 20.7) $\mathrm{s}$ } & {unsolvable}\\
{c.a. of ($\rho$, $\varepsilon$, $\beta$) under disturbance }& {(0.099 \ $\mathrm{m}$, 2.217 $ \mathrm{deg}$, 2.597 $ \mathrm{deg}$) } & {unsolvable}\\
{c.a. of ($x$, $y$, $z$) under disturbance }& {(0.215, 0.239, 0.198) $\mathrm{m}$ } & {unsolvable}\\
\bottomrule
\end{tabular*}
\end{table}
\end{center}

\end{document}